\documentclass[aps,prx,nofootinbib,twocolumn,superscriptaddress,longbibliography]{revtex4-2}
\usepackage{times}
\usepackage{hyperref}
\usepackage{graphicx}  
\usepackage{epstopdf}
\usepackage{mathtools}                    
\usepackage{amsfonts,amsthm,amssymb,latexsym,amsmath}
\usepackage{mathrsfs}
\usepackage{bbm}
\usepackage{units}  
\usepackage{textcomp} 
\usepackage{hhline}
\usepackage{multirow}   
\usepackage{makecell}
\usepackage{tabularx}
\usepackage{booktabs}
\usepackage{hyperref}
\hypersetup{
colorlinks = true,
linkcolor=magenta,
citecolor=[rgb]{0.15,0.65,0.13},
urlcolor = [rgb]{0.25, 0.41, 0.88}}
\usepackage{float}
\usepackage{cleveref}
\usepackage{tikz}
\usepackage{booktabs}
\usepackage{colortbl}
\usetikzlibrary{decorations.pathreplacing,shapes.misc, positioning,calc}

\makeatletter
\def\l@subsection#1#2{}
\def\l@subsubsection#1#2{}
\makeatother

\def\ket#1{\mathinner{|{#1}\rangle}}

\def\inner#1{\mathinner{\langle{#1}\rangle}}
\def\Inner#1{\mathinner{\left\langle{#1}\right\rangle}}
\def\bs#1{\boldsymbol{#1}}
\def\tbs#1{\tilde{\boldsymbol{#1}}}
\def\hbs#1{\hat{\boldsymbol{#1}}}

\newcommand{\GS}{\text{GS}}

\newcommand{\eqgs}{\mathrel{\mathop{=}\limits_{\text{GS}}}}
\newcommand{\proptogs}{\mathrel{\mathop{\propto}\limits_{\text{GS}}}}

\newcommand{\ob}{obstructor invariant}
\newcommand{\obs}{obstructor invariants}

\def\ZZ{\mathbb Z}

\tikzset{ vertex/.style={fill,circle,draw,scale=0.3}}

\hypersetup{
colorlinks = true,
linkcolor=magenta,
citecolor=[rgb]{0.15,0.65,0.13},
urlcolor = [rgb]{0.25, 0.41, 0.88}}

\begin{document}

\title{
A holographic view of topological stabilizer codes
}

\author{Thomas Schuster}
\affiliation{Walter Burke Institute for Theoretical Physics and Department of Physics, California Institute of Technology, Pasadena, CA 91125, USA}
\affiliation{Department of Physics, University of California, Berkeley, California 94720 USA}
\author{Nathanan Tantivasadakarn}
\affiliation{Walter Burke Institute for Theoretical Physics and Department of Physics, California Institute of Technology, Pasadena, CA 91125, USA}
\affiliation{Microsoft Quantum, Station Q, Santa Barbara, CA 93106, USA}
\affiliation{Department of Physics, Harvard University, Cambridge, MA 02138, USA}
\author{Ashvin Vishwanath}
\affiliation{Department of Physics, Harvard University, Cambridge, MA 02138, USA}
\author{Norman Y. Yao}
\affiliation{Department of Physics, Harvard University, Cambridge, MA 02138, USA}

\date{\today}

\begin{abstract}
The bulk-boundary correspondence is a hallmark feature of topological phases of matter.
Nonetheless, our understanding of the correspondence remains incomplete for phases with intrinsic topological order, and is nearly entirely lacking for more exotic phases, such as fractons.
Intriguingly, for the former, recent work suggests that bulk topological order  manifests in a non-local structure in the boundary Hilbert space; however, a concrete understanding of how and where this perspective applies remains limited. 
Here, we provide an explicit and general framework for understanding the bulk-boundary correspondence in Pauli topological stabilizer codes.
We show---for any boundary termination of any two-dimensional topological stabilizer code---that the boundary Hilbert space cannot be realized via local degrees of freedom, in a manner precisely determined by the anyon data of the bulk topological order.
We provide a simple method to compute this ``obstruction'' using a well-known mapping to polynomials over finite fields.
Leveraging this mapping, we generalize our framework to fracton models in three-dimensions, including both the X-Cube model and Haah's code. 
An important consequence of our results is that the boundaries of topological phases can exhibit emergent symmetries that are impossible to otherwise achieve without an unrealistic degree of fine tuning. For instance, we show how linear and fractal subsystem symmetries naturally arise at the boundaries of fracton phases. 
\end{abstract}

\maketitle 

\tableofcontents

Topological phases are stable, gapped phases of matter that do not exhibit a local order parameter---rather, they are  distinguished by the structure and pattern of their entanglement~\cite{Wenbook,sachdevbook}. 
Their classification has seen tremendous activity in the last decade, leading to a broad landscape that includes: symmetry-protected topological (SPT) phases~\cite{Gu09,Pollmann10,Fidkowski_2011,Turner11,Schuch11,ChenGuWen11A,ChenGuWen11B,Chen_2011,Chen_2013,pollmann_symmetry_2012,LuVishwanath12,SenthilLevin13,Levin_2012,VishwanathSenthil2013,ElseNayak2014,SenthilSPTreview}, intrinsic topological order~\cite{Read91,Wen_90,Fuchs02,Kitaev_2003,kitaev_anyons_2005} (both Abelian~\cite{leinaas_theory_1977,Goldin81,wilczek_quantum_1982} and non-Abelian~\cite{Goldin85,Moore89,moore_nonAbelions_1991,Wen91}), and fracton phases~\cite{Chamon2005,Haah2011,Yoshida2013,VijayHaahFu2015,VijayHaahFu2016,NandkishoreRev2019,PretkoRev2020}.  
A multitude of features distinguish between these classes, ranging from ground-state degeneracies to the nature of their excitations.

Nevertheless, a unifying expectation for all topological phases is that some form of a  \emph{bulk-boundary correspondence} should hold---i.e.~a general statement that any universal property of the bulk, can equivalently be probed by merely having access to the boundary~\cite{TIReview,pollmann_symmetry_2012,ChenGuWen11B,Kong14,Kong15,Kong17}. 
Despite this expectation, the precise nature of such a bulk-boundary correspondence has only been fully established for SPT phases~\cite{callan1985anomalies,ElseNayak2014,WangLinLevin16,TiwariChenShiozakiRyu18,Lu17,JiangChengQiLu19,ElseThorngren20,Burnell22,hsieh2022anomaly,witten2022anomaly,ChengSeiberg23}.
For intrinsic topological order, it is well-known that the statistics of bulk excitations determine the possible \emph{gapped} phases on the boundary~\cite{Kitaev2003,beigi2011quantum}. 
However, by restricting to gapped boundaries, this prescription does not provide a full correspondence between the bulk and the boundary theory. 
To this end, early conjectures identified a possible general correspondence, where the bulk topological order leads to an 
``obstruction'' on the boundary; in particular, it is impossible to realize the boundary theory using only local degrees of freedom (i.e. a local tensor product Hilbert space)~\cite{JiWen19}.
Recent work has verified this conjecture in the toric code, the simplest stabilizer model exhibiting topological order~\cite{Lichtmanetal20,JiWen19,JiWen20}.
However, proving this correspondence  as a general feature of intrinsic topological order remains an essential open question. 
Finally, in the context of fracton phases, to date, no explicit bulk-boundary correspondence has even been conjectured.

Our main results are three fold. 
First, we prove the bulk boundary correspondence conjectured above
for all stabilizer models exhibiting topological order. 
Our proof does not reference a particular choice of boundary, and only relies on certain universal invariants  of the boundary operator algebra, which are directly inherited from the bulk topological order. 
Second, for translationally-invariant stabilizer models, we introduce a simple framework to analyze the boundary theory  and to compute these universal ``obstructor'' invariants by using a well-known mapping to polynomials over finite fields~\cite{Haah2013}.
Crucially, this framework immediately allows us to generalize the conjecture to fracton phases. 
In particular, we provide an explicit bulk-boundary correspondence for both the X-cube model as well as Haah's code. 
We show that the bulk fracton order leads to emergent subsystem symmetries in the boundary theory, whose precise form is determined by the exchange statistics and mobilities of bulk excitations.
To the best of our knowledge, these represent the first examples of a complete bulk-boundary correspondence for fracton phases.

\section{Background and Summary of results}

Before proceeding to a full summary of our results, we first provide a brief overview of existing results on bulk-boundary correspondences in various classes of topological phases.

\subsection{Review of previous results}

\textit{Symmetry-protected topological phases.---}The bulk-boundary correspondence is perhaps best understood in the context of SPT phases.
SPTs are gapped phases of matter whose ground states can be adiabatically connected to product states by general unitary operations, but \emph{cannot} be adiabatically connected by unitary operations preserving a given symmetry~\cite{Gu09,Pollmann10,Fidkowski_2011,Turner11,Schuch11,ChenGuWen11A,ChenGuWen11B,Chen_2011,Chen_2013,pollmann_symmetry_2012,LuVishwanath12,SenthilLevin13,Levin_2012,VishwanathSenthil2013,ElseNayak2014}. 
The bulk-boundary correspondence in SPT phases is related to the action of the symmetry on the boundary of the system.
For example, in one-dimensional SPTs, each boundary of the system is acted upon by a \emph{projective} representation of the symmetry group~\cite{pollmann_symmetry_2012,ChenGuWen11B}.
This can be generalized to higher-dimensional SPTs, where the symmetry acts on the boundary in such a way that it cannot be consistently coupled to a dynamical gauge field. (This is understood more generally as an 't Hooft anomaly~\cite{thooft1980}.) 
This action of the symmetry on the boundary has important consequences for the boundary physics, as it places constraints on the allowed boundary Hamiltonians, which must respect the symmetry.
Perhaps the most notable example of this is the topological insulator, which exhibits gapless edge modes protected by charge conservation and time-reversal symmetry~\cite{TIReview,Kong14,Kong15,Kong17}.

\textit{Topological order in two dimensions.---}The bulk-boundary correspondence of intrinsic topological order is comparatively less understood than that of SPTs.
Nonetheless, a number of key features have been identified.
A central conjecture is that the bulk topological order leads to emergent symmetries in the boundary theory; more precisely, there exists a one-to-one correspondence between the set of allowed boundary Hamiltonians and the set of Hamiltonians that obey the emergent symmetry\footnote{This is referred to variously as ``categorical symmetry"~\cite{JiWen19,JiWen20,Kong20,KongZheng20,KongZheng21,JiWen21,chatterjee2022algebra,chatterjee2022holographic,chatterjee2022emergent} or SymTFT/SymTO~\cite{Lichtmanetal20,moradi2022topological,freed2022Ising,apruzzi2021symmetry,Freed2022topological,Bhardwaj23,Inamura2023} in the literature.}~\cite{freed2014relative,Kong14,Kong15,Kong17,JiWen19,JiWen20,Kong20,KongZheng20,KongZheng21,JiWen21,chatterjee2022algebra,chatterjee2022holographic,chatterjee2022emergent,Lichtmanetal20,moradi2022topological,freed2022Ising,apruzzi2021symmetry,Freed2022topological,Bhardwaj23,Inamura2023}.
The translation between the bulk topological order and the emergent boundary symmetry is known in many cases~\cite{Kong14,Kong15,Kong17}.
However, the correspondence has only been verified explicitly in a handful of microscopic models~\cite{Feiguin07,albert2021spin,Jones23,Jones2023local}.
Indeed, recent work suggests that some intrinisic topological orders may not fit into this framework (specifically, those with boundaries that cannot be gapped by any local perturbation)~\cite{Levin13,GaneshanLevin22}.

Building upon these ideas, several recent works have explored the implications of this correspondence for the structure of the boundary Hilbert space.
In particular, it is conjectured that the emergent symmetries arising from the bulk topological order imply that the boundary Hilbert space does not admit a local tensor product description~\cite{JiWen19}.
This has been analyzed for a particular boundary termination of toric code, which is easily shown to be equivalent to a local tensor product space augmented with a non-local ``Ising symmetry'' constraint (see Section~\ref{sec:TCreview} for a detailed review).
However, extending this mapping to more general models and boundary terminations, as well as proving that the boundary Hilbert space \emph{must not be} a local tensor product, remain open directions.

\textit{Topological order in higher dimensions.---} Even more open questions abound in higher dimensions. For example, the enumeration of all possible gapped boundaries for the simplest three-dimensional topological order, the toric code, remains an active direction of research~\cite{zhao2022string,luo2022gapped,Ji2022boundary}. As another example, one class of three-dimensional models where the bulk-boundary correspondence is particularly simple are the Walker-Wang models~\cite{WalkerWang2012}, where the three-dimensional topological order is explicitly constructed to give rise to some desired anyon theory on the boundary. Several additional interesting models have been constructed, and analyzed, using this approach~\cite{vKBS13,BCFV14,WangChen17,HaahFidkowskiHastings18,Shirley22}.

\textit{Fracton phases.---}As for other higher-dimensional topological orders, the bulk-boundary correspondence for fracton orders has remained relatively unknown.
There have been several studies of the possible gapped boundary Hamiltonians that can be realized for specific  terminations of the X-Cube and Chamon models~\cite{BulmashIadecola19,Luo22Xcubeboundary,Fontana23}, as well as further analysis of logical encodings in  Haah's code~\cite{Aitchison2023}.
In particular, Ref.~\cite{Luo22Xcubeboundary} and \cite{cao2023symmetry} finds that there exist emergent \emph{subsystem} symmetry constraints on the (100)-boundary of the X-Cube and 2-foliated fracton models respectively, and Ref.~\cite{LiuJi22} finds a similar connection when constructing bulk models given a desired 2D boundary theory. % (an analogue of Walker-Wang models for fractons).
However, a complete understanding of the bulk-boundary correspondence, even within these models, remains lacking.

\subsection{Summary of main results}

We now summarize our main results, in order of their appearance in the remainder of the paper.

We begin in Section~\ref{sec:TCreview} by reviewing known results for the bulk-boundary correspondence in the 2D toric code model~\cite{JiWen19}.
In particular, we describe the known mapping between the Hilbert space of the smooth toric code boundary and the symmetric sector of a 1D transverse-field Ising model.
We review how the ``emergent'' Ising symmetry on the boundary arises from a \emph{conservation law} of the bulk topological order, i.e.~a set of stabilizers that product to the identity in the bulk.
The termination of the conservation law on the boundary gives rise to a \emph{constraint} on the boundary operator algebra, which is interpreted as an emergent symmetry.

Inspired by this example, in Section~\ref{sec: toric boundary} we introduce our framework for the bulk-boundary correspondence in two-dimensional topological stabilizer codes.
Our framework applies quite broadly, since recent work has shown that two-dimensional stabilizer codes can realize any Abelian topological order with a gappable boundary~\cite{Ellison22}.
To do so, we extend the toric code analysis above to general two-dimensional topological stabilizer codes.
We show that each anyon in the bulk topological order is associated with a stabilizer conservation law, which gives rise to a non-local constraint on the boundary operator algebra.
To analyze the non-local structure of the boundary theory, we show that certain features of boundary operator algebra---corresponding to the commutation relations of these constraints---cannot be realized in any local tensor product Hilbert space.
We formalize this by defining ``obstructor invariants'' for each pair of  boundary constraints.
We show that the obstructor invariants are in one-to-one correspondence with the braiding statistics of anyons in the bulk topological order, and prove that any non-zero obstructor invariant implies that the boundary cannot be locally realized.

In Section~\ref{sec:poly}, we provide a different perspective on this bulk-boundary correspondence by utilizing a mapping~\cite{Haah2013} from translation-invariant stabilizer models to polynomials over finite fields.
We show that the bulk conservation laws correspond to zeros of associated polynomials, and similar for the boundary constraints (in particular, for a polynomial associated with the frustration graph of local boundary operators).
The obstructor invariants correspond to first derivatives of the polynomials, at the location of the zeros.

Finally, in Sections~\ref{sec: XCube} and~\ref{sec:fractal}, we extend our framework to three-dimensional fracton models.
In Section~\ref{sec: XCube}, we describe the bulk-boundary correspondence for Type-I fracton orders, focusing on the X-Cube model for concreteness.
We show that the bulk conservation laws give rise to linear ``subsystem'' constraints on the boundary operator algebra, which can be interpreted as emergent subsystem symmetries in the boundary theory.
Analogous to the two-dimensional setting, we define obstructor invariants associated to these constraints, and show that they are in one-to-one correspondence with the mutual~\cite{PaiHermele2019} and self~\cite{song2023fracton} statistics of the bulk fractons.
Intriguingly, we find that the structure of the boundary Hilbert space depends on the \emph{orientation} of the boundary considered; for example, the bulk self statistics lead to an obstructor invariant on the (111)-boundary, but not on the (001)- or (110)-boundaries.
In analogy to our results for two-dimensional stabilizer models, we prove that any non-trivial obstructor invariant implies that the boundary Hilbert space cannot be realized as a tensor product of \emph{one-dimensional Hilbert spaces}.
This provides a sharp distinction between the boundaries of Type-I fracton phases and, for example, the boundary of a stack of 2D toric codes.
Lastly, within the polynomial formalism, we show that the obstructor invariants correspond to \emph{multivariate} derivatives of associated polynomials.

In Section~\ref{sec:fractal}, we consider topological stabilizer codes with \emph{fractal} conservation laws, which include seminal Type-II fracton phases such as Haah's code~\cite{Haah2011}.
%and fractal spin liquids~\cite{Yoshida2013}.
Making heavy use of the polynomial formalism, we show that all of the core ideas of the previous sections generalize to these models.
In particular, the bulk conservation laws can be formulated using the mathematical notion of an \emph{ideal}, and similar for the constraints they impose upon the boundary.
These constraints can be viewed as emergent \emph{fractal} subsystem symmetries in the boundary theory.
The obstructor invariants, in turn, are related to quotients over these ideals.
We illustrate this explicitly in a large class of fracton stabilizer codes~\cite{Yoshida2013}, and, within a simpler subset of these codes, we show that the obstructor invariants are related to the exchange statistics of bulk fractons.
We conclude in Section~\ref{sec:outlook} with prospects for future work.

\section{Review of the toric code boundary}\label{sec:TCreview}

We begin by reviewing the boundary Hilbert space of the toric code model~\cite{Kitaev2003,JiWen19,JiWen20}.
In Sec.~\ref{sec: overview toric} we introduce the toric code model. 
In Sec.~\ref{sec: toric code boundary}, we turn to the boundary Hilbert space and review arguments that it does not obey a local tensor product structure~\cite{JiWen19}.

\subsection{Toric code model}\label{sec: overview toric}

\begin{figure}
    \centering
    \includegraphics[width=\columnwidth]{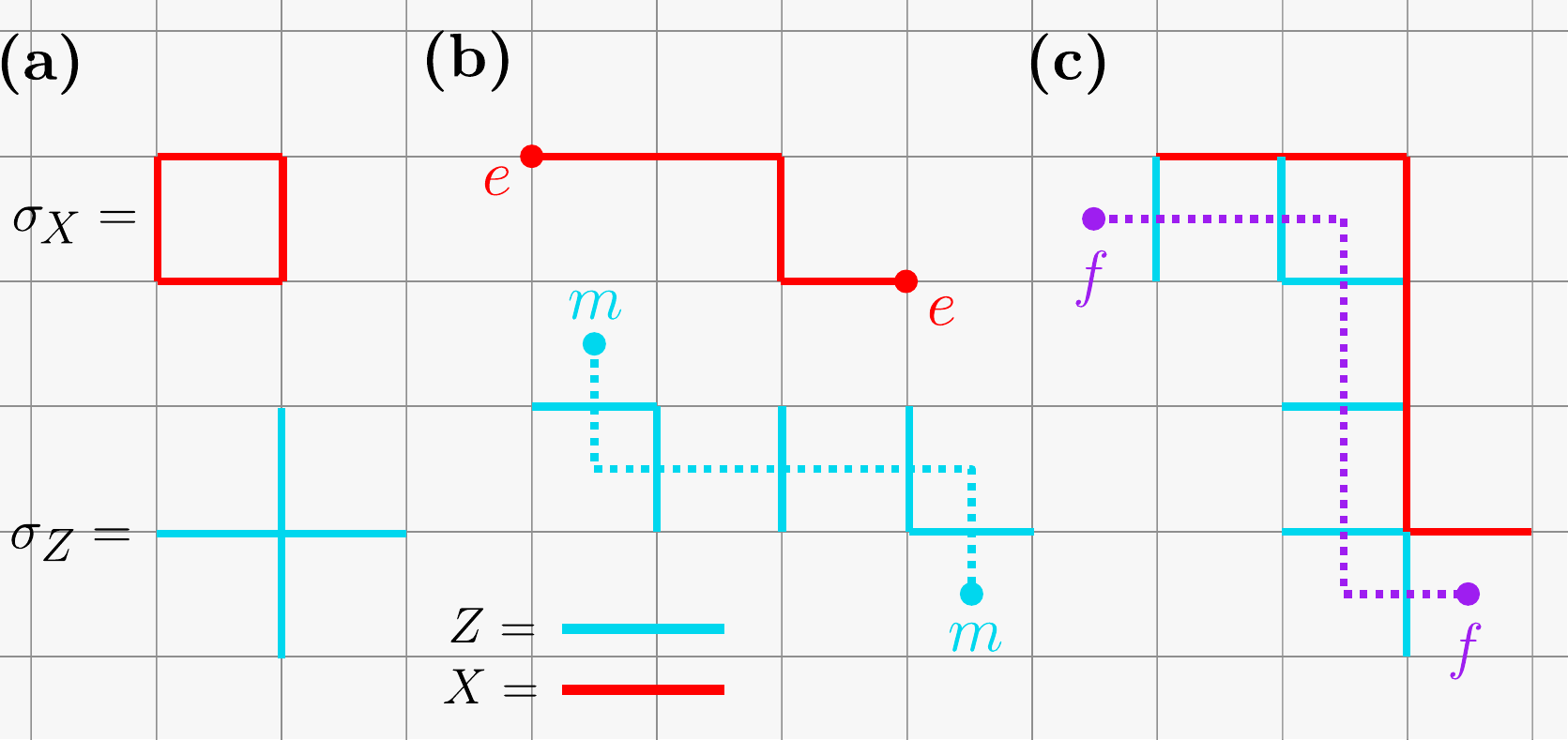}
    \caption{\textbf{(a)} Plaquette (red) and vertex (blue) stabilizers of the toric code. \textbf{(b)} String operators that violate the vertex and plaquette terms correspond to $e$ and $m$ quasiparticles, respectively. \textbf{(c)} The bound state of one $e$ and one $m$ quasiparticle corresponds to the fermionic quasiparticle $f$.}
    \label{fig: toric bulk}
\end{figure}

The toric code is defined on a 2D square lattice with spins residing on each bond (Fig.~\ref{fig: toric bulk}).
The spins are labelled by their unit cell, $\bs{i} = (i_x,i_y) \in \ZZ^2$.%,
The Hamiltonian takes the form:
\begin{equation}
    H_{TC} = - \sum_{\bs{i}} \sigma^{Z}_{\bs{i}} - \sum_{\bs{i}} \sigma^{X}_{\bs{i}},
    \label{eq:TCHam}
\end{equation}
where we define the 4-spin vertex and plaquette operators, $\sigma^{Z}_{\bs{i}}$ and $\sigma^{X}_{\bs{i}}$, as:
\begin{equation}
\begin{split}
    \sigma^Z_{\bs{i}} & = Z_{\bs{i}-\hbs{x},x} Z_{\bs{i}-\hbs{y},y} Z_{\bs{i},x} Z_{\bs{i},y}, \\
    \sigma^X_{\bs{i}} & = X_{\bs{i},x} X_{\bs{i}+\hbs{x},y} X_{\bs{i}+\hbs{y},x} X_{\bs{i},y}. \\
\end{split}
\label{equ:2DTCstabilizers}
\end{equation}
Here, subscripts denote the unit cell and orientation of single-qubit Pauli operators, $P_{\bs{i},m}$, for $P \in \{ \mathbbm{1}, X, Y, Z\}$, where $\hbs{x} =(1,0), \hbs{y}=(0,1)$ are unit vectors.
The Hamiltonian is fully commuting, thus its ground states satisfy, $\sigma^{Z}_{\bs{i}} \ket{\GS} = \sigma^{X}_{\bs{i}} \ket{\GS} = \ket{\GS}$ for all $\bs{i}$.

Excitations above the toric code ground state, or \emph{quasiparticles}, are labelled by the Hamiltonian terms that they violate.
They can be decomposed into three types: $e$-particles, which violate the plaquette terms, $\sigma^{X}_{\bs{i}}$; $m$-particles, which violate the vertex terms, $\sigma^{Z}_{\bs{i}}$; and $f$-particles, corresponding to a bound state of $e$- and $m$-particles.
The toric code quasiparticles possess a few notable features.
First, the parity of each quasiparticle is conserved.
This arises from \emph{conservation laws} of the stabilizer operators, i.e. extensive sets of stabilizers that product to the identity.
In the toric code, we have:
\begin{equation} \label{eq: toric conservation laws}
    \prod_{\bs{i}} \sigma^{X}_{\bs{i}} = \mathbbm{1}, \,\,\,\,\, \prod_{\bs{i}} \sigma^{Z}_{\bs{i}} = \mathbbm{1},
\end{equation}
which hold exactly under periodic boundary conditions, and enforce the parity of $e$- and $m$-particles (and thus, $f$-particles as well) to be even.
Parity conservation implies that local perturbations can create only pairs of quasiparticles. Individual quasiparticles are obtained by separating these pairs via non-local string operators (Fig.~\ref{fig: toric bulk}b).

\begin{figure}
    \centering
    \includegraphics[width=0.85\columnwidth]{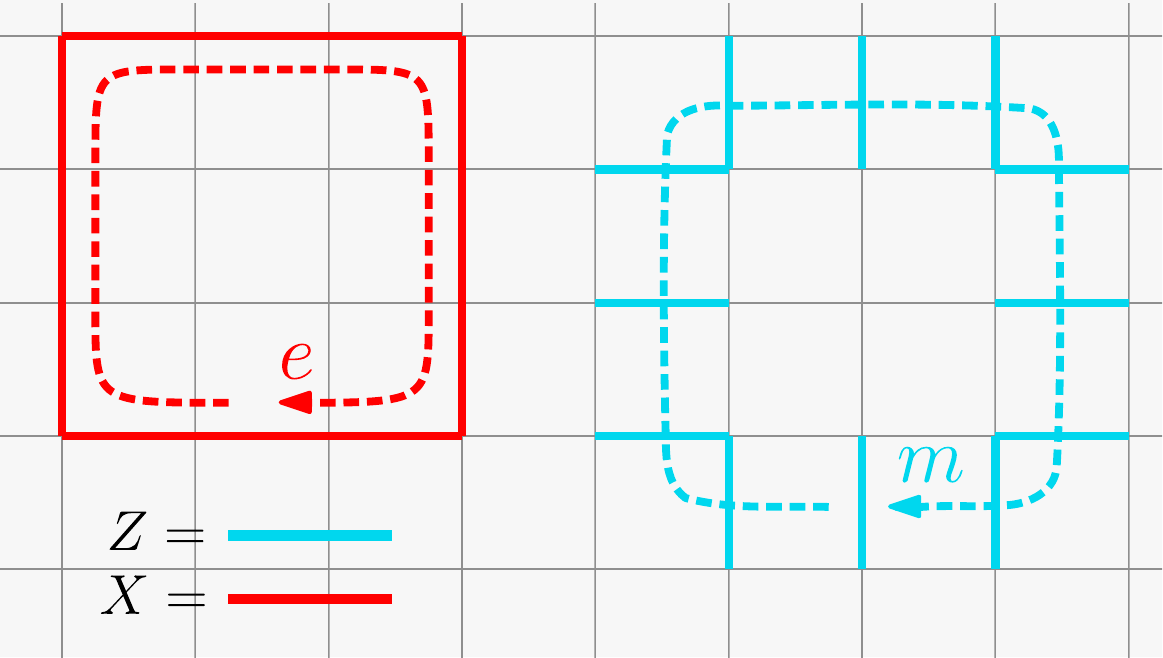}
    \caption{The conservation laws of the toric code are formed by products over plaquette (red) or vertex (blue) stabilizers. When taking the product over a finite region, one obtains a string operator moving $e$ or $m$ quasiparticles along the boundary.}
    \label{fig:ToricCodeString}
\end{figure}

These string operators can in fact be obtained from the conservation laws themselves.
Consider the product of all vertex stabilizers in a finite region (Fig.~\ref{fig:ToricCodeString}).
From Eq.~(\ref{eq: toric conservation laws}), the stabilizers product to the identity on spins inside the region.
The two-dimensional product of stabilizers is thus reducible to a one-dimensional product of operators along the region's boundary.
By ``cutting'' this one-dimensional product in half, one obtains a string operator that can create excitations only at its ends, since the middle of the string commutes with all stabilizers.
In the toric code, performing this procedure for vertex or plaquette stabilizers produces a string that excites $m$- or $e$-particles, respectively.

\begin{figure}
    \centering
    \includegraphics[width=0.75\columnwidth]{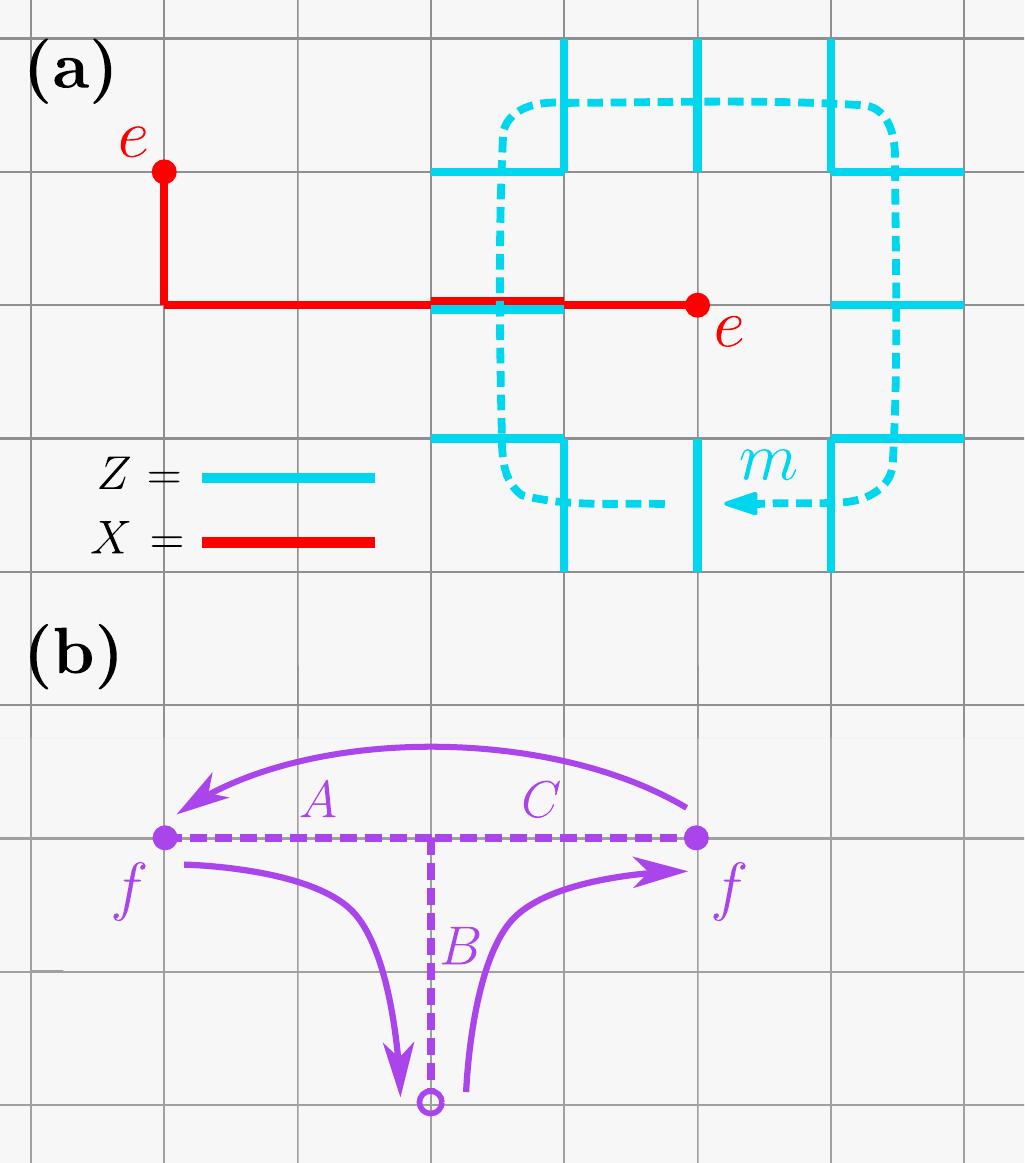}
    \caption{\textbf{(a)} The $e$ and $m$ quasiparticles exhibit mutual braiding statistics, as seen from the fact that their string operators anti-commute where they intersect. \textbf{(b)} The $f$ quasiparticle exhibits non-trivial self statistics, as seen from the overall minus sign accumulated during the three-prong exchange process.}
    \label{fig:ToricCodeBraiding}
\end{figure}

Finally, the quasiparticles possess non-trivial braiding statistics.
The \emph{mutual statistics} of two quasiparticles are defined as the phase acquired by the many-body wavefunction when one particle is transported in a loop about the other.
This can be computed from the commutator of two string operators (one for each type of quasiparticle) that intersect, see Fig.~\ref{fig:ToricCodeBraiding}.
In the toric code, the $e$- and $m$-particles have non-trivial mutual statistics, acquiring a phase $e^{i\pi} = -1$ (and similar the $e$- and $f$-particles, and $m$- and $f$-particles).
Quasiparticles can also possess non-trivial \emph{self statistics}, defined as the phase acquired when exchanging the locations of two quasiparticles of the same species.
On the lattice however, the exchange process must be carefully designed so that non-universal phase factors do not contribute. An example of such a process is the three-prong exchange process shown in Fig.~\ref{fig:ToricCodeBraiding}~\cite{LevinWen2003}. In the toric code, only the $f$-particle has non-trivial self statistics, with phase $e^{i\pi} = -1$.

\subsection{Boundary Hilbert space} \label{sec: toric code boundary}

\begin{figure}[t]
    \centering
    \includegraphics[width=\columnwidth]{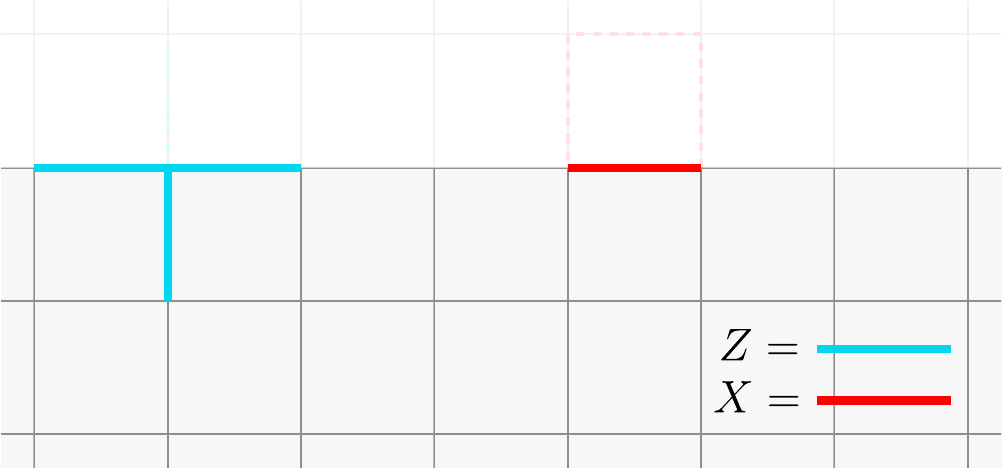}
    \caption{Local boundary operators in the toric code are obtained by truncating stabilizers that cross the boundary. The truncated boundary operators commute with all bulk stabilizers, but not necessarily with one another.}
    \label{fig:ToricCodeBoundary}
\end{figure}

We now turn to the boundary of the toric code model. We are interested in the structure of the boundary degrees of freedom when the bulk is in the ground state.
For concreteness, we specify to a smooth boundary along the $i_y = 0$ edge of the lattice (Fig.~\ref{fig:ToricCodeBoundary}).
We define the \emph{bulk stabilizers} of the model as the vertex and plaquette operators whose spins lie entirely within the boundary.
This corresponds to vertex operators ($\sigma_{\bs{i}}^{Z}$) with $i_y > 0$, and plaquette operators ($\sigma_{\bs{i}}^{X}$) with $i_y \geq 0$ [as defined in Eq.~(\ref{equ:2DTCstabilizers})].
The \emph{boundary Hilbert space} is the manifold of states where all bulk stabilizers have eigenvalue one
\begin{equation}
    \mathcal{H}_{\text{bndry}} \equiv \left\{ \ket{\psi} \, | \, \sigma^{Z}_{\bs{i}} \ket{\psi} = \sigma^{X}_{\bs{i}} \ket{\psi} = \ket{\psi} \, \forall \, \bs i  \text{ with } i_y \geq 0 \right\},
\end{equation}
i.e.~where the bulk is in the ground state.

In our work, we will study the boundary Hilbert space through the set of \emph{operators} that act upon it\footnote{A gapped boundary can be obtained by choosing a maximal set of commuting boundary operators, which correspond to a Lagrangian subgroup. Often in the literature, there is a canonical choice of such operators for different boundary terminations (e.g.~on a smooth or rough boundary). However, this choice is in some ways arbitrary, and in fact a single boundary termination already contains many non-commuting boundary operators.}. %;
To construct these operators, we first observe that any boundary operator must commute with every bulk stabilizer, in order to leave the bulk in the ground state.
Now, note that such an operator is easily obtained by \emph{truncating} the $i_y < 0$ components of any stabilizer on the infinite lattice (i.e.~the lattice that would exist if there were no boundary). This is shown in Fig.~\ref{fig:ToricCodeBoundary}.
On the lattice spins, $i_y \geq 0$, the boundary operator is equal to what would have been a bulk stabilizer had the boundary not existed.
The mutual commutation of stabilizers on the infinite lattice  guarantees that the truncated boundary operator and non-truncated bulk stabilizers mutually commute.

In the toric code, this produces two types of boundary operators\footnote{On the specific boundary considered here, these operators in fact form a generating set for the entire boundary operator algebra. However, this is not guaranteed in general. As a trivial example, consider an isolated one-dimensional chain of spins (e.g.~the uppermost horizontal spins in Fig.~\ref{fig:ToricCodeBoundary}) and view it as the boundary of a fictional bulk toric code model. The vertex stabilizers truncate to two-site operators, $Z_i Z_{i+1}$ while the plaquette stabilizers truncate to (two copies of the) single-site operators $X_i$. The single-site operator $Z_i$ is allowed on the boundary, but is not generated by any truncated bulk operator.}, which we denote by $\tilde \sigma$.
First, we have the three-spin operator  $\tilde{\sigma}^Z_{i_x}$, which is equal to a vertex operator with its uppermost spin truncated,
\begin{equation}
    \tilde{\sigma}^Z_{i_x} = 
    Z_{(i_x-1,0),x} Z_{(i_x,0),y} Z_{(i_x ,0),x},
     \label{equ:sigmaZ1D}
\end{equation}
where we have taken $i_y=0$.
Second, we have the single-spin operator $\tilde{\sigma}^X_{i_x}$, which is obtained from the plaquette operator with its upper three spins truncated,
\begin{equation}
    \tilde{\sigma}^X_{i_x} = 
    X_{(i_x,0),x}.
    \label{equ:sigmaX1D}
\end{equation}
As a result of the truncation, the boundary operators do not necessarily commute with one another.

%%% Toric code global constraints
The boundary operator algebra generated by the truncated operators above is augmented by global \emph{constraints} arising from the bulk conservation laws.
Specifically, the product of all bulk stabilizers of a given type [as in Eq.~(\ref{eq: toric bulk conservation})] is not equal to the identity on a finite lattice, but rather to the product of all boundary operators of the same type:
\begin{align}
    \prod_{\bs i \in \text{bulk}} \sigma^Z_{\bs{i}} &= \prod_{i_x \in \text{bndry}} \tilde{\sigma}^Z_{i_x},
   &
    \prod_{\bs i \in \text{bulk}} \sigma^X_{\bs{i}} &= \prod_{i_x \in\text{bndry}} \tilde{\sigma}^X_{i_x}.
   \label{equ:bulkproduct1D}
\end{align}
When the bulk is in the ground state this product is equal to one,
\begin{align}
    \prod_{i_x \in\text{bndry}} \tilde{\sigma}^Z_{i_x} &\eqgs \mathbbm{1},& \prod_{i_x \in\text{bndry}} \tilde{\sigma}^X_{i_x} &\eqgs \mathbbm{1},
    \label{equ:constraint1D}
\end{align}
which enforces global constraints on the boundary operator algebra.

\begin{figure}
    \centering
    \includegraphics[width=\columnwidth]{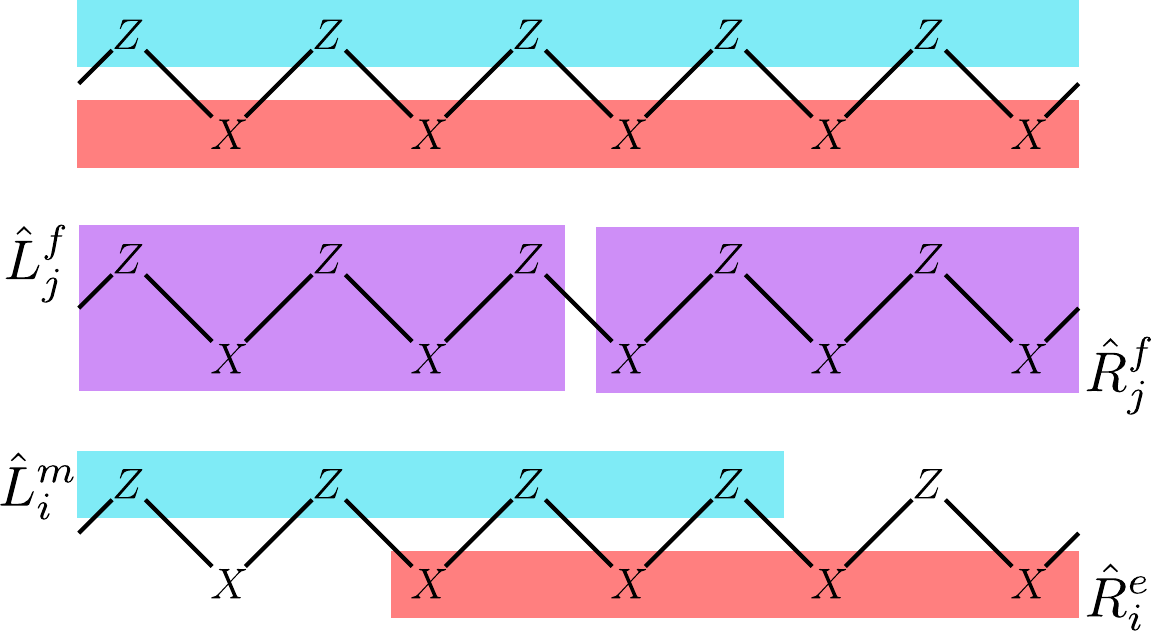}
    \caption{Boundary operator algebra of the toric code. The operators $\tilde \sigma_Z$ and  $\tilde \sigma_X$ are abbreviated $Z$ and $X$, respectively. A solid line connecting two operators signifies that the operators anti-commute. (Top) The $e$ (red) and $m$ (blue) constraints are formed from products of all $\tilde \sigma_Z$ and  $\tilde \sigma_X$ operators, respectively. (Middle) The self-obstructor invariant for the $f$ constraint (purple) is given by the commutation of two adjacent patch operators, $\hat{L}^f_j$ and $\hat{R}^f_j$. The patches anti-commute because they share a single line between them. (Bottom) The mutual-obstructor invariant for the $e$ and $m$ constraints is given by the commutation of two overlapping patch operators, $\hat{L}^m_i$ and $\hat{R}^e_j$, which also anti-commute.}
    \label{fig:frustation_TC}
\end{figure}

%%% Picture as Z2 symmetric sector
This construction leads to a convenient physical picture for the boundary Hilbert space of the toric code model~\cite{Lichtmanetal20,JiWen19}.
Namely, the boundary Hilbert space is equivalent to the Hilbert space of a one-dimensional spin chain restricted to a global $\ZZ_2$ symmetry sector, $\prod_{i_x} X_{i_x} \ket{\psi} = \ket{\psi}$.
To see this, observe that a generating set of operators that commute with the symmetry is given by  $Z_{i_x} Z_{i_x+1}$ and $X_{i_x}$, familiar from the transverse field Ising model.
The algebra generated by such terms is exactly equivalent to the boundary operator algebra of the toric code.
In particular, the frustration graphs of the operators (i.e.~the pattern of how pairs of operators commute or anti-commute, Fig.~\ref{fig:frustation_TC}) are identical: neighboring $X$- and $Z$-operators anti-commute. The constraints are identical as well, since in the Ising model, the product of all $Z$-type operators is trivially the identity, $\prod_{i_x} Z_{i_x} Z_{i_x+1} = \mathbbm{1}$,  while the product of $X$-type operators is equal to one within the symmetric sector.

\section{Obstructor invariants in 2D stabilizer models}\label{sec: toric boundary}

We now turn to the question: What, if any, are the distinguishing features of the boundary Hilbert space of stabilizer models with topological order?
This question is motivated by the mapping in the previous section, where we saw that the boundary Hilbert space of the toric code model is isomorphic to the symmetric sector of a local tensor product space.
This is in contrast to the boundary Hilbert space of a model without topological order, where the boundary is a simple local tensor product space (since the bulk can be disentangled from the boundary by a finite-depth unitary circuit).
From this observation, Ref.~\cite{JiWen19} conjectured that \emph{any} boundary Hilbert space of the toric code cannot be realized as a 1D local tensor-product Hilbert space (LTPS).

In this section, we provide a framework for understanding the structure of the boundary Hilbert space of stabilizer models.
Our framework centers on two features of the boundary operator algebra, introduced in the previous section: the frustration graph of local boundary operators (i.e.~the pattern of how they commute and anti-commute), and the global constraints enforced on these operators by the bulk conservation laws.
We will show that by taking finite ``patches'' of the global constraints~\cite{JiWen20,chatterjee2022algebra}, and considering their commutation relations, we can construct invariants that quantify the precise obstruction to realizing the boundary Hilbert space as a local tensor product. 
We call these \emph{\obs{}}.
In particular, we can use the \obs{} to prove the conjectured bulk-boundary correspondence in Ref.~\cite{JiWen19}, for both the specific toric code boundary previously considered, and, more generally, for any boundary termination of any two-dimensional stabilizer model with topological order.

The section proceeds as follows.
In Sec.~\ref{sec:selfTC} we introduce the notion of a patch operator as well as our first example of an \ob{}---the self-\ob{}---on the toric code boundary.
We show that a non-trivial self-\ob{} arises as a consequence of the self statistics of the fermionic quasiparticle of the toric code.
In Sec.~\ref{sec:mutualTC}, we introduce a second class of invariants------the mutual-\obs{}---and show they arise from the mutual statistics of bulk quasiparticles.
In Sec.~\ref{sec:2Dgeneral} we construct the boundary operator algebra for generic 2D stabilizer models, and in Sec.~\ref{sec:obstructor} we generalize our results on \obs{} to this context.

\begin{figure*}[t!]
\centering
\includegraphics[width=0.9\textwidth]{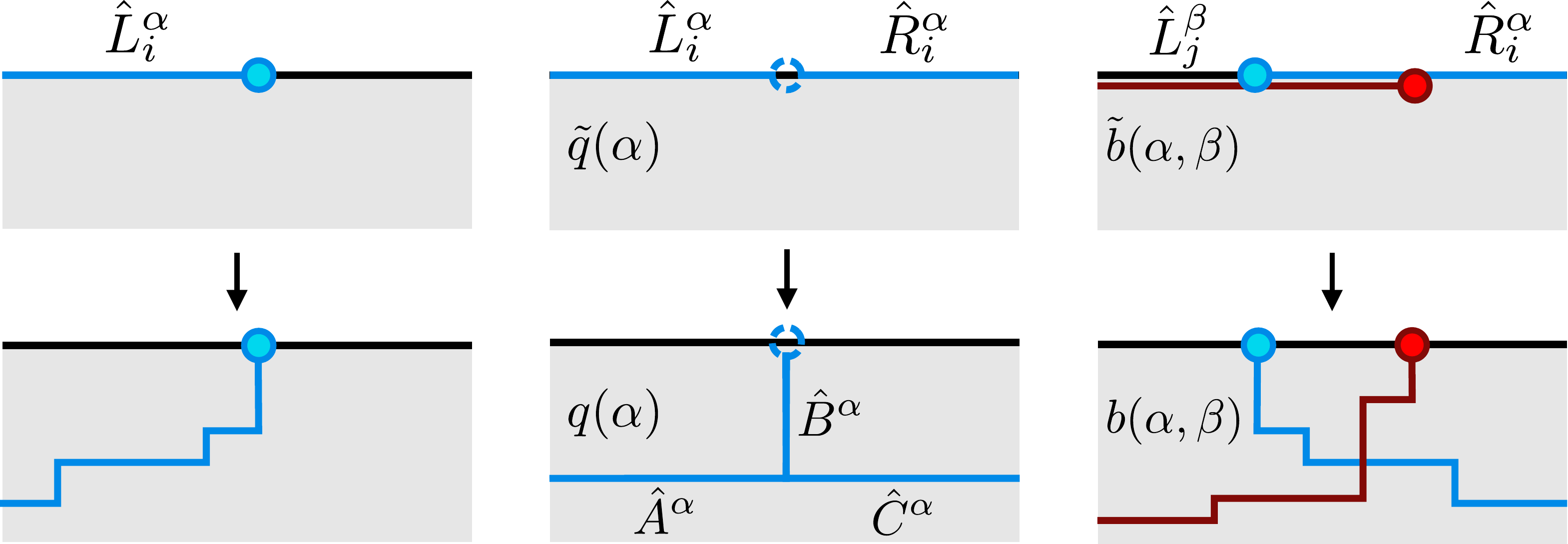}
\caption{
(Left) A patch operator on the boundary can be deformed into a string operator in the bulk, via multiplication with bulk stabilizers. (Middle) The self-obstructor invariant $\tilde q(\alpha)$ [Eq.~(\ref{eq:mutual-obstructordef})], is equal to the commutator of the two patch operators $\hat{L}^\alpha_i$, $\hat{R}^\alpha_i$. This commutator can be re-written in the bulk as a three-prong exchange process for the quasiparticle $\alpha$, and is thus equal to the self statistics $q(\alpha)$. (Right) The mutual-obstructor invariant $\tilde b(\alpha,\beta)$ [Eq.~(\ref{eq:mutual-obstructordef})] is equal to the commutator of the patch operators shown, and can similarly be re-written in terms of the mutual statistics of the bulk quasiparticles $\alpha$ and $\beta$.
}
\label{fig: bulk equals boundary statistics}
\end{figure*}

\subsection{Self-obstructor invariants in the toric code model}\label{sec:selfTC}

To motivate our construction, we begin with a short proof that the toric code boundary discussed previously cannot be realized in any 1D local tensor product space (LTPS).
Our proof follows directly from the \textit{frustration graph} of  local boundary operators, depicted in Fig.~\ref{fig:frustation_TC}.
The graph contains an edge between each pair of local boundary operators that anti-commute. (This method can be extended to qudits by labelling each edge by an element of $\mathbbm{Z}_d$.)
Note that the commutation of any product of boundary operators with another product is given by the parity of the number of lines extending from one product to the other.

We need one additional ingredient to show that the boundary is not a 1D LTPS: the global constraints [Eq.~(\ref{equ:constraint1D}),  top of Fig.~\ref{fig:frustation_TC}].
In particular, the product of the $Z$- and $X$-constraints contains all boundary operators in the frustration graph.
From Eq.~(\ref{equ:constraint1D}), this product is equal to the identity within the boundary Hilbert space.
Now, consider splitting this constraint in half, into the product of a left and a right ``patch operator''~\cite{JiWen19,JiWen20,chatterjee2022algebra}, as shown in the middle panel of Fig.~\ref{fig:frustation_TC}.
Specifically, we can take the left patch operator, $\hat{L}$, to be equal to the product of all boundary operators with $i_x < 0$ and the right patch operator, $\hat{R}$, similarly with $i_x \geq 0$.
Now, the constraint implies that the two patch operators product to the identity (up to a possible sign), $\hat{L} \hat{R} \proptogs \mathbbm{1}$.
At the same time, the two patch operators must anti-commute since a single edge extends between them in the frustration graph (Fig.~\ref{fig:frustation_TC}).

Now, if the toric code boundary operator algebra could be realized in a 1D LTPS, then the patch operators in the LTPS would instead obey the strict equality, $\hat{L} \hat{R} \propto \mathbbm{1}$.
However, this is inconsistent with the anti-commutation of the patch operators, since a matrix and its inverse must commute.
We conclude that the boundary operator algebra of the toric code cannot be realized in any 1D LTPS.

In what follows, we generalize the above argument by introducing the notion of a self-\ob{}.
To formulate the self-\ob{}, let us first define the patch operators more generally.
For each bulk conservation law $\alpha$ (which can be associated to an anyon type, $\alpha \in \{1,e,m,f\}$), we define the left and right patch operators at site $j$ as follows:
\begin{align}
   \hat{L}^m_j &= \prod_{i_x=-\infty}^{j-1} \tilde \sigma^Z_{i_x} &  \hat{R}^m_j &= \prod_{i_x=j}^{\infty} \tilde \sigma^Z_{i_x}\\
    \hat{L}^e_j &= \prod_{i_x=-\infty}^{j-1} \tilde \sigma^X_{i_x} &  \hat{R}^e_j &= \prod_{i_x=j}^{\infty} \tilde \sigma^X_{i_x}\\
\hat{L}^f_j  &= \prod_{i_x=-\infty}^{j-1} \tilde \sigma^Z_{i_x} \tilde \sigma^X_{i_x} &  \hat{R}^f_j &= \prod_{i_x=j}^{\infty} \tilde \sigma^Z_{i_x} \tilde \sigma^X_{i_x}.
\end{align}
We also have the trivial patch operators $\hat{L}^1_j = \hat{R}^1_j = \mathbbm{1}$. The patch operators $\hat{L}$ and $\hat{R}$ defined in our previous argument correspond to $\hat{L}^f_0$ and $\hat{R}^f_0$, respectively. %
The global constraints imply that the two patch operators product to the identity (again, up to a sign),
\begin{equation}
    \hat{L}^\alpha_j \hat{R}^\alpha_j  \proptogs \mathbbm{1},
\end{equation}
for every conservation law $\alpha$ and every site $j$.

Now, for each boundary constraint $\alpha$, we define the \textit{self-obstructor invariant}, $\tilde q(\alpha)$, as the phase acquired when commuting the left and right patch operators:
\begin{align}
\exp \left( \frac{2\pi i}{d}\tilde{q}(\alpha) \right)  = [\hat{R}^\alpha_j,\hat{L}^\alpha_j].
\label{eq:self-obstructordef}
\end{align}
Here $[A,B] = ABA^{-1}B^{-1}$ is the group commutator, in contrast to the usual commutator for quantum mechanical operators.
In translation-invariant models, the commutator is automatically independent of the site $j$; this will also follow from our later arguments linking the commutator to the bulk statistics.
In the toric code model, the $e$ and $m$ constraints give zero self-\ob{} while the $f$ constraint gives
\begin{align}
    \exp \left( \frac{2\pi i}{d}\tilde{q}(f) \right) = -1.
\end{align}
This is precisely the anti-commutation in the proof at the beginning of this section.
Following the above proof, we see that whenever any self-\ob{} is not equal to one, then the boundary Hilbert space cannot be a 1D LTPS.

As our naming suggests, the self-obstruction invariant is related to the self statistics of the bulk topological order: specifically, the self statistics of the quasiparticle labelling the patch operator of interest.
This is visualized in Fig.~\ref{fig: bulk equals boundary statistics}.
To derive this correspondence, recall that the boundary constraint is formed by an extensive product of bulk stabilizers [Eq.~(\ref{equ:bulkproduct1D})].
By multiplying the patch operators $\hat{L}^\alpha_j$ and $\hat{R}^\alpha_j$ by adjacent bulk stabilizers, we can deform the patch operators into string operators that travel through the bulk and terminate on the boundary at site $j$.
Crucially, this multiplication does not change the patch operators' commutation because the bulk stabilizers commute amongst themselves, and with all boundary operators.

The  string operators obtained above move quasiparticles of type $\alpha$ through the bulk. %
By arranging these string operators as in the middle panel of Fig.~\ref{fig: bulk equals boundary statistics}, we see that the commutation of the boundary patch operators---i.e.~the self-\ob{}---is exactly equal to the \emph{self statistics} of the corresponding bulk quasiparticle.
Specifically, considering $\alpha = f$, we have
\begin{equation}
\begin{split}
    \exp \left( \frac{2\pi i}{d}\tilde{q}(f) \right) & =  \hat{R}^f_j \hat{L}^f_j  (\hat{R}^f_j)^\dagger (\hat{L}^f_j)^\dagger \\
    & = 
    \hat{C}^f \hat{B}^f \hat{A}^f (\hat{C}^f)^\dagger (\hat{B}^f)^\dagger (\hat{A}^f)^\dagger \\
    & = -1,
\end{split}
\end{equation}
where the strings $\hat{A}^f, \hat{B}^f, \hat{C}^f$ are as shown in Fig.~\ref{fig: bulk equals boundary statistics}.
The multiplication by bulk stabilizers deforms the left patch operator into the product $\hat{L}^f \proptogs \hat{A}^f \hat{B}^f$, and the right patch operator into the product $\hat{R}^f \proptogs \hat{C}^f \hat{B}^f$.
The equality above follows from plugging these expressions into the LHS above and cancelling a factor of $\hat{B}^f (\hat{B}^f)^\dagger = \mathbbm{1}$ in the center of the product.

\subsection{Mutual-obstructor invariants in the toric code model} \label{sec:mutualTC}

A variation of the construction above allows us to connect to the bulk mutual statistics as well.
Namely, we again consider two patch operators, but now for different boundary constraints $\alpha$ and $\beta$.
Moreover, instead of taking the patch operators to meet at a single point, we will take them to overlap in the manner shown in the right panel of Fig.~\ref{fig: bulk equals boundary statistics}.
Specifically, we consider the patch operators $\hat{R}^\alpha_i$ and $\hat{L}^\beta_j$, where $j > i + K$ and  $K$ is the maximum range of any bulk stabilizer (i.e.~$K=1$ in the toric code).
The commutator of these patch operators constitutes the \textit{mutual-obstructor invariant}, $\tilde b(\alpha,\beta)$:
\begin{equation}
\exp \left( \frac{2\pi i}{d} \tilde{b}(\alpha,\beta) \right) = [\hat{R}^\alpha_i  ,\hat{L}^\beta_j].
\label{eq:mutual-obstructordef}
\end{equation}
This quantity is independent of $i,j$ within the regime $j > i + K$, since any two left patch operators within this regime differ only by local boundary operators that commute with the right patch operator (since they are contained entirely inside of it), and vice versa.

As an example, consider the mutual-\ob{} for patch operators $e$ and $m$.
This is given by the group commutator of $\hat{R}^e_i$ and $ \hat{L}^m_j $. Observing the frustration graph (bottom panel of Fig.~\ref{fig:frustation_TC}), we see that for $j > i + K$ the number of anti-commutations is always odd. We therefore find
\begin{align}
    \exp \frac{2\pi i}{d} \tilde{b}(e,m) =-1.
\end{align}
Similarly, when deforming the patch operators into the bulk as shown in the right panel of Fig.~\ref{fig: bulk equals boundary statistics}, we find a single crossing of the anyon strings for $\alpha$ and $\beta$. The mutual-obstructor invariant is therefore given by mutual statistics of the bulk $e$ and $m$-quasiparticles.

Like the self-obstructor invariant, any non-zero value of the mutual-obstructor invariant implies that the boundary theory cannot be realized as a 1D local tensor product space. To show this, note that the following quantity is proportional to the identity,
\begin{align}
     \hat{L}^e_i \hat{L}^m_j  \hat{R}^e_i \hat{R}^m_j\proptogs \mathbbm{1}
     \label{eq:LemRem}
\end{align}
in the boundary Hilbert space. For $j j > i + K$, the commutation between the right patch operators, $\hat{R}^e_i \hat{R}^m_j$, and the left patch operators, $ \hat{L}^e_i\hat{L}^m_j$, is simply given by the commutation of $\hat{R}^m_j$ and $\hat{L}^e_i$.
This follows because all other pairs of left and right patch operators are mutually commute.
We thus find,
\begin{align}
    [\hat{R}^e_i \hat{R}^m_j,\hat{L}^e_i\hat{L}^m_j ] = [\hat{R}^m_j,\hat{L}^e_i] = \exp \frac{2\pi i}{d} \tilde{b}(e,m) =-1.
\end{align}
By the same arguments we applied to the self-obstructor invariant, the above result, combined with Eq.~\eqref{eq:LemRem}, shows that the boundary is not a 1D local tensor product space.

Finally, as for the self-\ob{}, the mutual-\ob{} is directly given by the mutual statistics of the corresponding bulk quasiparticles.
Deforming the two patch operators into the bulk strings as in the right panel of Fig.~\ref{fig: bulk equals boundary statistics}, we find that their commutation is given by the mutual statistics of the $e$ and $m$-quasiparticles.
Interestingly, in the bulk, we know that the mutual statistics of the $e$ and $m$-quasiparticles and the self statistics of the $f$-quasiparticles are in fact the same quantity (since we can obtain the $f$ anyon by fusing $e$ and $m$).
This implies that a similar relation should hold for the boundary \obs{}.
In Sec.~\ref{sec:obstructor}, we prove such relations directly for the boundary \obs{} without reference to the bulk.

\subsection{Boundary operator algebra in generic 2D stabilizer models}\label{sec:2Dgeneral}

We now extend the our framework to generic translation-invariant stabilizer models.
We focus for now on two-dimensional systems, and turn to three-dimensions in Sections~\ref{sec: XCube} and~\ref{sec:fractal}.
Our results show that any 2D translation-invariant stabilizer model with bulk topological order cannot have a local tensor product boundary, as quantified by the obstructor invariants.

%%% Define generic models and notation
We consider translation-invariant stabilizer models in two-dimensions with $n$-dimensional qudits and $M$ stabilizers per unit cell.
Under these conditions, the Hamiltonian can be written as a sum of commuting stabilizers
\begin{equation}
    H = - \sum_{\bs i} \sum_{m=1}^M \sigma_{\bs i}^m.
\end{equation}
We assume the stabilizers, $\sigma_{\bs i}^m$, are geometrically local, in the sense that their support is contained within a $K \times K$ grid of unit cells about site $\bs{i}$.
We also assume that the stabilizers are maximal, in the sense that there are no further independent stabilizers $\sigma_{\bs i}^{M+1}$ that can be added to the model that mutually commute with all current stabilizers, $\sigma_{\bs i}^{1}, \ldots, \sigma_{\bs i}^{M}$.
In what follows, we outline how each aspect of the previous section extends to such models.

%%% Define conservation laws
\emph{Bulk conservation laws.}---We begin by addressing the bulk conservation laws.
These correspond to products of stabilizers that equal the identity (up to phase factor) on an infinite lattice\footnote{We note that here, we have assumed that all conservation laws involve products of stabilizers over every unit cell [as in Eq.~(\ref{eq: toric bulk conservation})].
For periodic conservation laws, this can always be ensured by enlarging the unit cell to encompass the given periodicity.
In Section~\ref{sec:fractal}, we will extend this formalism to include fractal conservation laws.}:
\begin{equation} \label{eq: general conservation law}
    \prod_{\bs i} \prod_{m=1}^M \left( \sigma^m_{\bs i} \right)^{c^\alpha_m} \propto \mathbbm{1}.
\end{equation}
Here, each conservation law, indexed by $\alpha$, is specified by the powers, $c^\alpha_m \in \ZZ_n$, of each stabilizer involved.
Note that the set $\mathcal{C}$  of conservation laws forms an Abelian group, i.e.~given two conservation laws $\alpha, \beta \in \mathcal{C}$, we can define a third conservation law $\alpha \beta \in \mathcal{C}$ via $c^{\alpha\beta}_m = c^{\alpha}_m + c^{\beta}_m$.

%%% Bulk conservation law --onto--> quasiparticles
As in the toric code, each conservation law $\alpha$ is naturally mapped to a quasiparticle by noting that the restriction of a conservation law to a finite region is equal to a loop operator that moves some quasiparticle around the region's boundary.
Cutting the loop at two points produces a string operator that commutes with the stabilizers in its center, and thus has a well-defined quasiparticle type at either end.
This mapping is in fact onto, i.e.~each quasiparticle in turn generates a conservation law\footnote{To see this, construct a loop operator $\ell^{\bs i}$ that transports a quasiparticle $\alpha$ in a $1 \times 1$ square beginning at site $\bs i$. By definition $\ell^{\bs i}$ obeys a conservation law, since taking the product of $\ell^{\bs i}$ over a finite region gives a loop operator acting only on the boundary. To write this conservation law in the form Eq.~(\ref{eq: general conservation law}), note the loop operator can be written as a product of stabilizers, $\ell^{\bs i} = \prod_{\bs j} (\sigma_{\bs j}^m)^{c^{\bs j}_m}$. The conservation law described by $c^\alpha_m = \sum_{\bs j} c^{\bs j}_m$ corresponds to the quasiparticle $\alpha$.
}. Therefore, moving forward, we will use the label $\alpha$ for both the quasiparticles and conservation laws interchangeably.

%%% Truncation -> Boundary operators
\emph{Boundary operator algebra.}---Turning to the boundary Hilbert space, we note that the procedure for obtaining boundary operators by truncating bulk stabilizers is also entirely general.
Specifically, we decompose a given bulk stabilizer as a product of operators at each value of the $y$-coordinate,
\begin{equation}
    \sigma_{\bs i}^m = \prod_{j=i_y}^{i_y+K-1} \left[ \sigma_{\bs i}^m \right]_{j},
\end{equation}
where each term on the right has support only within unit cells at $y$-coordinate $j$.
The product is over $K$ values of $j$, where $K$ is maximum range of the stabilizer. We assume these values run from $i_y$ to $i_y+K-1$ (since we can always shift the stabilizers such that this is the case).
Truncating translations of this operator along the boundary, $i_y = 0$, produces $K-1$ boundary operators labeled by the initial $i_y$ value, $k$, of the truncated operator,
\begin{equation} \label{eq: general boundary truncation}
    \tilde{\sigma}_{i_x}^{m,k} = \prod_{j=k}^{K-1} \left[ \sigma_{(i_x,-k)}^m \right]_{-k+j},
\end{equation}
Here $k$ runs from $0$ to $-K+2$.

%%% Conservation law -> Boundary constraints
The termination of bulk conservation laws onto boundary constraints proceeds similarly.
Namely, the conservation law Eq.~(\ref{eq: general conservation law}) is generalized to:
\begin{equation} \label{eq: general conservation law boundary}
    \prod_{\bs i \in \text{ bulk}} \prod_{m=1}^M \left( \sigma_m^{\bs i} \right)^{c^\alpha_m} = \prod_{i_x \in \text{ bndry}} \prod_{m=1}^M \prod_{k=1}^{K} \left( \tilde{\sigma}^{m,k}_{i_x} \right)^{c^\alpha_m},
\end{equation}
where the product includes the boundary operators for all ``initial $i_y$ values'', as discussed in above.
When the bulk is in the ground state, this equality implies the following constraint on the boundary operator algebra,
\begin{equation} \label{eq: general boundary constraint}
    \prod_{i_x \in \text{ bndry}} \left[ \prod_{m=1}^M \prod_{k=1}^{K} \left( \tilde{\sigma}^{m,k}_{i_x} \right)^{c^\alpha_m} \right] \proptogs \mathbbm{1}.
\end{equation}
%%% Also local constraints
In addition to these global constraints, we may also have \emph{local constraints} on the truncated boundary operators of the form $\prod_j ( \sigma_{m_j}^{{\bs i}_j} )^{c_j} \proptogs 1$, whenever a local product of  boundary operators is equal to a local product of bulk stabilizers.
This will not be the case for the models we consider in the main text, but does arise in other models, such as the stabilizer double-semion model~\cite{Ellison22}, which we address in Appendix~\ref{app:DS}.

\subsection{Obstructor invariants in generic 2D stabilizer models}\label{sec:obstructor}

Our construction of the patch operators for generic stabilizer models again resembles our construction for the toric code.
For each conservation law $\alpha$, we  define the left and right patch operators  at a boundary site $j$,
\begin{align}
     \hat{L}^\alpha_j &= \prod_{i_x=-\infty}^{j-1} \prod_{m,k} \left( \sigma_{i_x}^{m,k} \right)^{c^\alpha_m}, & \hat{R}^\alpha_j &= \prod_{i_x=j}^{\infty} \prod_{m,k} \left( \sigma_{i_x}^{m,k} \right)^{c^\alpha_m}.   
\end{align}
The boundary constraints imply that the two operators product to one,
\begin{equation}
    \hat{L}^\alpha_j \hat{R}^\alpha_j  = \prod_{i_x=-\infty}^{\infty} \prod_{m,k} \left( \sigma_{i_x}^{m,k} \right)^{c^\alpha_m} \proptogs \mathbbm{1}.
\end{equation}

%%% Define \ob
Patch operators in hand, we define the \obs{} as follows:
\begin{enumerate}
    \item For each constraint $\alpha$ on the boundary operator algebra, we define the self-\ob:
\begin{equation}
\exp \frac{2\pi i}{d}\tilde{q}(\alpha)  = \hat{R}^\alpha_i  \hat{L}^\alpha_i   (\hat{R}^\alpha_i)^\dagger (\hat{L}^\alpha_i)^\dagger .
\end{equation}
\item For each pair of constraints $\alpha$ and $\beta$, we define the mutual-\ob:
\begin{equation}
\exp \frac{2\pi i}{d} \tilde{b}(\alpha,\beta) = \hat{R}^\alpha_i \hat{L}^\beta_j  (\hat{R}^\alpha_i)^\dagger (\hat{L}^\beta_j)^\dagger .
\end{equation}
\end{enumerate}

The obstructor invariants possess a number of convenient properties, many of which we observed in our discussion of the toric code boundary.
First, as mentioned in Sec.~\ref{sec:mutualTC}, the obstructor invariants have a convenient property that  $\tilde{q}$ defines a quadratic form on the boundary constraints, and $\tilde{b}$ is its associated bilinear form. That is, one can confirm that they satisfy:
\begin{enumerate}
    \item $\tilde{q}(\alpha^n) = n^2 \tilde{q}(\alpha)$,
    \item  $\tilde{b}(\alpha\gamma,\beta) = \tilde{b}(\alpha,\beta)+\tilde{b}(\gamma,\beta)$ (and similarly for the second argument),
    \item  $\tilde{b}(\alpha,\beta) = \tilde{q}(\alpha \beta)-\tilde{q}(\alpha)-\tilde{q}(\beta)$,
\end{enumerate}
We prove these properties in Appendix~\ref{app:quadraticform}.

%%% Invariant under tensor product

\emph{Relation to bulk statistics.}---As in the toric code, the values of the obstructor invariants are inherited from the bulk topological order.
This arises directly from the definition of the constraints (and in turn, the patch operators) as the boundaries of bulk conservation laws.
As before, using this correspondence we can always multiply boundary patch operators by products of bulk stabilizers, in order to express the obstructor invariant as the commutator of string operators that overlap only in the bulk (Fig.~\ref{fig: bulk equals boundary statistics}).
The commutation of the bulk string operators is equal to the exchange statistics of the anyons that each string operator creates at its ends.
Thus, the obstructor invariants on the boundary Hilbert space are directly determined by the bulk anyon data.
We provide a detailed proof of this equality for translation-invariant stabilizer models in Appendix~\ref{sec: bulk boundary computation}.

\emph{Invariance under local tensor products.}---We will now show that the \obs{} are indeed \emph{invariants}: that is, they are unchanged upon taking tensor products of the boundary theory with any local tensor product space.
To begin, let us first show that the \obs{} are equal to zero in any 1D LTPS.
Consider a LTPS and suppose that, similar to the boundary Hilbert space, it contains infinite collections of $K$-local operators $\sigma_i^{m,\text{TP}}$ that product to the identity,
\begin{equation} \label{eq: LTPS strings}
    \prod_{i = -\infty}^\infty \prod_{m} \left( \sigma_i^{m,\text{TP}} \right)^{c^\alpha_m} \propto \mathbbm{1}.
\end{equation}
This contains a strict equality (up to a phase) since we are working in a LTPS (i.e.~we assume there are no global constraints arising from a bulk).

To formulate the \obs{}, consider splitting the infinite product into two patch operators\footnote{The patch operators can be finite instead of semi-infinite, as long as the strings extend for a distance at least $2K$ away from the cut. In this case, the strict equality in Eq.~(\ref{eq: LTPS strings}) implies that the string contains support only within a region of width $K$ about either of the endpoints. The support at the endpoint far from the cut commutes with all operators near the cut due to spatial locality.}, $\hat{L}^{\alpha,\text{TP}}_i$ and $\hat{R}^{\alpha,\text{TP}}_i$, at some site $i$.
By spatial locality, the left string can have support only on sites less than $i+K$, and the right string only for sites greater than $i-K$.
Since the equality in Eq.~(\ref{eq: LTPS strings}) is strict, this implies that both $\hat{L}^{\alpha,\text{TP}}_i$ and $\hat{R}^{\alpha,\text{TP}}_i$ contain support only within the region $[i-K,i+K]$.
%Naively, one might think that the self-\obs{} can be non-zero, since the left and right string now share support.
Since the two patch operators product to the identity, they also must product to the identity within the region $[i-K,i+K]$.
However, any two operators that product to the identity necessarily commute, since conjugating $\hat{A}^{-1} \hat{A} = \mathbbm{1}$ by $\hat{A}$ gives $\hat{A} \hat{A}^{-1} = \mathbbm{1}$ as well.
Therefore the self-\obs{} are zero.
An even simpler argument implies that the mutual-\obs{} are zero, since the support of patch operators $\hat{L}^{\beta,\text{TP}}_j$ and $\hat{R}^{\alpha,\text{TP}}_i$ is non-overlapping whenever $j > i + K$.

These arguments can be extended to show that taking tensor products with a LTPS does not change the \obs{}.
Consider a tensor product $\mathcal{H}_{\text{bndry}}' = \mathcal{H}_{\text{bndry}} \otimes \mathcal{H}_{\text{TP}}$.
Note that Pauli operators on $\mathcal{H}_{\text{bndry}}'$ are equal to tensor products of Pauli operators on the individual Hilbert spaces, i.e.~$\tilde{\sigma}' = \tilde{\sigma} \otimes \sigma^{\text{TP}}$.
An operator $\tilde{\sigma}'$ on $\mathcal{H}_{\text{bndry}}'$ features a global constraint, $\prod_{\bs i} \tilde{\sigma}'_{\bs i} \proptogs \mathbbm{1}$, if and only if both $\prod_{\bs i} \tilde{\sigma}_{\bs i} \proptogs \mathbbm{1}$ and $\prod_{\bs i} \sigma^{\text{TP}}_{\bs i} = \mathbbm{1}$.
The tensor product structure implies that the commutation of patch operators involving $\tilde{\sigma}'$ are equal to the product of the commutation of patch operators involving $\tilde{\sigma}'$ and the commutation of those involving $\sigma^{\text{TP}}$.
The latter are zero by the arguments above.
Thus the \obs{} are unchanged by the tensor product.
%
%

%%% Connecting self-obstructor invariant to bulk mutual statistics

%%% Since model is generic, choice of boundary termination is generic
\emph{Generic boundary terminations.}---Before proceeding, we pause to note the arguments of this section imply that our simple choice of boundary termination at $i_y \geq 0$ is in fact quite generic.
Specifically, consider instead an arbitrary linear boundary represented by the integers, $(b_x,b_y)$, where the qudits contained in the bulk obey, $b_x i_x + b_y i_y \geq 0$.
By transforming to coordinates, $i_x' = -b_y i_x + b_x i_y$ and $i_y' = b_x i_x + b_y i_y$, we obtain a new stabilizer model with an enlarged unit cell [now containing $M(b_x^2 + b_y^2)$ instead of $M$ stabilizers] and the ``simple'' boundary condition, $i_y' \geq 0$.
While the specific operator algebra of the $j_y \geq 0$ boundary may differ from the $i_y \geq 0$ boundary, the structure of the constraints placed upon it and the commutation relations of the patch operators will remain identical since their properties are  inherited from the bulk topological order (as mentioned before, we prove this explicitly in Appendix~\ref{sec: bulk boundary computation}).

These relations complete our characterization of the \obs{} of 2D translation invariant stabilizer models in terms of the bulk topological order.
In the following section we will see that this characterization takes a particularly simple, computable form within a polynomial formalism for analyzing these models.

\section{Polynomial formalism for boundaries and \obs{}}\label{sec:poly}

In this section, we utilize a mapping~\cite{Haah2013} between stabilizer models and polynomials over finite fields to characterize the \ob.
This framework provides an alternate algebraic perspective on the results of the previous section, and will carry over naturally to higher-dimensional models in Sections~\ref{sec: XCube} and~\ref{sec:fractal}.

We begin by reviewing the polynomial formalism (Sections~\ref{sec: intro algebraic} and ~\ref{sec: conservation laws algebraic}).
In Section~\ref{sec: commutations algebraic} we turn to the boundary commutation relations, introduced in Section~\ref{sec:obstructor} to characterize the \ob{} and bulk anyon data.
We show that these correspond to simple \emph{derivatives} in the polynomial formalism.
In Appendix~\ref{sec: bulk boundary computation}, we utilize this result to explicitly compute the commutation relations for arbitrary translation-invariant stabilizer models, and verify that they are indeed directly equal to the bulk mutual and self statistics.
This complements our pictorial arguments in the previous sections.

\subsection{Review of the polynomial formalism}\label{sec: intro algebraic}

We consider translation-invariant stabilizer models on a $D$-dimensional hypercubic lattice with $M$ qudits of Hilbert space dimension $n$ per unit cell.
We denote the number of stabilizers per unit cell as $T$ and again assume the stabilizers are geometrically $K$-local.

To represent the Pauli operators of the system using polynomials, first note that any Pauli operator, $\sigma$, can be uniquely decomposed (up to an overall phase) as a product of single-qudit Paulis, $X_{\bs i,m}$ and $Z_{\bs i,m}$:
\begin{equation}
    \sigma = \bigotimes_{\substack{\bs i \in \ZZ^D \\ m \in \{1, \ldots, N \}}} \left(Z_{\bs i,m}\right)^{a_{\bs i,m}} \left(X_{\bs i,m}\right)^{b_{\bs i,m}}.
    \label{equ:Pauliform}
\end{equation} 
Here, $\bs i, m$ index the unit cell and sublattice of the single-qudit Pauli operator, respectively, and the exponents, $a_{\bs i,m}, b_{\bs i,m}$, lie in $\ZZ_n$. 
Using this decomposition, we can equivalently represent the Pauli operator as a $2M$-component vector, $\bs \sigma \in \ZZ_n[x_1,\ldots,x_d,\bar x_1,\ldots,\bar x_d]^{\otimes 2M}$, of multivariate (Laurent) \emph{polynomials} over $\ZZ_n$, where $\bar x_d \equiv x_d^{-1}$:
\begin{equation}
%\bs{\sigma}_t(x_1,\ldots, x_d) \propto
\bs{\sigma}(x_1,\ldots,x_D) =
 \begin{pmatrix}
    \sum_{\bs i}  a_{\bs i,1}  x_1^{i_1} \cdots x_{D}^{i_D} \\
    \vdots\\
    \sum_{\bs i}  a_{\bs i,M}  x_1^{i_1} \cdots x_{D}^{i_D} \\
    \sum_{\bs i}  b_{\bs i,1}  x_1^{i_1} \cdots x_{D}^{i_D}\\
        \vdots\\
    \sum_{\bs i}  b_{\bs i,M}  x_1^{i_1} \cdots x_{D}^{i_D} \\
    \end{pmatrix}.
    \label{equ:polynomialform}
\end{equation}
In the future, we will suppress the dependence on $x_d$, $\bs \sigma(x_1,\ldots,x_D) \rightarrow \bs \sigma$, when clear from context. 
Here, the $m^{\text{th}}$ component of $\bs{\sigma}$ is a polynomial in $x_1,\ldots, x_D$ with coefficients $a_{\bs{i},m} \in \ZZ_n$ (corresponding to Pauli $Z$ operators), and the $(M+m)^{\text{th}}$ component is a  polynomial with coefficients $b_{\bs{i},m} \in \ZZ_n$ (corresponding to Pauli $X$ operators).
It will often be convenient to expand these polynomials power-by-power, in which case we denote $\bs{\sigma} = \sum_{\bs{i}} [ \bs{\sigma} ]_{\bs{i}} \, x_1^{i_1} \cdots x_n^{i_n}$, where $[ \bs{\sigma} ]_{\bs{i}} \in (\ZZ_n)^{\otimes 2M}$ is a vector over $\ZZ_n$. % 

A number of elementary matrix operations are represented easily within this formalism.
For instance, translation of a Pauli operator, ${\sigma} \rightarrow T_{\bs j}[{\sigma}]$, by a lattice vector $\bs j$ corresponds to \emph{multiplication} of $\bs{\sigma}$ by the monomial, $x_1^{j_1}\ldots x_D^{j_D}$.
Inversion about the site $\bs{i} = (0,\ldots,0)$ corresponds to exchanging $x_d \leftrightarrow \bar x_d$ for all $d=1,\ldots,D$.
Finally, multiplication of two Pauli operators, ${\sigma}_1$ and ${\sigma}_2$, corresponds to \emph{addition} of their polynomial vectors, ${\sigma}_1 \cdot {\sigma}_2 \leftrightarrow \bs{\sigma}_1 + \bs{\sigma}_2$.

We can also use these elementary operations to compute the commutation of two Pauli operators,
\begin{equation}
    {\sigma}_1 {\sigma}_2 = \exp \left[ i \frac{2 \pi}{n} \theta({\sigma}_1,{\sigma}_2) \right] \cdot {\sigma}_2 {\sigma}_1,
 \end{equation}
Here, $\theta(\hat{\sigma}_1,\hat{\sigma}_2) \in \ZZ_n$ determines the overall phase gained by commuting $\hat{\sigma}_1$ past $\hat{\sigma}_2$.
In fact, the polynomial formalism naturally computes the commutation of \emph{all translations} of $\bs{\sigma}_1$, $T_{\bs j}[\bs{\sigma}_1]$, with $\bs{\sigma}_2$.
These are calculated via the following inner product of polynomial vectors:
\begin{equation}
    \inner{\bs \sigma_1,\bs \sigma_2} = \bs \sigma_1^\dagger \bs \lambda_N \bs \sigma_2,
\end{equation}
which we refer to as the \textit{commutation polynomial}. 
 Here, $\bs \sigma^\dagger$ denotes a combination of matrix transposition and spatial inversion applied to $\bs \sigma$:
 \begin{equation}
    \bs \sigma^\dagger \equiv
 \begin{pmatrix}
    \sum_{\bs i}  a_{\bs i,1} \bar x_1^{i_1} \cdots x_{D}^{-i_D} \\
    \vdots\\
    \sum_{\bs i}  a_{\bs i,M}  \bar x_1^{i_1} \cdots x_{D}^{-i_D} \\
    \sum_{\bs i}  b_{\bs i,1} \bar x_1^{i_1} \cdots x_{D}^{-i_D}\\
        \vdots\\
    \sum_{\bs i}  b_{\bs i,M}  \bar  x_1^{i_1} \cdots x_{D}^{-i_D} \\
    \end{pmatrix}^T,
\end{equation}
 and $\bs \lambda_N$ is the symplectic matrix:
\begin{align}
\bs \lambda_N = \begin{pmatrix} 0_{N\times N} & \mathbbm 1_{N\times N}\\-\mathbbm{1}_{N\times N} & 0_{N\times N}\end{pmatrix}.
\end{align}
The coefficients of the commutation polynomial correspond to the desired commutators:
\begin{equation}
    \inner{\bs \sigma_1,\bs \sigma_2} = \sum_{\bs j} \theta(T_{\bs j}[{\sigma}_1],{\sigma}_2) \cdot x_1^{j_1} \ldots x_D^{j_D}.
\end{equation}

We illuminate this correspondence with a simple example.
Consider the commutation polynomial of an Pauli $X$ operator at the origin, ${\sigma_1} = X_{\bs{0}}$, with a Pauli $Z$ operator at a site $\bs{i}$, ${\sigma}_2 = Z_{\bs{i}}$ (taking $M=1$ for simplicity).
These are represented by polynomial vectors, $\bs \sigma_1 = (1, 0)^T$ and $\bs \sigma_2 = (0, x_1^{i_1} \cdots x_{D}^{i_D})^T$, respectively.
Their anti-commutation polynomial is $\inner{\bs \sigma_1,\bs \sigma_2} = x_1^{i_1} \cdots x_{D}^{i_D}$.
This has only a single non-zero term, indicating that the translation, $T_{\bs i}[\hat{\sigma}_1]$, anti-commutes with $\hat{\sigma}_2$, while all other translations commute.

In what follows, it will be convenient to consider the commutation relations within \emph{sets} of multiple Pauli operators, e.g. a set of $T$ operators, $\{ \hat{\sigma}_1 , \ldots, \hat{\sigma}_T \}$.
We can represent such a set by a polynomial \emph{matrix}:
\begin{align}
    \bs \Sigma = (\bs \sigma_1  \cdots \bs \sigma_T)
\end{align}
where the $t^{\text{th}}$ column is equal to the polynomial vector corresponding to $\hat{\sigma}_t$.
The commutation relations between each pair of elements in the set are now represented by an \textit{adjacency matrix}:
\begin{align}
    \bs A = \inner{\bs \Sigma,\bs \Sigma},
\end{align}
with entries, $A_{ij}=\inner{\bs \sigma_i,\bs \sigma_j}$, equal to the pairwise commutation polynomials. By definition, $\bs A$ is skew-Hermitian with respect to the dagger operation.

A translation-invariant stabilizer model is specified by a set of local stabilizer operators $\bs \Sigma$.
The stabilizers must mutually commute, i.e.~$\bs A = \inner{\bs \Sigma,\bs \Sigma}=0$.
The set of linear combinations of stabilizer vectors generates a stabilizer module, $S = \text{span}\left( \bs \Sigma \right) \in \mathbbm{Z}_d^{2M}[x_1,\ldots,x_D]$.
We further assume the set of local stabilizers $\bs \Sigma$ is complete, in the sense that every local Pauli operator $\bs \tau$ that commutes with all stabilizers, $\inner{\bs \tau,\bs \Sigma} = 0$, is itself already contained in the stabilizer module, $\bs \tau \in S$.

We can illustrate these concepts in the $\ZZ_2$ toric code.
Each lattice site contains $m=2$ spins (the horizontal and vertical edges), so the stabilizers are polynomial vectors with $2m = 4$ indices.
The bulk plaquette and vertex stabilizers [Eq.~(\ref{equ:2DTCstabilizers})] take the form~\cite{Haah2013}:
\begin{equation} \label{eq: toric bulk terms}
\bs \sigma_Z = 
\begin{pmatrix}
     1+ \bar x \\
    1+\bar y\\
     0 \\
   0 
   \end{pmatrix}, \,\,\,\,\,\,\,\,\, \bs \sigma_X = 
\begin{pmatrix}
    0\\
     0\\
      1+ y \\
    1+x
   \end{pmatrix},
\end{equation}
which for convenience we collect into the polynomial matrix,
\begin{equation}
    \bs \Sigma = (\bs \sigma_Z ,  \bs \sigma_X).
\end{equation}
Here we replace $x_1,x_2 \rightarrow x,y$ for clarity.
%
%We denote this compactly as  $\bs \Sigma = (\bs \sigma_Z ,  \bs \sigma_X)$. 
Recalling that the polynomial coefficients are now binary, it is straightforward to verify that all translations of the stabilizers mutually commute, i.e. $\inner{\bs \Sigma,\bs \Sigma}=0$.

\subsection{Bulk conservation laws}\label{sec: conservation laws algebraic}

We now address how bulk conservation laws appear in the polynomial formalism.
We restrict for now to two-dimensions, and for simplicity we again only consider conservation laws that involve products of stabilizers over every unit cell.

We begin in the toric code. 
The conservation law [Eq.~(\ref{eq: toric conservation laws})] corresponds to the fact that the product of vertex operators over all unit cells is equal to the identity, and similarly for the plaquette operators.
In the polynomial formalism, this takes the form,
\begin{align} \label{eq: toric bulk conservation}
    \sum_{i, j} x^i y^j \bs \sigma_Z(x,y) &= 0,& \sum_{i, j} x^i y^j \bs \sigma_X(x,y) &= 0,
\end{align}
where $i,j$ are summed over the integers. 
This can be neatly re-expressed using the following identity for infinite sums:
\begin{equation} \label{infinite sum evaluate}
\begin{split}
    \sum_{\bs{i}} x_1^{i_1} \ldots x_M^{i_M} & \cdot \bs \sigma(x_1, \ldots,x_M) \\
     & = \sum_{\bs{i}} x_1^{i_1} \ldots x_M^{i_M} \cdot \bs \sigma(1,\ldots,1),
\end{split}
\end{equation}
which is derived by expanding $\bs{\sigma}$ power-by-power and re-indexing the sum\footnote{For example in 2D we have, $\sum_{ij} x^{i} y^{j} \bs{\sigma}_Z(x,y) = \sum_{ijkl} [ \bs{\sigma}_Z ]_{kl} x^{i+k} y^{j+l} = \sum_{ijkl} [ \bs{\sigma}_Z ]_{kl} x^{i} y^{j} = \sum_{ij}  x^{i} y^{j} \bs{\sigma}_Z(1,1)$, where we re-index $i,j \rightarrow i-k,j-l$ in the third step.}.
The conservation laws [Eq.~(\ref{eq: toric bulk conservation})] are thus equivalent to the conditions that both $\bs \sigma_Z$ and $\bs \sigma_X$ have zeros at $(x,y) = (1,1)$:
\begin{align}
     \bs \sigma_Z(1,1)&=0, &  \bs \sigma_X(1,1)&=0.
\end{align}
This property is easily verified from Eq.~(\ref{eq: toric bulk terms}) for both plaquette and vertex operators.
%.

In generic systems, it may be the case that only certain subsets of the bulk stabilizers feature conservation laws.
To this end, we adopt a more general notation, in which each conservation law is labelled by a vector, $\bs c_\alpha \in \ZZ_d^T$, such that
\begin{align}
     \bs \Sigma(1,1) \cdot \bs c_\alpha = 0.
\end{align}
Recall that $T$ is number of stabilizers per unit cell, so $c_\alpha$ encodes combinations of stabilizers that product to the identity over the entire lattice.
In the case of the toric code, we have the following conservation laws for the three non-trivial anyons:
\begin{align} 
\bs c_{e} = 
\begin{pmatrix}
     1 \\
    0\\
   \end{pmatrix},  \,\,\,\,\,\,\,\,\,
 \bs c_{m} = 
\begin{pmatrix}
    0\\
     1\\
   \end{pmatrix}, \,\,\,\,\,\,\,\,\,
 \bs c_{f}  = 
\begin{pmatrix}
    1\\
     1\\
   \end{pmatrix},
\end{align}
We provide a more general algebraic formulation of conservation laws in Section~\ref{sec:fractal}.

\subsection{Boundary operator algebra and constraints}\label{sec: boundary algebraic}

We now turn to stabilizer models in the presence of open boundary conditions. 
For concreteness, we again specify to the toric code with a boundary at $i_y = 0$ (Fig.~\ref{fig:ToricCodeBoundary}).
In the polynomial formalism, this boundary entails truncating all terms with negative powers of $y$: in effect, setting $\bar y \rightarrow 0$ while leaving $y$.
We find the boundary operators [Eqs.~(\ref{equ:sigmaZ1D},\ref{equ:sigmaX1D})],
\begin{equation} \label{eq: toric Sigma boundary}
    \tbs \Sigma = (\tbs \sigma_Z ,  \tbs \sigma_X),
\end{equation}
with
\begin{align} \label{equ:truncatedopTC}
\tbs \sigma_Z  = \bs \sigma_Z \big|_{\bar y \rightarrow 0} = 
\begin{pmatrix}
     1+ \bar x \\
    1\\
     0 \\
   0 
   \end{pmatrix}, \,\,\,\,
\tbs \sigma_X  = \bar y \bs \sigma_X \big|_{\bar y \rightarrow 0} =
\begin{pmatrix}
    0\\
     0\\
      1 \\
    0
   \end{pmatrix}.
\end{align}
Note that the boundary operators are now described by \emph{single variable} polynomial vectors, $\tbs \Sigma = \tbs \Sigma(x)$, since the boundary is one-dimensional.
For more general models, we can systematically compute all allowed boundary operators by considering the truncation, $\bar y \rightarrow 0$, of each $y$-translation of the bulk stabilizers (i.e. $\tbs \sigma_j \equiv (\bar y^{j} \bs \sigma)|_{\bar y \rightarrow 0}$ for $j \in [0,\ldots,K-1]$, see Appendix~\ref{sec: bulk boundary computation}).

From Eqs.~(\ref{eq: toric Sigma boundary}-\ref{equ:truncatedopTC}), we can compute the adjacency matrix of the toric code boundary operators:
\begin{align}
    \bs A = \inner{\tbs \Sigma,\tbs \Sigma} = \begin{pmatrix}
  0  & 1+  x \\
     1+  \bar x  &0
   \end{pmatrix}.
\end{align}
The zero diagonal elements signify that boundary operators of the same type commute amongst themselves, while the non-zero off-diagonal elements signify that $\tbs \sigma_X$ operators anti-commute with each of their neighboring $\tbs \sigma_Z$ operators (corresponding to translations of $i_x = 0$ and $i_x=1$). 

We now turn to the constraints placed on the boundary Hilbert space by the bulk conservation laws.
As in Eq.~(\ref{equ:constraint1D}), the product of all bulk operators that feature a conservation law, $\bs c_\alpha$, is equal to the product of all boundary operators corresponding to truncations of the given bulk operators.
When the bulk is in the ground state, this enforces\footnote{Formally, this can implemented by taking the \emph{quotient} of the polynomial ring, $(\ZZ_n[x,\bar x])^{\otimes 2M}$, by the elements, $\sum_i x^i \bs c_\alpha$/ However, we will not need this formalism until Section~\ref{sec:fractal}.}
\begin{equation} \label{eq: boundary constraint algebraic}
    \sum_i x^i \tbs \Sigma(x) \cdot \bs c_\alpha \eqgs 0.
\end{equation}

These constraints restrict the form of the boundary adjacency matrix. %
To see this, note that the product of boundary operators in Eq.~(\ref{eq: boundary constraint algebraic}) necessarily commutes with every operator in the boundary Hilbert space (since the product of boundary operators is equal to a product of bulk operators, each of which by definition commutes with every boundary operator).
Therefore, we must have
\begin{align}
    \Inner{\tbs \Sigma(x), \sum_i x^i \tbs \Sigma(x) \bs c_\alpha} = \sum_i x^i \bs A(x) \bs c_\alpha = 0.
\end{align}
Using Eq.~(\ref{infinite sum evaluate}), this is equivalent to the condition,
\begin{align} \label{eq: consrv law algebraic}
    \bs A(1) \cdot \bs c_\alpha = 0,
\end{align}
i.e. the vector, $\bs A(x) \cdot \bs c_\alpha$, has a zero at $x=1$.
This property is clearly obeyed on the toric code boundary.

\subsection{Mutual- and self-\obs{}}\label{sec: commutations algebraic}

We now turn to the \obs{}.
With a modest effort, we will show that the \obs{} are given simply by \emph{derivatives} of the adjacency matrix $\bs A(x)$.

We begin with the mutual-obstructor invariant (right panel of Fig.~\ref{fig: bulk equals boundary statistics}). First, let us express the patch operators $L^\alpha_{j}$ and $R^\beta_{i}$ algebraically,
\begin{align}
\bs{L}_\alpha &= \sum_{i=-\infty}^{i_0} x^i \tbs{\Sigma} \bs{c}_\alpha, &
\bs{R}_\alpha &= \sum_{i=1}^\infty x^i \tbs{\Sigma} \bs{c}_\alpha,&
\end{align}
For simplicity, we have set the right endpoint to $1$ and the left end point to an integer $i_0 > K$. 

The commutation of the patch operators can be computed as
\begin{equation} \label{eq: compute mutual commutation 1}
\begin{split}
    \Inner{\bs{R}_\alpha,\bs{L}_\beta} & = \Inner{\sum_{i=1}^{\infty} x^i \tbs{\Sigma} \bs{c}_\alpha, \sum_{j=-\infty}^{i_0} x^j \tbs{\Sigma} \bs{c}_\beta} \\
    & = \sum_{i=1}^\infty \sum_{j=-i_0}^{\infty} x^{-i-j} \bs{c}_\alpha^\dagger \bs{A} \bs{c}_\beta \\
    & = x^{i_0} \sum_{k=1}^\infty k \cdot x^{-k} \bs{c}_\alpha^\dagger \bs{A} \bs{c}_\beta, \\
\end{split}
\end{equation}
where in the final line we use the identity, $\sum_{i=1
}^\infty \sum_{j=0}^\infty x^{-i-j} = \sum_{k=1}^\infty k \cdot x^{-k}$.
The mutual-obstructor invariant corresponds to the $x^0$-component of this inner product (since this corresponds to zero relative translation between the strings). We can compute this by expanding power-by-power, $\bs{A}(x) = \sum_i x^i [\bs A]_i$, and restricting to the zero-component,
\begin{equation} \label{eq: compute mutual commutation}
\begin{split}
    \tilde{b}(\alpha,\beta) & = \left[ x^{i_0} \sum_{k=0}^\infty k \cdot x^{-k} \bs{c}_\alpha^\dagger \bs{A} \bs{c}_\beta \right]_0 \\
    & = \bs{c}_\alpha^\dagger \left( \sum_{k=0}^\infty k \left[\bs{A}\right]_{k-i_0} \right) \bs{c}_\beta \\
    & = \bs{c}_\alpha^\dagger \left( \sum_{k=-\infty}^\infty k \left[\bs{A}\right]_{k} \right) \bs{c}_\beta. \\
\end{split}
\end{equation}
In the final step we extend the summation to negative infinity, since $[\bs{A}]_{k-i_0}$ is only non-zero for $k - i_0 > K$, and use Eq.~(\ref{eq: consrv law algebraic}), in the form $\bs{A} (1)\bs{c}_\beta = \sum_i [\bs{A}]_i \bs{c}_\beta = 0$, to shift the summation from $k \rightarrow k + i_0$.

Now, we notice that the above sum closely resembles the derivative of the adjacency matrix with respect to $x$.
Specifically, the Hasse derivative of a polynomial is defined power-by-power as
\begin{align}
     D_x[x^k] = k x^{k-1}.
\end{align}
Observing Eq.~(\ref{eq: compute mutual commutation}), we see that the commutation of boundary strings is equal to the derivative of the adjacency matrix evaluated at $x=1$,
\begin{equation}
    \tilde{b}(\alpha,\beta) = D_x[ \bs{c}_\alpha^\dagger \bs A \bs{c}_\beta](1).
\end{equation}
As a simple example, applying this to the toric code shows (taking $\bs{c}_Z$ and $\bs{c}_X$ as basis vectors) gives
\begin{align}
    \tilde{\bs b} = D_x[\bs A](1) = \begin{pmatrix} 0  & 1 \\ 1&0 \end{pmatrix}.
\end{align}
The off-diagonal elements signify that the $e$ and $m$ patch operators mutually anti-commute.

The self-obstructor invariant can be calculated similarly.
Consider the same commutation as above with $i_0 = 0$ and $\beta =\alpha$. 
Taking the $x_0$-component, we find 
\begin{equation} \label{eq: compute self commutation1}
\begin{split}
    \tilde{q}(\alpha) & = \left[ \sum_{k=1}^\infty k \cdot x^{-k} \bs{c}_\alpha^\dagger \bs{A} \bs{c}_\alpha \right]_0 \\
    & =  \sum_{k=1}^\infty k \cdot \left[ \bs{c}_\alpha^\dagger \bs{A} \bs{c}_\alpha \right]_k. \\
\end{split}
\end{equation}
Unlike the mutual-\ob{}, the self-\ob{} involves only the positive ``half'' (i.e. powers $k > 0$) of the polynomial $a_\alpha(x) \equiv [\bs{c}_\alpha^\dagger \bs{A} \bs{c}_\alpha](x)$.
We can re-express this using the following two properties: first, $a_\alpha(x) = -a_\alpha(\bar x)$, since this simply flips the ordering of the commutator, and second, $[a_\alpha]_0 = 0$, since any operator commutes with itself.
Together, these allow one to decompose $a_\alpha(x)$ as the sum of a \emph{positive part}, $a_\alpha^+(x)$, and its Hermitian conjugate, $(a^+_\alpha)^\dagger(x) = - a^+_\alpha(\bar x)$,
\begin{equation} \label{eq: decompose pos part}
    a_\alpha(x) \equiv [\bs{c}_\alpha^\dagger \bs{A} \bs{c}_\alpha](x) = a_\alpha^+(x) - a_\alpha^+(\bar x).
\end{equation}
For example, one could define $a_\alpha^+$ to contain only the positive powers within $a_\alpha$, $a_\alpha^+(x) = \sum_{k=1}^{\infty} [a_\alpha]_k x^k$, while its conjugate contains the negative powers.
The self-\ob{} is equal to the derivative of the positive part,
\begin{equation} \label{eq: compute self commutation2}
\begin{split}
    \tilde{q}(\alpha) & =  \sum_{k=1}^\infty k \cdot \left[ a_\alpha^+(x) - a_\alpha^+(\bar x) \right]_k \\
    & =  \sum_{k=1}^\infty k \cdot \left( \left[ a_\alpha^+ \right]_k - \left[ a_\alpha^+ \right]_{-k} \right) \\
    & =  \sum_{k=-\infty}^\infty k \cdot \left[ a_\alpha^+(x) \right]_k \\
    & =  D_x \left[ a_\alpha^+ \right](1). \\
\end{split}
\end{equation}
This result holds regardless of the particular decomposition, i.e.~for any valid choice of $a^+_\alpha(x)$. Suppose we had chosen a different decomposition $a_\alpha(x) = {a_\alpha^+(x)}' - {a_\alpha^+(\bar x)}'$, then such a choice can only differ up to a symmetric polynomial, so ${a_\alpha^+}' - a_\alpha^{+} = \sum_n s_n (x^n + \bar x^n)$ for some coefficients $s_n$. However, $D_x[x^n + \bar x^n]_{x=1} = [nx^{n-1} - n\bar x^{n+1}]_{x=1}=0 $, thus  $D_x [ {a_\alpha^+}' ](1) =  D_x [ a_\alpha^+ ](1).$

We have shown that the \obs{} can be computed simply as derivatives of the adjacency matrix on the boundary.
We can now provide an algebraic perspective for why the invariants obstruct a local tensor product realization, by showing that in a LTPS these derivatives must always vanish.
Consider a 1D LTPS with local operators, $\bs \Sigma_{\text{TP}}$.
Suppose that these operators obey a set of conservation laws, $\bs c_\alpha$,
\begin{equation} \label{eq: TP constraint algebraic}
    \sum_i x^i \cdot \bs \Sigma_{\text{TP}}(x) \bs c_\alpha = 0.
\end{equation}
Using Eq.~(\ref{infinite sum evaluate}), the conservation law is equivalent to the condition $\bs \Sigma_{\text{TP}}(1) \bs c_\alpha = 0$.
Now, this implies that $\bs \Sigma(x) \bs c_\alpha$ contains a factor of $x-1$. That is, we can write
\begin{equation}
    \bs \Sigma_{\text{TP}}(x) \bs c_\alpha = (x-1) \bs \tau_\alpha(x),
\end{equation}
for some polynomial vector $\bs \tau_\alpha$.
However, this implies that inner products of conservation laws with the adjacency matrix, $\bs A_{\text{TP}} = \inner{\bs \Sigma_{\text{TP}}, \bs \Sigma_{\text{TP}}}$, contain \emph{two} factors of $x-1$,
\begin{equation}
    \bs{c}_\alpha^\dagger \bs{A}_{\text{TP}} \bs{c}_\beta = (x-1)(\bar x-1) \inner{\bs \tau_\alpha, \bs \tau_\beta}
\end{equation}
and thus have derivative zero at $x=1$,
\begin{equation}
    D_x[\bs{c}_\alpha^\dagger \bs{A}_{\text{TP}} \bs{c}_\beta ](1) = 0.
\end{equation}
Hence, any non-zero mutual-\ob{} signifies that the boundary is not a LTPS.

A similar argument applies to the self-\ob{}. In a 1D LTPS, we have
\begin{equation}
\begin{split}
    \bs{c}_\alpha^\dagger \bs{A}_{\text{TP}} \bs{c}_\alpha & = (x-1)(\bar x - 1) \inner{\bs \tau_\alpha, \bs \tau_\alpha} \\
    & = (x-1)(\bar x - 1) \left( t^+_\alpha(x)- t^+_\alpha(\bar x) \right)   \\
\end{split}
\end{equation}
where we decompose $\inner{\bs \tau_\alpha, \bs \tau_\alpha} = t^+_\alpha(x) - t^+_\alpha(\bar x)$, similar to Eq.~(\ref{eq: decompose pos part}).
The positive part of the commutation thus also has derivative zero:
\begin{equation}
    D_x[ (x-1)(\bar x - 1)t^+_\alpha](1) = 0.
\end{equation}
Hence, any non-zero self-\ob{} signifies that the boundary is not a LTPS.

In Appendix~\ref{app: 1D LTPS}, we go further and show that the polynomials $[\bs c^\dagger_\alpha \bs A \bs c_\beta](x)$ are in fact \emph{fully} characterized by the values of the \obs{}, up to tensor products with stabilizers in a 1D LTPS.

\section{Boundaries of Type-I fracton models}\label{sec: XCube}

Stabilizer codes in higher dimensions can realize a vastly greater variety of topological phases compared to two dimensions.
We find that conventional topological orders in higher dimensions, such as higher-dimensional toric codes and their relatives, display a qualitatively similar bulk-boundary correspondence as in two-dimensional topological orders.
Thus, we relegate their analysis to Appendix \ref{sec:higherdimtoriccode} for interested readers.

In the remainder of the work, we instead focus on new bulk-boundary phenomena that occur in three-dimensional fracton orders. In this section, we will focus on Type-I fracton orders, specifying for concreteness to the X-Cube model.
We will show that the bulk-boundary correspondence of Type-I fracton boundaries depends on the \emph{orientation} of the boundary considered, in stark contrast to our results on conventional topological orders. 
We construct \obs{} for fracton models, and show that they allow one to distinguish the X-Cube boundary from both a LTPS, and, more generally, the boundary of any stack of 2D topological orders. 
In the following Section~\ref{sec:fractal}, we extend this analysis to fracton models with fractal conservation laws, including Type-II fracton orders such as Haah's code.

\subsection{Review of X-Cube model}

\begin{figure}[b]
    \centering
    \includegraphics[width=\columnwidth]{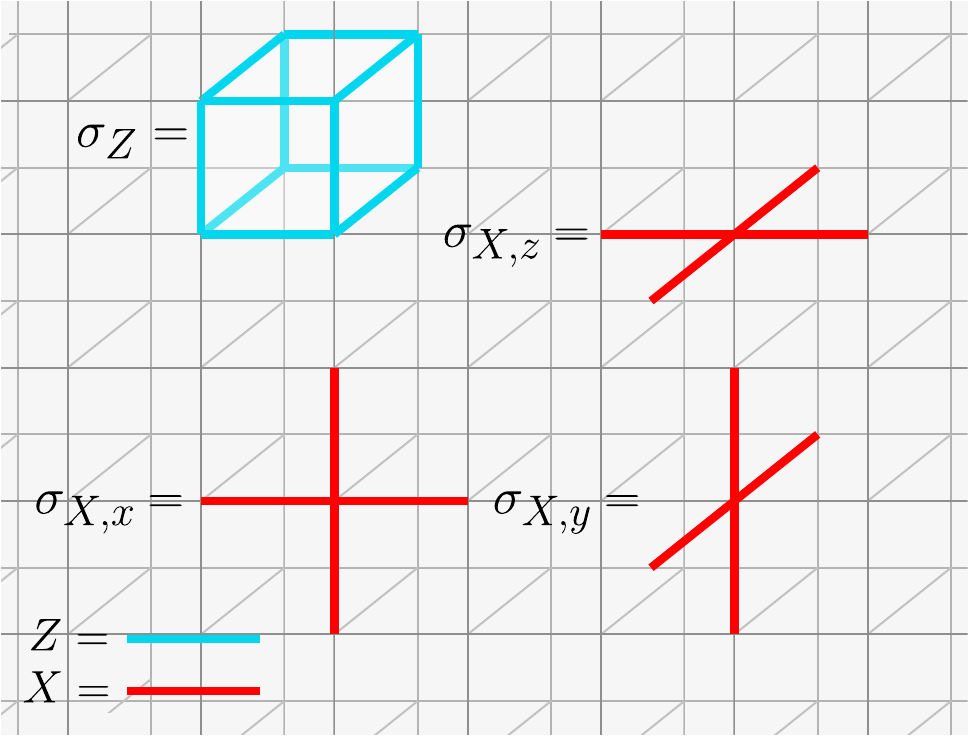}
    \caption{Cube (blue) and vertex (red) stabilizers of the X-Cube model.}
    \label{fig:Xcube}
\end{figure}
\begin{figure*}[t]
\centering
\includegraphics[width=0.9\textwidth]{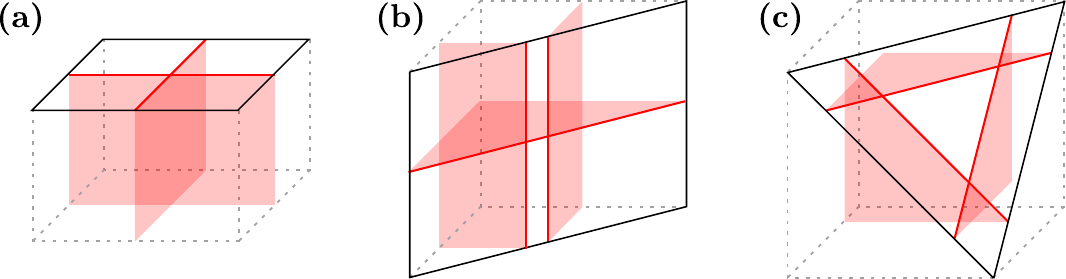}
\caption{
Bulk conservation laws over the $xy$-,$yz$-, and $zx$-planes terminate to form linear constraints on the boundary of the X-Cube model.
\textbf{(a)} The (001)-boundary has two constraints, from the $yz$- and $zx$-planes. The $xy-$plane does not contribute a constraint, since it is parallel to the boundary.
\textbf{(b)} The (110)-boundary similarly has only two constraints. The $yz$- and $zx$-planes terminate in an identical constraint (up to tensor products with stacks of 2D toric codes, see text), while the $xy$-plane forms a second independent constraint.
\textbf{(c)} The (111)-boundary has three independent constraints in the $xy$-,$yz$-, and $zx$-planes.
} 
\label{fig: terminations}
\end{figure*}

The Hamiltonian of the X-Cube model,
\begin{equation}
    H_{TC} =  - J \sum_i \sigma^{Z}_{\bs{i}} - J \sum_{\bs{i}}\sum_{r=x,y,z} \sigma^{X,r}_{(\bs{i})},
\end{equation}
consists of a 12-spin cube stabilizer,
\begin{equation}
\begin{split}
    \sigma^Z_{\bs{i}} =  &Z_{\bs{i},x}Z_{\bs{i}-\hat{\bs{y}},x}Z_{\bs{i}-\hat{\bs{z}},x}Z_{\bs{i}-\hat{\bs{y}}-\hat{\bs{z}},x}\\
   & Z_{\bs{i},y}  Z_{\bs{i}-\hat{\bs{x}},y}Z_{\bs{i}-\hat{\bs{z}},y}Z_{\bs{i}-\hat{\bs{z}}-\hat{\bs{x}},y}\\
    & Z_{\bs{i},z} Z_{\bs{i}-\hat{\bs{x}},z}Z_{\bs{i}-\hat{\bs{y}},z}  Z_{\bs{i}-\hat{\bs{x}}-\hat{\bs{y}},z},
\end{split}
\label{equ:XcubeZ}
\end{equation}
 and three 4-spin vertex stabilizers, 
\begin{equation}
\begin{split}
    \sigma^{X,x}_{\bs{i}} = &  X_{\bs{i},y} X_{\bs{i},z} X_{\bs{i}+\hat{\bs{y}},y} X_{\bs{i}+\hat{\bs{z}},z} \\
    \sigma^{X,y}_{\bs{i}} = &  X_{\bs{i},z} X_{\bs{i},x} X_{\bs{i}+\hat{\bs{z}},z} X_{\bs{i}+\hat{\bs{x}},x} \\
    \sigma^{X,z}_{\bs{i}} = &  X_{\bs{i},x} X_{\bs{i},y} X_{\bs{i}+\hat{\bs{x}},x} X_{\bs{i}+\hat{\bs{y}},y}, \\
\end{split}
\label{equ:XcubeX}
\end{equation}
as depicted in Fig.~\ref{fig:Xcube}.
In the polynomial formalism, the stabilizers can be written as
\begin{equation} \label{X Cube bulk stabilizers}
\begin{split}
    \bs \Sigma 
    & =
\begin{pmatrix} (1+\bar y)(1+\bar z) & 0&0\\
(1+\bar x)(1+\bar z) &0&0\\
(1+\bar x)(1+\bar y)&0&0\\
0 & 1+x &0\\
 0&0 &1+y\\
 0&1+z&1+z
 \end{pmatrix} \\
 & = \begin{pmatrix} \bs \sigma_Z & \bs \sigma_{X,y} & \bs \sigma_{X,x} \end{pmatrix},
\end{split}
\end{equation}
where we use the fact that the three vertex operators product to the identity,
\begin{equation}
    \bs \sigma_{X,x} + \bs \sigma_{X,y} + \bs \sigma_{X,z} = 0,
\end{equation}
to neglect $\bs \sigma_{X,z}$.

\begin{figure*}
    \centering
    \includegraphics[width=\textwidth]{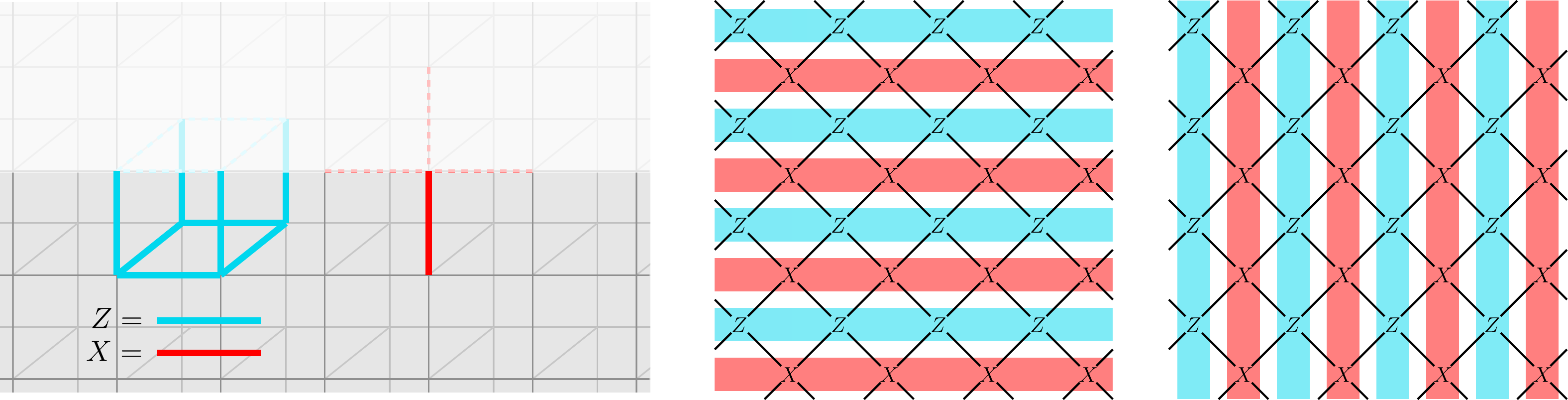}
    \caption{(Left) Truncated operators at the (001)-boundary of X-cube. (Right) Frustration graph and constraints of the boundary operators. Each boundary operator participates in a linear constraint in both the $x$- and $y$-directions.}
    \label{fig:001Xcubeboundaryop}
\end{figure*}

The conservation laws of the X-Cube model correspond to products of bulk stabilizers over the $xy$-, $yz$-, and $zx$-planes.
The cube stabilizer has conservation laws in all three orientations,
\begin{align} \label{cube conservation laws}
    \sum_{j,k}y^jz^k \bs \sigma_Z = \sum_{i,k}x^iz^k \bs \sigma_Z = \sum_{i,j}x^iy^j \bs \sigma_Z=0
\end{align}
while the vertex stabilizers have conservation laws in one orientation each,
\begin{align} \label{3 vertex conservation laws}
   \sum_{j,k}y^jz^k \bs \sigma_{X,x}=  \sum_{i,k}x^iz^k \bs \sigma_{X,y}  = \sum_{i,j}x^iy^j \bs \sigma_{X,z}=0.
\end{align}
Here, each sum runs from negative to positive infinity.

At a formal level, the story of quasiparticles in the X-Cube model proceeds similarly to the toric code, even though the behavior of the quasiparticles differs greatly.
The X-Cube model has $e$-quasiparticles that violate the cube stabilizers $\bs{\sigma}_Z$.
The conservation laws in Eq.~(\ref{cube conservation laws}) enforce that the parity of $e$-quasiparticles in each plane of the lattice is conserved.
These conservation laws imply that a lone $e$-quasiparticle is \emph{fractonic}: it is not free to move in any direction, since any movement will change the parity in at least one plane.
The model also has $m_z$-quasiparticles that violate the vertex stabilizers $\bs{\sigma}_{X,y}$ and $\bs{\sigma}_{X,x}$ (and similar for $m_x$ and $m_y$-quasiparticles).
The conservation laws Eq.~(\ref{3 vertex conservation laws}) enforce conservation of the parity of $m_z$-quasiparticles in the $xz$- and $yz$-planes.
This implies that a lone $m_z$-quasiparticle is a \emph{lineon} that is free to move only in the $z$-direction.
An $m_x$- and $m_y$-quasiparticle can fuse to form an $m_z$-quasiparticle (and similar for other permutations).

As in the toric code, we can use the conservation laws to formulate exchange operations for the quasiparticles.
Consider the product of all cube stabilizers over a finite rectangular prism.
The conservation laws, Eq.~(\ref{cube conservation laws}), imply that the resulting operator is the identity within the bulk of the prism as well as on each face of the prism.
This ``cage'' operator is non-identity only along the edges of the prism, which can be viewed as transporting lineons along the edge (at each corner, an $m_x$- and $m_y$-lineon fuse to form an $m_z$-lineon). 
Meanwhile, an analogous product of $\bs \sigma_{X,y}$ vertex operators is the identity within the bulk of the prism as well as on the $xz$-face [observing Eq.~(\ref{3 vertex conservation laws})].
The resulting operator is non-identity along four of the faces, which can be viewed as transporting pairs of distant fractons around a closed loop.
The cage and net operators allow us to formulate mutual statistics between the fractons and lineons.
In particular, we can consider transporting a lineon around a fracton excitation via the cage operator, which results in a minus sign being applied to the many-body wavefunction.

Recent work has also introduced a ``windmill'' self-exchange operation for the fracton quasiparticles~\cite{song2023fracton}.
This exchange process gives trivial statistics for $e$ quasiparticle in the X-Cube model, but can be non-zero for bound states of $e$ and $m$ particles, as well as in other fracton models.
The process involves exchanging two triplets of fractons via a third triplet of locations, in a manner similar to the self statistics exchange process in two-dimensional topological orders.

\subsection{Boundary operator algebra and subsystem constraints}

We now turn to the boundary Hilbert space of the X-Cube model.
We will show that a central feature of the boundary Hilbert space is that the planar conservation laws of the bulk Hamiltonian terminate as linear \emph{subsystem constraints} on the boundary (Fig.~\ref{fig: terminations}).
Thus, much as the boundary of the toric code can be thought of as the symmetric sector of a one-dimensional spin chain, the boundary  of the X-Cube model can be thought of as the subsystem-symmetric sector of a two-dimensional spin lattice.
Intriguingly however, the precise subsystem constraints depend on the orientation of the boundary chosen.
In what follows, we explore this in (001)-, (110)-, and (111)-boundaries.

\emph{(001)-boundary}---The boundary operators of the (001)-boundary are shown in Fig.~\ref{fig:001Xcubeboundaryop}.
They can be derived algebraically as in the previous section, by translating each stabilizer such that it traverses the boundary and setting powers of $\bar z$ to zero.
This gives:
\begin{align}
    \tbs \Sigma &= 
 \begin{pmatrix} 1+ \bar y & 0\\
1+\bar x &0&\\
(1+ \bar x)(1+ \bar y)&0\\
0 & 0 \\
 0&0 \\
 0&1
 \end{pmatrix} = \begin{pmatrix} \tbs \sigma_Z & \tbs \sigma_{X} \end{pmatrix}.
\end{align}
Note that the two bulk operators $\bs \sigma_{X,y}$ and $\bs \sigma_{X,x}$ give rise to the same boundary operator since their product, $\bs \sigma_{X,z}$, remains a bulk stabilizer.
Eliminating this redundancy leaves us with two independent local boundary operators, as above. The adjacency matrix of the boundary operators is
\begin{align}
    \bs A(x,y) =  \begin{pmatrix}0 &(1+x)(1+y)\\
(1+\bar x)(1+\bar y) &0
 \end{pmatrix}.
\end{align}
The $X$- and $Z$-boundary operators anti-commute in a checkerboard pattern as shown in Fig.~\ref{fig:001Xcubeboundaryop}. The algebra matches that of the subsystem-symmetric subspace of the Xu-Moore (plaquette Ising) model~\cite{XuMoore2004}.

The bulk conservation laws of the X-Cube model [Eqs.~(\ref{cube conservation laws},\ref{3 vertex conservation laws})] enforce linear subsystem constraints on the boundary Hilbert space (Fig.~\ref{fig: terminations}).
In the polynomial formalism, these are:
\begin{align}
    \sum_i \sum_{k=0}^\infty x^iz^k   \bs \sigma_{Z} 
& = \sum_i x^i \tbs \sigma_{Z} \eqgs 0,\\
       \sum_{j} \sum_{k=0}^\infty y^iz^k  \bs \sigma_{Z}  
&= \sum_j y^j \tbs \sigma_{Z} \eqgs 0,\\
%\sum_{i,j}x^iy^j \bs \sigma_{1} &=0\\ 
\sum_{i} \sum_{k=0}^\infty x^iz^k  \bs \sigma_{X,y} 
&=\sum_i x^i \tbs \sigma_{X} \eqgs 0,\\
\sum_{j} \sum_{k=0}^\infty y^iz^k  \bs \sigma_{X,x} 
&=\sum_j y^j \tbs \sigma_{X}\eqgs 0.
\end{align}
Each boundary operator is involved in two constraints, one in each of the $x$- and $y$-directions.
Note that there is no constraint arising from the bulk $xy$-conservation laws, since they run parallel to the (001)-boundary.
As in Section~\ref{sec: boundary algebraic}, the constraints must commute with all local boundary operators.
The analogue of Eq.~(\ref{eq: consrv law algebraic}) becomes
%dictate that the $Z$-type constraints must commute with the truncated $X$-type operator, we obtain
\begin{align}
    \bs A(x,1)= \bs A(1,y)=0.
\end{align}
This equation contains four constraints, for each combination of a basis vector $\bs c_X$, $\bs c_Z$ and a direction $x,y$.

\emph{(110)-boundary}---Unlike the (001)-boundary, all three conservation laws of the bulk topological order terminate non-trivially on the (110)-boundary.
However, as depicted in Fig.~\ref{fig: terminations}, the conservation laws in the $xz$- and $yz$-planes terminate in parallel lines, along the $z$-direction on the (110)-boundary.
Thus, it is not immediately clear whether these terminations give rise to independent or redundant constraints.
In what follows, we show that, to a large extent, the latter is the case.
Specifically, we find that the (110)-boundary operator algebra is equivalent to the tensor product of the (001)-boundary operator algebra and the boundary operator algebra of a stack of toric codes along the (110)-direction.

To calculate the truncated boundary operators, we first define the new coordinate $w = x \bar y$, which runs parallel to the (110)-boundary.
Following our usual procedure, the truncated boundary operators are
\begin{equation}
    \tbs \Sigma =
    \begin{pmatrix} 
        \bar w ( 1 + \bar z) & 0 & 0 & 0 \\
        1 + \bar z & 0 & 0 & 0 \\
        1 + \bar w & 1 & 0 & 0 \\
        1 & 0 & 0 & 0 \\
        0 & 0 & 1 & 0 \\
        0 & 0 & 0 & 1 \\
        0 & 0 & 1+z & 1+z \\
        0 & 0 & 0 & 0 \\
    \end{pmatrix}.
\end{equation}
The eight rows correspond to four spins: the first three correspond to the $x$-, $y$-, and $z$-bonds on sites of the form $w^i z^j$, the fourth corresponds to $z$-bonds on sites along $x^{-1} w^i z^j$.
The first two columns correspond to two different truncations of the cube stabilizer $\bs \sigma_Z$, one acting on seven spins and one acting on only a single spin.
The second two columns correspond to truncations of the vertex stabilizers $\bs \sigma_{X,x}$ and $\bs \sigma_{X,y}$.

The constraints on the boundary Hilbert space are as follows.
Along the $z$-direction, we have constraints for  $\bs c = ( 1, \bar w, 0, 0 )^T , ( 1,1,0,0 )^T, ( 0, 0,1,0)^T , ( 0, 0,0,1)^T$ and all linear combinations thereof.
Meanwhile, along the $w$-direction we have constraints for $\bs c = ( 1, 1,0,0)^T ,  ( 0,0,1,1 )^T$.

To diagnose the structure of the boundary Hilbert space, we analyze the matrix elements of the commutator $\bs A$ under the constraints.
From the above description, we see that there are two vectors with constraints in both the $z$- and $w$-directions, $\bs c_Z = (1,1,0,0)^T$ and $\bs c_X = (0,0,1,1)^T$.
Each constraint has zero commutator with itself, and the two have a mutual commutator $\bs c_Z \bs A \bs c_X = (1+w)(1+z)$.
This is an identical commutation structure as the (001)-boundary.

We now turn to the vectors $\bs c$ that only have constraints in the $z$-direction.
We can form a basis for these vectors by defining $\bs c^z_Z = (1, \bar w , 0 , 0)^T$ and $\bs c^z_X = ( 0,0 , 1 , 0)^T$.
Both of these entirely commute with the constraints, $\bs c_Z$ and $\bs c_X$, in the previous paragraph.
They also both have zero commutator with themselves.
Their mutual commutator is $\bs c^z_Z \bs A \bs c^z_X = (1+w)(1+z)$.
This has a non-zero first derivative corresponding to the $z$-direction constraint, $D_z[ \bs c^z_Z \bs A \bs c^z_X ](z=1) = 1$, as would be found at the boundary of a stack of toric codes.
Indeed, the same constraints and commutator can be achieved at the boundary of a stack of toric codes, by `pairing' $X$ operators in adjacent stacks (i.e.~taking $\bs c_X \rightarrow (1+w) \bs c_X$ where here $\bs c_X$ denotes the usual boundary operator in a toric code boundary along the $z$-direction).
We conclude that the (110)-boundary is equivalent to the (001)-boundary tensored with the boundary of a stack of toric codes.

\begin{figure*}[t]
    \centering
    \includegraphics[width=0.87\textwidth]{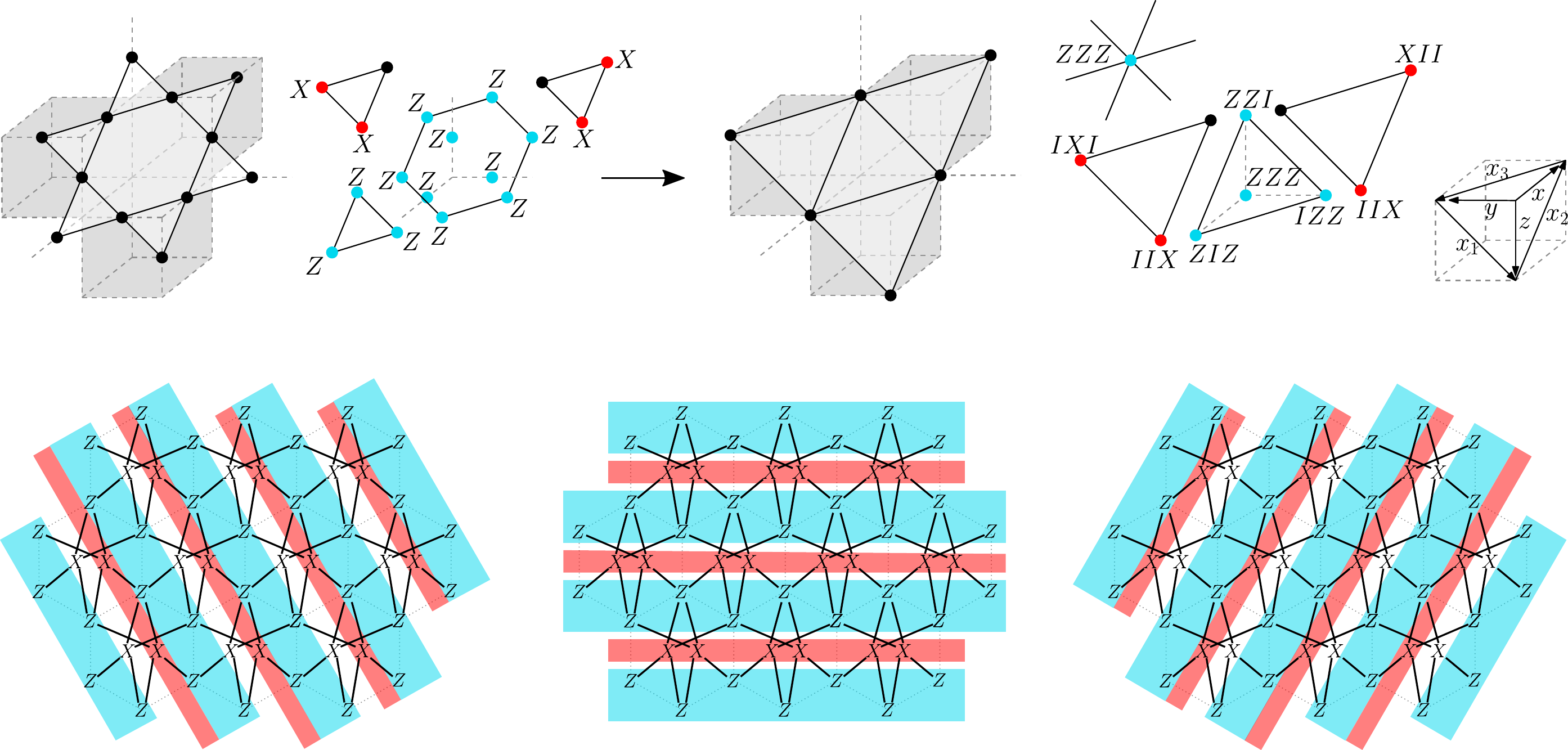}
    \caption{(Top left) Truncated operators of the (111)-boundary of the X-cube model on the Kagome lattice. (Top right) The same operators on a triangular lattice with three sites per unit cell. Two out of the  three coordinates, $x_1 = z\bar y$, $x_2=x\bar z$, $x_3= y\bar x$, span the boundary. (Bottom) Commutation graph and constraints of the (111)-boundary. Each operator participates in three constraints, in the $x_1$-, $x_2$-, and $x_3$-directions.}
    \label{fig:111Xcube}
\end{figure*}

\emph{(111)-boundary}---The edges of a cubic lattice naturally form vertices of a Kagome lattice on the (111)-boundary.
The corresponding truncated boundary operators are shown in Fig.~\ref{fig:111Xcube}.
As in the bulk, it is convenient to label the boundary spins by the cubic lattice vertex that they extend from.
This leads us to group each trio of spins on upward triangles on the Kagome lattice.
The resultant lattice is triangular with three spins per unit cell (Fig.~\ref{fig:111Xcube}).

To derive the boundary operators algebraically, we define the new monomials $x_1 = z\bar y$, $x_2 = x\bar z$, $x_3 = y\bar x$.
Note that the third monomial is redundant with the first two since $x_1 x_2 x_3 = 1$.
Translations by any of these monomials are parallel to the (111)-boundary.
We can therefore use $x_1$ and $x_2$ to parameterize the boundary operators, while the independent monomial $z$ parameterizes translations into the bulk.
Substituting these into Eq.~(\ref{X Cube bulk stabilizers}), we can rewrite the bulk stabilizers as
\begin{equation}
\begin{split}
\bs \Sigma & = 
\begin{pmatrix} 
 (x_1+ z)(1+ z) & 0&0\\
(\bar x_2+ z)(1+ z) &0&0\\
(\bar x_2+  z)(x_1 + z)&0&0\\
0 & 1 + x_2 z &0\\
 0&0 &1+ \bar x_1 z\\
 0&1+z&1+z
\end{pmatrix}\\
& = \begin{pmatrix} \bs \sigma_Z & \bs \sigma_{X,y} & \bs \sigma_{X,x} \end{pmatrix}.
\end{split}
\end{equation}

The boundary operators are obtained by taking various translations of the bulk stabilizers, and setting negative powers of $z$ to zero.
This gives two boundary operators per unit cell for the cube stabilizer (corresponding to truncations of $z^{-1} \bs \sigma_Z$ and $z^{-2} \bs \sigma_Z$), and a single boundary operator for each vertex stabilizer (corresponding to $z^{-1} \bs \sigma_{X,\alpha}$):
\begin{equation}
\begin{split}
\tbs \Sigma & =\begin{pmatrix}
    1 & z + x_1 + 1 &0 &0\\
    1 & z + \bar x_2 + 1  &0 &0\\
    1 & z + x_1 + \bar x_2 &0 &0 \\
    0&0&x_2&0\\
    0&0&0& \bar x_1\\
    0&0&1&1
\end{pmatrix}\\
& = \begin{pmatrix} \tbs \sigma_{Z,1} & \tbs \sigma_{Z,2} & \tbs \sigma_{X,y} & \tbs \sigma_{X,x} \end{pmatrix}.
\end{split}
\end{equation}
This is shown in Fig.~\ref{fig:111Xcube}.
It is important to note that in the boundary theory, the $z$ variable should be understand as an extension of the unit cell, and so shifts of the above operators by powers of $z$ are not physical.
For example, the first order in $z$ terms correspond to spins lying one layer ``into the bulk'' in Fig.~\ref{fig:111Xcube}.
An alternate way to denote this would be to eliminate the $z$ variable entirely and introduce new row vectors corresponding to such terms.
In the current notation, the adjacency matrix of the boundary operators is obtained by taking the $z^0$-component of the inner product $\inner{\tbs \Sigma, \tbs \Sigma}$, which gives
\begin{align}
\bs A = 
 \begin{pmatrix}
 0&0&   1 + x_2       & 1 +  \bar x_1\\
  0&0&   \bar x_1 (1 + x_2)    & x_2 (1 + \bar x_1)\\
1 + \bar x_2  & x_1 (1 + \bar x_2)  &0&0\\
1 + x_1 & \bar x_2 (1 + x_1)  &0&0
 \end{pmatrix}.
\end{align}

The bulk conservation laws of the X-Cube model terminate into three orientations of line constraints on the (111)-boundary [Fig.~\ref{fig: terminations}(c) and Fig.~\ref{fig:111Xcube}].
To derive these constraints explicitly, let us first re-write the bulk conservation laws in the $x_1, x_2, z$ coordinates. Focusing on the cube stabilizers, we have:
\begin{equation}
\begin{split}
    \sum_{i,j} x^i y^j \cdot \bs \sigma_Z & = \sum_{i,j} z^{i+j} x_2^i \bar x_1^j \cdot \bs \sigma_Z = 0  \\
    \sum_{i,j} y^i z^j \cdot \bs \sigma_Z & = \sum_{i,j} z^{i+j} \bar x_1^i  \cdot \bs \sigma_Z = 0 \\
    \sum_{i,j} z^i x^j \cdot \bs \sigma_Z & = \sum_{i,j} z^{i+j} x_2^j   \cdot \bs \sigma_Z = 0.
\end{split}
\end{equation}
The termination of the conservation laws on the boundary involves terms $\tbs \sigma_{Z,1}, \tbs \sigma_{Z,2}$, which correspond to powers $z^{-1}, z^{-2}$ respectively.
Isolating such terms, we find the boundary constraints
\begin{equation}
\begin{split}
    \sum_{i} x_1^i \cdot ( \tbs \sigma_{Z,1} + \tbs \sigma_{Z,2} ) \equiv  \sum_{i} x_1^i \cdot (\tbs \Sigma \cdot \bs c_{Z,1}) & \eqgs  0 \\ %( x_1 Z^e_0 + Z^e_1 ) \cdot \sum_{i} x_1^i \\
    \sum_{i} x_2^i \cdot ( \tbs \sigma_{Z,1} + x_2 \tbs \sigma_{Z,2} ) \equiv  \sum_{i} x_2^i \cdot (\tbs \Sigma \cdot \bs c_{Z,2}) & \eqgs  0 \\ 
    \sum_{i} x_3^i \cdot ( \tbs \sigma_{Z,1} + x_1 \tbs \sigma_{Z,2} ) \equiv  \sum_{i} x_3^i \cdot (\tbs \Sigma \cdot \bs c_{Z,3}) & \eqgs  0, \\ 
\end{split}
\end{equation}
where we define the vectors $\tbs c_{Z,1} = (1,1,0,0)^T$, $\tbs c_{Z,2} = (1,x_2,0,0)^T$, $\tbs c_{Z,3} = (1,x_1,0,0)^T$ to compactify our notation.
Performing a similar procedure for the vertex stabilizers gives:
\begin{equation}
\begin{split}
    \sum_{i}  x_1^i  \cdot \bs \sigma_{X,x} 
    \equiv  \sum_{i} x_1^i \cdot (\tbs \Sigma \cdot \bs c_{X,1}) 
    & \eqgs 0 \\
    \sum_{i}  x_2^i  \cdot \bs \sigma_{X,y} 
    \equiv  \sum_{i} x_2^i \cdot (\tbs \Sigma \cdot \bs c_{X,2}) 
    & \eqgs 0 \\
    \sum_{i}  x_3^i  \cdot (\bs \sigma_{X,x}+\bs \sigma_{X,y}) 
    \equiv  \sum_{i} x_3^i \cdot (\tbs \Sigma \cdot \bs c_{X,3}) 
    & \eqgs 0. \\
\end{split}
\end{equation}
for $\tbs c_{X,1} = (0,0,0,1)^T$, $\tbs c_{X,2} = (0,0,1,0)^T$, $\tbs c_{Z,3} = (0,0,1,1)^T$.
Each local boundary operator is thus involved in three constraints, one in each of the $x_1$-, $x_2$- and $x_3$-directions.
The constraints must again commute with all local boundary operators, which enforces:
\begin{equation}
\begin{split}
    \bs A(1,x_2) \cdot \tbs c_{P,1} & = 0 \\
    \bs A(x_1,1) \cdot \tbs c_{P,2} & = 0 \\
    \bs A(x_1,\bar x_1) \cdot \tbs c_{P,3} & = 0
\end{split}
\end{equation}
for $P = \{X,Z\}$.
Note that the final expression corresponds to setting $x_3 = \bar x_1 \bar x_2 = 1$.

\subsection{Patch operators and \obs{}}

\begin{figure*}
\centering
\includegraphics[width=0.95\textwidth]{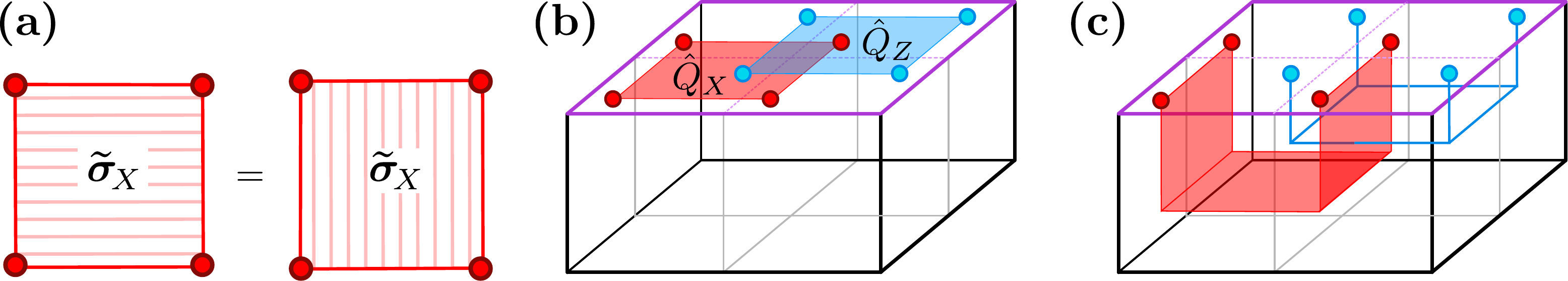}
\caption{
Patch operators and the mutual-\ob{} on the (001)-boundary of the X-Cube model.
\textbf{(a)} The patch operator can be viewed either as the product of $x$-oriented line constraints along the $y$-direction (left), or as the product of $y$-oriented line constraints along the $x$-direction (right).
This implies that the patch operator commutes with all boundary operators except at its corners (red circles).
\textbf{(b)} The mutual-\ob{} is given by the commutation of overlapping rectangular patch operators.
\textbf{(c)} By multiplying the patch operators with bulk conservation laws, the mutual-\ob{} is found to equal the cage-net statistics of bulk quasiparticles.
} 
\label{fig: XCube braiding}
\end{figure*}

We now turn to the \obs{} of the X-Cube boundaries.
We begin with the (001)-boundary, where we introduce rectangular patch operators that generalize the string-like patch operators from the toric code boundary.
From these, we define intrinsically-two-dimensional mutual-\obs{}, and relate them to the cage-net mutual statistics of the bulk fracton order~\cite{PremHuangSongHermele2019}.
The (110)-boundary displays similar features, owing to its similar subsystem symmetry constraints.
However, we do not find any intrinsically-two-dimensional \emph{self}-\ob{} on the (001)- or (110)-boundaries.
This changes on the (111)-boundary.
Here, we introduce hexagonal patch operators whose commutation gives rise to a self-\ob{} that is inherited from the ``windmill'' self statistics of the bulk fracton quasiparticles~\cite{song2023fracton}.

\emph{(001)-boundary}---We begin with (001)-boundary.

Let us first observe that the (001)-boundary operator algebra (Fig.~\ref{fig:001Xcubeboundaryop}) already contains within it the same patch operator commutators as we saw on the toric code boundary.
Indeed, viewing Fig.~\ref{fig:001Xcubeboundaryop}, if we restrict our attention to two adjacent lines of $X$- and $Z$-boundary operators (in either the $x$- or $y$-direction), we have an identical boundary operator algebra as in the toric code boundary. 
Applying our previous arguments, we have that the (001)-boundary Hilbert space is not a 1D local tensor product space.

In the remainder of this Section, we will show a stronger statement, namely that X-Cube boundary cannot be written as a tensor product of the boundary Hilbert spaces of two-dimensional toric codes.
To do so, we introduce the rectangular patch operators shown in Fig.~\ref{fig: XCube braiding}.
The two patches, $\hat{Q}_X$ and $\hat{Q}_Z$, are formed from products of $\tbs \sigma_X$ and $\tbs \sigma_Z$ boundary operators, respectively.
As shown in Fig.~\ref{fig: XCube braiding}(a), the patch operator can be equivalently viewed as \emph{either} a product of line constraints in the $x$- or the $y$-direction.
Since the line constraint segments commute with all boundary operators except at their endpoints, this equivalence implies that the rectangular patch operator commutes with all boundary operators except at its four corners.
% %

We define a ``rectangular'' mutual-\ob{} as the commutator of the two patch operators:
\begin{equation}
    \exp \left( \frac{2\pi i}{d} \tilde{b}_r(Z,X) \right)  = \hat{Q}_Z \hat{Q}_X \hat{Q}_Z^\dagger \hat{Q}_X^\dagger.
\end{equation}
Note that any non-trivial commutation indicates that the boundary cannot be realized as a local tensor product space, by similar arguments as in Section~\ref{sec:TCreview}.
We calculate the commutation in the polynomial formalism, using the same manipulations as in Eqs.~(\ref{eq: compute mutual commutation 1},\ref{eq: compute mutual commutation}).
Taking the corners of the patch operators to be separated by $(i_0,j_0)$ in the $xy$-plane, we have
\begin{equation} \label{eq: mutual comm 001}
\begin{split}
    \Inner{\bs{Q}_{Z}, \bs{Q}_{X}} & = 
    \Inner{
    \sum_{i=1}^{\infty} \sum_{j=1}^{\infty} x^i y^j \tbs{\Sigma} \bs{c}_Z
    , 
    \sum_{i'=-\infty}^{i_0} \sum_{j'=-\infty}^{j_0} x^{i'} y^{j'} \tbs{\Sigma} \bs{c}_X
    } \\
    & = 
    \sum_{i=1}^\infty \sum_{i'=-i_0}^{\infty}
    \sum_{j=1}^\infty \sum_{j'=-j_0}^{\infty}
    x^{-i-i'} y^{-j-j'} \bs{c}_Z^\dagger \bs{A} \bs{c}_X \\
    & = x^{i_0} y^{j_0} \sum_{k=1}^\infty \sum_{l=1}^\infty k \cdot l \cdot x^{-k} y^{-l} \bs{c}_Z^\dagger \bs{A} \bs{c}_X. \\
\end{split}
\end{equation}
Taking the $x^0 y^0$ component and assuming $i_0,j_0 > K$, the mutual-\ob{} is equal to a double derivative over $x$ and $y$ evaluated at $(x,y) = (1,1)$,
\begin{equation} \label{eq: compute mutual commutation XC}
\begin{split}
    \tilde{b}_r(Z,X) & = \left[ x^{i_0} y^{j_0} \sum_{k=0}^\infty \sum_{l=0}^\infty k \cdot l \cdot \cdot x^{-k} y^{-l} \bs{c}_\alpha^\dagger \bs{A} \bs{c}_\beta \right]_{0,0} \\
    %& = \bs{c}_\alpha^\dagger \left( \sum_{k=0}^\infty \sum_{l=0}^\infty k \cdot l \left[\bs{A}\right]_{k-i_0,l-j_0} \right) \bs{c}_\beta \\
    & = \bs{c}_\alpha^\dagger \left( \sum_{k=-\infty}^\infty \sum_{l=-\infty}^\infty k \cdot l \cdot \left[\bs{A}\right]_{k,l} \right) \bs{c}_\alpha, \\
    & = D_x \left[ D_y \left[ \bs{c}_\alpha^\dagger \bs{A} \bs{c}_\beta \right] \right](1,1). \\
\end{split}
\end{equation}
In the X-Cube model we have $\bs{c}_Z^\dagger \bs{A} \bs{c}_X = (1+x)(1+y)$, and thus
\begin{align}
    \tilde{b}_r(Z,X) = D_x \left[ D_y \left[ \bs{c}_Z^\dagger \bs{A} \bs{c}_X \right] \right](1,1) = 1.
\end{align}
i.e.~the patch operators anti-commute.

\begin{figure}[b]
 \centering
 \includegraphics[width=\columnwidth]{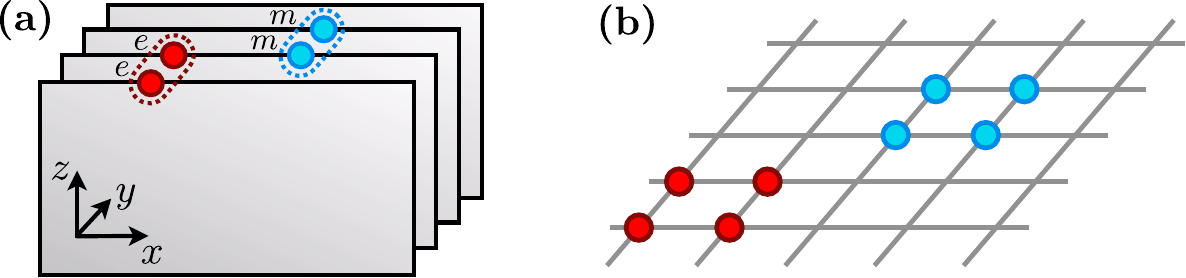}
 \caption{
 Two models that display the same constraints as the X-Cube (001)-boundary, but with different adjacency matrices and \obs{} [Fig.~\ref{fig: anomaly graphs}(d-e)].
 \textbf{(a)} Stacked $xz$-toric codes with `paired' operators in the $y$-direction.  The paired operators product to the identity in the $x$-direction due to the bulk topological order, and in the $y$-direction as a result of the pairing.
  \textbf{(b)} A 2D local tensor product space with `quadrupled' operators. The operators product to the identity in the $x$- and $y$-directions as a result of the quadrupling.
 } 
 \label{fig: stacked toric}
 \end{figure}

  \begin{figure*}
\centering
\includegraphics[width=\textwidth]{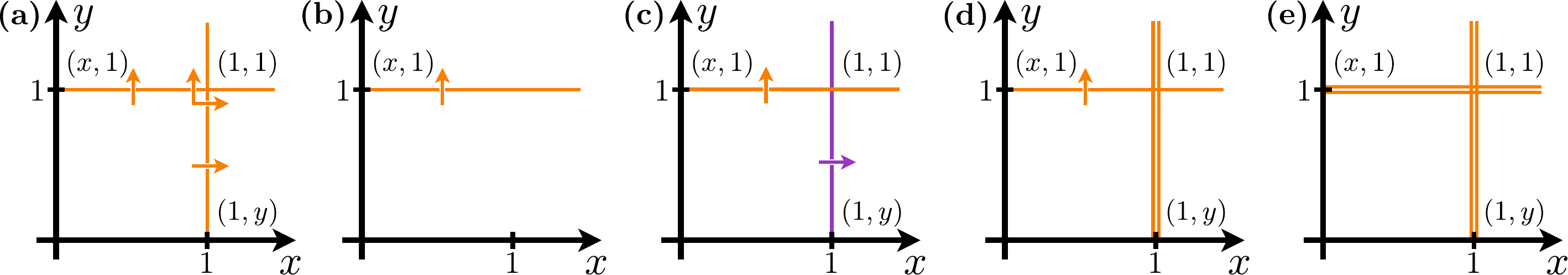}
\caption{
Schematic of the (001)-boundary commutator, $\bs c_Z^\dagger \bs A \bs c_X$, in various three-dimensional models.
A line of zeros of multiplicity 1 (single lines) corresponds to a linear boundary constraint arising from the bulk topological order.
A line of multiplicity 2 (double lines) can arise in a LTPS.
The lack of a LTPS is detected by a non-zero first derivative perpendicular to the line (single arrow).
The lack of stacked toric codes is detected by a non-zero second derivative at the intersection of two constraints that involve the same operator (double arrows).
\textbf{(a)} The X-Cube model has zeros of multiplicity 1 along $x = 1$ and $y=1$, and a non-zero double derivative at $x = y = 1$.
\textbf{(b)} A stack of $yz$-toric codes has zeros only along $y=1$.
\textbf{(c)} A tensor product of $yz$-plane and $xz$-plane toric codes has zeros along $x = 1$ and $y=1$, but for different operators.
\textbf{(d)} A stack of $yz$-toric codes with paired  operators [Fig.~\ref{fig: stacked toric}(a)] has zeros of multiplicity 2 along $x=1$.
\textbf{(e)} A LTPS with quadrupled  operators [Fig.~\ref{fig: stacked toric}(b)] has zeros of multiplicity 2 along $x = 1$ and $y=1$.
} 
\label{fig: anomaly graphs}
\end{figure*}

As in the toric code, we can relate the mutual-\ob{} to the bulk quasiparticle statistics.
Specifically, recall that the net operator in the bulk model was formed by a product of cube stabilizers $\bs \sigma_Z$ over a rectangular prism.
The termination of such a prism on the boundary is equal to the product of $\tbs \sigma_Z$ operators over a rectangular region, i.e.~the patch operator $\hat{Q}_Z$.
Similarly, a cage operator formed from either $\bs \sigma_{X,x}$ or $\bs \sigma_{X,y}$ stabilizers in the bulk terminates to the patch operator $\hat{Q}_X$, formed of $\tbs \sigma_X$ operators, on the boundary.
As shown in Fig.~\ref{fig: XCube braiding}, by multiplying the respective patch operators with cage and net operators we can deform them into the bulk.
By doing so, we see that commutation of the patch operators is equal to the cage-net statistics described in the previous section.

We can also show that the commutation of patch operators is invariant upon taking tensor products of the boundary Hilbert space with 2D toric code boundaries.
For example, consider boundary operators $\bs{t}^x_\alpha$ associated with a toric code(s) in the $xz$-plane and $\bs{t}^y_\alpha$ associated with a toric code(s) in the $yz$-plane.
We can form patch operators that involve $\bs{t}^x_\alpha$, $\bs{t}^y_\alpha$ by performing the operator multiplication, $\bs{c}_\alpha \rightarrow \bs{c}_\alpha + (y-1) \bs{t}^x_\alpha  + (x-1) \bs{t}^y_\alpha$.
Note that we need to multiply the $x$-oriented toric code operators by $(y-1)$ in order for them to product to the identity in \emph{both} the $x$- and $y$-direction, which is required so that the patch operators product to the identity in the bulk ground state (and similar for the $y$-oriented toric code operators).
Performing a similar procedure for $\beta$, the patch operator commutation becomes $\tilde{b}_r(\alpha,\beta) \rightarrow \tilde{b}_r(\alpha,\beta) + D_x[D_y[(y-1)(\bar y-1) \langle \bs{t}^x_\alpha,\bs{t}^x_\beta \rangle]](1,1)  + D_x[D_y[(x-1)(\bar x-1) \langle \bs{t}^y_\alpha,\bs{t}^y_\beta \rangle]](1,1)$.
The duplicate factors of $(y-1)$ in the second term cause it to evaluate to zero, and similar for the factors of $(x-1)$ in the third term.
The commutation of the patch operators $\tilde{b}_r(\alpha,\beta)$ is therefore unchanged.
This implies that the X-Cube boundary cannot be written as a tensor product of toric code boundaries, since any  tensor product would  have trivial commutation (Fig.~\ref{fig: stacked toric}). We further summarize the obstructor invariants that appear in various boundaries in Fig.~\ref{fig: anomaly graphs}.

It is natural to wonder whether there is a self-\ob{} for the X-Cube patch operators.
The most obvious candidate would be the commutation of two patch operators that touch at a single point, and are formed from the same boundary constraint $\alpha$.
Let us take the two patch operators to lie in the second and fourth quadrants of the $xy$-plane (i.e. the upper left and lower right quadrants, respectively).
We can perform algebraic manipulations identical to Eq.~(\ref{eq: compute self commutation1}) to find that the commutator, $\tilde{q}^{24}_r(\alpha)$, is equal to:
\begin{equation}
    \tilde{q}^{24}_r(\alpha) = \sum_{k=1}^\infty \sum_{l=1}^\infty k \cdot l \cdot \left[ \bs{c}_\alpha^\dagger \bs{A} \bs{c}_\alpha \right]_{k,l}.
\end{equation}
We can also consider the analogous commutator of the first and third quadrants,
\begin{equation}
    \tilde{q}^{13}_r(\alpha) = \sum_{k=1}^\infty \sum_{l=-\infty}^0 k \cdot l \cdot \left[ \bs{c}_\alpha^\dagger \bs{A} \bs{c}_\alpha \right]_{k,l}.
\end{equation}
In the X-Cube model we have $\tilde{q}^{24}_r(\alpha) = 1$, $\tilde{q}^{13}_r(\alpha) = 0$ for $\bs{c}_\alpha = (1, 1)$.

 \begin{figure*}
\centering
\includegraphics[width=0.9\textwidth]{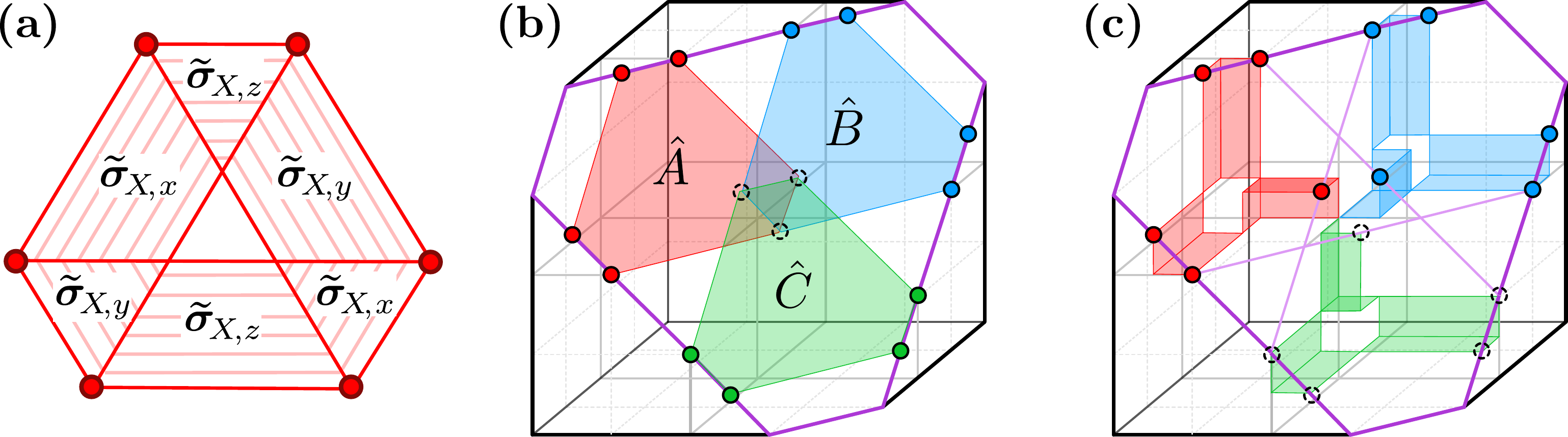}
\caption{
Hexagonal patch operators and the self-\ob{} on the (111)-boundary of the X-Cube model.
\textbf{(a)} The hexagonal patch operator is formed from the product of line constraints in all three directions, as shown.
It commutes with all boundary operators except at its corners (red circles).
\textbf{(b)} The self-\ob{} is given by the threefold commutator of the patch operators $\hat{A}, \hat{B}, \hat{C}$, arranged as shown.
Here the colors denote the different patch operators which each involve the same boundary constraint (as opposed to previous figures, where  colors denoted different boundary constraints).
\textbf{(c)} By multiplying the patch operators with bulk conservation laws, the self-\ob{}  is found to equal the windmill statistics of bulk fracton quasiparticles.
This can be viewed as an exchange process where the 5 red quasiparticles are exchanged with the 5 blue quasiparticles through the 5 intermediate locations denoted by open circles.
} 
\label{fig: 111 statistics}
\end{figure*}

However, unlike the mutual-\ob{} [Eq.~(\ref{eq: mutual comm 001})], these quantities are not invariant upon taking tensor products with toric code boundaries.
In particular, consider the same scenario as above and set $\bs{t}^x_\alpha = (1,1)$ and $\bs{t}^y_\alpha = (0,0)$, i.e.~we fuse the given conservation law with a fermion operator from the $x$-oriented toric code and nothing from the $y$-oriented toric code.
This adds a term $(x-1)(\bar x - 1)(y-1)$ to $\bs{c}_\alpha^\dagger \bs{A} \bs{c}_\alpha$ and thus modifies the self commutators via $\tilde{q}^{24}_r(\alpha) \rightarrow \tilde{q}^{24}_r(\alpha) + 1, \tilde{q}^{13}_r(\alpha) \rightarrow \tilde{q}^{13}_r(\alpha) + 1$.
Meanwhile, setting $\bs{t}^x_\alpha = (0,0)$ and $\bs{t}^y_\alpha = (1,1)$ instead gives $\tilde{q}^{24}_r(\alpha) \rightarrow \tilde{q}^{24}_r(\alpha) + 1, \tilde{q}^{13}_r(\alpha) \rightarrow \tilde{q}^{13}_r(\alpha) - 1$.
Sequences of these two moves can therefore adjust $\tilde{q}^{24}_r,\tilde{q}^{13}_r$ to take any pair of values that conserve the parity $\tilde{q}^{24}_r + \tilde{q}^{13}_r \text{ mod } 2$.
However, even this parity is not entirely invariant.
For example, consider taking a tensor product with a $\mathbbm{Z}_{2n}$-toric code (where the original model is over $\mathbbm{Z}_n$).
The original patch commutators are promoted to $\tilde{q}^{24}_r, \tilde{q}^{13}_r \rightarrow 2\tilde{q}^{24}_r, 2\tilde{q}^{13}_r$ in $\mathbbm{Z}_{2n}$.
These are even over $\mathbbm{Z}_{2n}$ and thus can be reduced to zero via toric code tensor products.

\emph{(111)-boundary}---We now turn to the (111)-boundary, and introduce a new self-\ob{} related to the windmill statistics of bulk fractons.
We consider hexagonal patch operators as shown in Fig.~\ref{fig: 111 statistics}(a).
Similar to the rectangular patch operators on the (001)-boundary, the hexagonal patch operators can be written as products of finite strings of the boundary constraints.
One of way doing so is depicted in Fig.~\ref{fig: 111 statistics}(a). 
Another way of doing so would be to eliminate the line constraint in each triangular region of Fig.~\ref{fig: 111 statistics}(a), and extend the line constraints in the trapezoidal regions to run over the two adjacent triangles instead. In this case, each triangle consists of a product of two constraints from its neighboring trapezoid. This product is equal to the single constraint shown in the triangles in Fig.~\ref{fig: 111 statistics}(a), since $\tbs \sigma_{X,x} + \tbs \sigma_{X,y} + \tbs \sigma_{X,z} = 0$.
Since the line constraints commute with all boundary operators except at their end, these two pictures imply that the hexagonal patch operator commutes with all boundary operators except those that overlap with its six vertices.

To formulate the self-\ob{}, we consider three hexagonal patch operators, $\hat{A}, \hat{B}, \hat{C}$, arranged as in Fig.~\ref{fig: 111 statistics}(b).
Each pair of patch operators share exactly one vertex.
Moreover, the product of any pair of patch operators is proportional to a boundary constraint near the shared vertex (i.e.~the product commutes with all boundary operators near the vertex).
We define the ``hexagonal'' self-\ob{} as the threefold commutator of the patch operators,
\begin{equation}
    \exp \left( \frac{2\pi i}{d} \tilde{q}_h(X) \right) = \hat{A} \hat{B} \hat{C} \hat{A}^\dagger \hat{B}^\dagger \hat{C}^\dagger.
\end{equation}

As for previous \obs{}, the hexagonal self-\ob{} is zero whenever the boundary Hilbert is a LTPS.
To see this, note that in a LTPS the hexagonal patch operators can be written as a product of local operators at each of the six vertices.
By locality, the commutator of a pair of patch operators is equal to the commutator of the local operators at the shared vertex.
However, since the pair product to the identity, the local operators do as well, and hence they, and the patch operators, mutually commute.

We can evaluate the hexagonal self-\ob{} by connecting it to the windmill statistics of the bulk fracton quasiparticles.
We begin by re-writing the invariant as $\tilde{q}_h = (\hat{A} \hat{C}^\dagger) (\hat{C} \hat{B}) ( \hat{C} \hat{A}^\dagger) (\hat{B}^\dagger \hat{C}^\dagger)$.
This corresponds to the commutator between the product of patch operators $\hat{A} \hat{C}^\dagger$, and the product $\hat{C} \hat{B}$.
Now, we can view each product as transferring 5 ``quasiparticles'' from a region near the first patch operator to a region near the second.
This can be seen in Fig.~\ref{fig: 111 statistics}(c), where the product $\hat{A} \hat{C}^\dagger$ transfers the 5 red quasiparticles near patch operator $\hat{A}$ to the 5 green locations near patch operator $\hat{C}$.
This is analogous to the picture presented in Fig.~\ref{fig: bulk equals boundary statistics}, where we view a semi-infinite string operator on the boundary as transferring a quasiparticle from infinity to the end of the string.

This motivates a picture of the hexagonal self-\ob{} in terms of the quasiparticle self statistics.
The total commutator, $(\hat{A} \hat{C}^\dagger) (\hat{C} \hat{B}) ( \hat{C} \hat{A}^\dagger) (\hat{B}^\dagger \hat{C}^\dagger)$, serves to exchange the 5  quasiparticles (red) near patch operator $\hat{A}$ with the 5 quasiparticles (blue) near patch operator $\hat{B}$, through an intermediary set of 5 locations (green) near patch operator $\hat{C}$.
To express this exchange process entirely in the \emph{bulk}, we can multiply the boundary operators $\hat{A} \hat{C}^\dagger$ and $\hat{C} \hat{B}$ by bulk stabilizers.
This is analogous to our procedure for deforming the semi-infinite boundary strings into the bulk in Fig.~\ref{fig: bulk equals boundary statistics}.
The result is shown in Fig.~\ref{fig: 111 statistics}(c).
The boundary operator $\hat{A} \hat{C}^\dagger$ can be written as a product of bulk operators shown in red and green, and the boundary operator $\hat{C} \hat{B}$ as a product of bulk operators shown in green and blue.
The commutator of the boundary operators is equal to the commutator of the corresponding bulk operators.
Viewing Fig.~\ref{fig: bulk equals boundary statistics}, this is equal to the threefold commutator of the red, blue, and green bulk plaquette operators at the point at which they intersect.
This is precisely the windmill statistics introduced in Ref.~\cite{song2023fracton}.

\section{Boundaries of fractal models}\label{sec:fractal}

We now turn to the boundaries of stabilizer models with fractal conservation laws.
The most well-known example is the Haah's code~\cite{Haah2011}, with stabilizers:
\begin{equation}
\begin{split}
\bs \Sigma & =\begin{pmatrix}
    1+ \bar x + \bar y + \bar z  &0\\
    1+ \bar x \bar y + \bar y \bar z + \bar z \bar x &0\\
    \hline
    0& 1 + xy + yz + zx \\
   0& 1+ x + y + z
    \end{pmatrix}. 
\end{split}
\label{eq:Haahscode}
\end{equation}
In contrast to the simple geometric conservation laws in our previous examples, the conservation laws of Haah's code take ``fractal'' patterns in real space.
These patterns are elegantly captured within the polynomial formalism.
We will show that the fractal conservation laws lead to fractal constraints on the boundary operator algebra, and introduce generalizations of the self- and mutual-\obs{} in these settings.

\subsection{Conservation laws in the polynomial formalism}

We begin by developing a more robust framework for analyzing conservation laws in stabilizer models using the language of module theory in mathematics~\cite{Haah2013}.

%%% Bulk conservation laws
\emph{Bulk conservation laws.}---Thus far, the conservation laws we've considered have corresponded to infinite products of stabilizers over a plane of the lattice.
They are defined by conditions of the form:
\begin{equation}
    \Xi(x,y,z) \cdot \bs \sigma(x,y,z)  = 0,
\end{equation}
where $\bs \sigma \in S$ is a stabilizer and $\Xi$ describes an infinite sum of all monomials within the plane, for example $\Xi_{xy} = \sum_{i,j} x^i y^j$.
As previously discussed, the effect of multiplication by $\Xi_{xy}$ can be captured by setting the $x$ and $y$ coordinates of the polynomial vector to $1$.
Specifically, we have
\begin{align}
    \Xi_{xy} \cdot \bs \sigma = \Xi_{xy} \cdot \bs \sigma' \,\,\,\,\, & \Leftrightarrow \,\,\,\,\, \bs \sigma(1,1,z) = \bs \sigma'(1,1,z),
\intertext{
i.e.~the polynomial vectors $\bs \sigma$ and $\bs \sigma'$ are equal after multiplication by $\Xi_{xy}$ if and only if they are equal when evaluated at $x,y=1$.
This is understandable because $\Xi_{xy}$ involves a product of stabilizers over the entire $xy$-plane, and hence translations by $x$ or $y$ do not affect the result. %
It will be illuminating to recast this in the following, equivalent form
}
    \Xi_{xy} \cdot \bs \sigma = \Xi_{xy} \cdot \bs \sigma' \,\,\,\,\, & \Leftrightarrow \,\,\,\,\, \bs \sigma = \bs \sigma' + (x-1) \bs \delta_x + (y-1) \bs \delta_y.  \label{eq: Xi map}
\end{align}
for some $\bs \delta_x, \bs \delta_y \in \mathbbm{Z}_n^{2M}[x,y,z]$.
Multiplication by $\Xi_{xy}$ thus serves to ``set the polynomials $(x-1)$ and $(y-1)$ to zero''.

We can formalize this using the mathematical notions of a polynomial ideal and a quotient ring.
First, we define the set of all linear combinations of a set of polynomials $p_1,\ldots,p_n$ as the \emph{ideal} generated by the polynomials, denoted $\mathfrak{p}=(p_1,\ldots,p_n)$.
Note that the coefficients of the linear combinations are allowed to be polynomials themselves.
We can also define the multiplication of an ideal $\mathfrak{p}$ with a module $M$ as the set generated by multiples of elements of $\mathfrak{p}$ with elements of $M$, $\mathfrak{p} M = \text{span}( \{ p \, \bs m | p \in \mathfrak{p}, \bs m \in M \} ) \subseteq M$.
Note that $\mathfrak{p} M$ is sub-module of $M$.
Finally, we can define the \emph{quotient module} $M/N$ of a module $M$ by a sub-module $N \subseteq M$ as the set of equivalence classes of $M$ under addition by elements of $N$.
That is, two equivalence classes $[\bs m], [\bs m'] \in M/N$ are equal if there exists $\bs n \in N$ such that
\begin{align} \label{eq: quotient map}
     [\bs m] = [\bs m'] \,\,\,\,\, & \Leftrightarrow \,\,\,\,\, \bs m = \bs m' + \bs n.
\end{align}
We can also define the \emph{quotient map}, $\pi: M \rightarrow M/N$ via $\pi(\bs m) = [\bs m]$.

Returning to the subject at hand, comparing Eq.~(\ref{eq: Xi map}) and Eq.~(\ref{eq: quotient map}) we see that multiplication by $\Xi_{xy}$ is naturally viewed as quotienting the module $\mathbbm{Z}^{2M}_n[x,y,z]$ by the sub-module
\begin{align}
    \mathfrak{c}_{xy} & \mathbbm{Z}^{2M}_n[x,y,z]  \\
    & = \{  (x-1) \bs \delta_x + (y-1) \bs \delta_y \, | \, \bs \delta_x, \bs \delta_y \in \mathbbm{Z}^{2M}_n[x,y,z] \}, \nonumber
\end{align}
where we define the ideal $\mathfrak{c}_{xy} = (x-1,y-1)$.
We immediately have
\begin{equation}
    \Xi_{xy} \mathbbm{Z}^{2M}_n[x,y,z] \cong \mathbbm{Z}^{2M}_n[x,y,z] / \mathfrak{c}_{xy} \mathbbm{Z}^{2M}_n[x,y,z].
\end{equation}
This follows from Eq.~(\ref{eq: Xi map}), because two elements on the left are equal if and only if their difference lies in $\mathfrak{c}_{xy} \mathbbm{Z}^{2M}_n[x,y,z]$.
We can therefore view multiplication by $\Xi_{xy}$ as equivalent to the quotient map,
\begin{align}
    \pi_{\mathfrak{c}_{xy}}: \mathbbm{Z}^{2M}_n[x,y,z] & \rightarrow \mathbbm{Z}^{2M}_n[x,y,z] / \mathfrak{c}_{xy} \mathbbm{Z}^{2M}_n[x,y,z]. 
\end{align}

This formulation immediately suggests a generalization to arbitrary polynomial ideals $\mathfrak{c}$.
Suppose we have an infinite pattern $\Xi_{\mathfrak{c}}$ that annihilates exactly the set of polynomials in the ideal $\mathfrak{c}$,
\begin{equation}
    \Xi_{\mathfrak{c}}(x,y,z) \cdot c(x,y,z) = 0 \,\,\,\,\,  \Leftrightarrow \,\,\,\,\,  c(x,y,z) \in \mathfrak{c}.
\end{equation}
Then multiplication by $\Xi_{\mathfrak{c}}$ naturally induces a quotient map,
\begin{align}
    \pi_{\mathfrak{c}}: \mathbbm{Z}^{2M}_n[x,y,z] & \rightarrow \mathbbm{Z}^{2M}_n[x,y,z] / \mathfrak{c} \mathbbm{Z}^{2M}_n[x,y,z]. 
\end{align}

We are now in position to state our more general formulation of conservation laws.
We say that a stabilizer $\bs \sigma \in S$ \emph{features a conservation law over the ideal} $\mathfrak{c}$ if
\begin{equation}
    \pi_{\mathfrak{c}}( \bs \sigma ) = 0,
\end{equation}
where $\pi_{\mathfrak{c}}$ is the quotient map defined above restricted to the stabilizer module $S \subset \mathbbm{Z}^{2M}_n[x,y,z]$.
The set of conservation laws for a pattern $\mathfrak{c}$  corresponds to the kernel of $\pi_{\mathfrak{c}}$.

We can further refine this definition by noting that any stabilizer multiplied with a polynomial $c \in \mathfrak{c}$ trivially obeys the condition above.
The set of such stabilizers is the sub-module $\mathfrak{c} S \subset S$.
To `mod out' these trivial elements, we define the \emph{conservation law module} of the ideal $\mathfrak{c}$ as,
\begin{equation}
    \mathcal{C} = \text{ker}(\pi_\mathfrak{c})/\mathfrak{c}S.
\end{equation}
The conservation laws of a stabilizer model are thus characterized by pairs
\begin{equation}
    \bigg( \mathfrak{c} \, , \, \mathcal{C} \bigg)
\end{equation}
of an ideal $\mathfrak{c}$ and its associated conservation law module $\mathcal{C}$.

\emph{Boundary constraints.}---The set of all linear combinations of truncated boundary operators forms a sub-module on $\mathbbm{Z}^{2\tilde{M}}_n[x,y]$, where $\tilde{M} = M(K-1)$ is the number of spins in the boundary unit cell.
The bulk stabilizers provide equivalence relations, ``$\eqgs$", on $\mathbbm{Z}^{2\tilde{M}}_n[x,y]$.
We thus define the module $\tilde{S}$ of truncated boundary operators via the quotient, $\tilde{S} = \text{span}( \tbs \Sigma ) /\eqgs$, with respect to this equivalence relation.
To characterize the global boundary constraints, we again note that multiplication by a pattern, $\Xi_{\mathfrak{b}}$ corresponding to an ideal $\mathfrak{b}$, induces a map
\begin{equation}
    \pi_\mathfrak{b}: \text{span}(\tbs \Sigma) \rightarrow \mathbbm{Z}^{2\tilde{M}}_n[x,y]/\mathfrak{b} \mathbbm{Z}^{2\tilde{M}}_n[x,y],
\end{equation}
and similarly,
\begin{equation}
    \pi_\mathfrak{b}^{\tilde{S}}: \tilde{S} \rightarrow \tilde{S}/\mathfrak{b} \tilde{S}.
\end{equation}
The boundary constraints correspond to the kernel of $\pi^{\tilde{S}}_\mathfrak{b}$, i.e.~operators that are equal to identity on the boundary Hilbert space after multiplication by the infinite pattern $\Xi_\mathfrak{b}$, but not before.
To isolate the constraints from potential conservation laws in the boundary theory, we define the \emph{constraint module} of the ideal $\mathfrak{b}$ as,
\begin{equation}
    \mathcal{B} = \text{ker} \big(\pi^{\tilde{S}}_\mathfrak{b} \big) \big/ \big( \text{ker}(\pi_\mathfrak{b}) / \eqgs \big).
\end{equation}
The constraints of the boundary operator algebra are thus characterized by pairs
\begin{equation}
    \bigg( \mathfrak{b} \, , \, \mathcal{B} \bigg)
\end{equation}
of an ideal $\mathfrak{b}$ and its associated constraint module $\mathcal{B}$.

\emph{Examples.}---We can illustrate these definitions using our previous examples.
In 2D stabilizer models, we focused on conservation laws over the ideal $\mathfrak{c}_{xy} = (x-1,y-1)$.
The quotient module obeys the isomorphism $\mathbbm{Z}^{2M}_n[x,y]/\mathfrak{c}_{xy}\mathbbm{Z}^{2M}_n[x,y] \cong \mathbbm{Z}^{2M}_n$, while the corresponding quotient map is obtained by evaluating the polynomial at one, $\pi_{\mathfrak{c}_{xy}}(\bs \sigma) = \bs \sigma(1,1) \in \mathbbm{Z}^{2M}_n$.
In the $\mathbbm{Z}_n$ toric code, the kernel of the quotient map is equal to the entire set of stabilizers, $\text{ker}(\pi_{\mathfrak{c}_{xy}}) = S$.
The quotient by $\mathfrak{c}_{xy} S$ serves to ``evaluate the coefficients of the stabilizer generators at one'', giving the conservation law module
\begin{equation}
    \mathcal{C}_{xy} = \{ d_X \bs \sigma_X + d_Z \bs \sigma_Z | d_X,d_Z \in \mathbbm{Z}_n \} \cong \mathbbm{Z}_n \times \mathbbm{Z}_n.
\end{equation}
Denoting $\bs \sigma_X = \bs \Sigma \bs c_X$ for $\bs c_X = (0,1)^T$ and similar for $\bs c_Z$, we said that $\bs c_X$ and $\bs c_Z$ featured a conservation law.

Meanwhile, in the X-Cube model we had non-trivial conservation laws for each of the ideals $\mathfrak{c}_{xy} = (x-1,y-1)$, $\mathfrak{c}_{yz} = (y-1,z-1)$, $\mathfrak{c}_{zx} = (z-1,x-1)$.
Focusing on $\mathfrak{c}_{xy}$ for specificity, we have the isomorphism $\mathbbm{Z}^{2M}_n[x,y,z]/\mathfrak{c}_{xy}\mathbbm{Z}^{2M}_n[x,y,z] \cong \mathbbm{Z}^{2M}_n[z]$ and the quotient map $\pi_{\mathfrak{c}_{xy}}(\bs \sigma) = \bs \sigma(1,1,z) \in \mathbbm{Z}^{2M}_n[z]$.
The corresponding conservation law module is thus
\begin{equation}
\begin{split}
    \mathcal{C}_{xy} & = \{ d_X(z) \bs \sigma_{X,z} + d_Z(z) \bs \sigma_Z | d_X,d_Z \in \mathbbm{Z}_n \} \\
    & \cong \mathbbm{Z}_n[z] \times \mathbbm{Z}_n[z],
\end{split}
\end{equation}
and similar for $(\mathfrak{c}_{yz},\mathcal{C}_{yz})$ and $(\mathfrak{c}_{zx},\mathcal{C}_{zx})$.

\subsection{Fractal stabilizer models}

%%% Introduce fractal models
We now turn to stabilizer models with fractal conservation laws.
Specifically, we consider models of the so-called ``fractal spin liquids" of the form~\cite{Yoshida2013},
\begin{equation} \label{LCA 2}
\bs \Sigma=\begin{pmatrix}
    1 - \bar z  g(\bar x,\bar y)  & 0\\
    h(\bar x,\bar y) &0\\
    \hline
    0& h(x,y) \\
   0& 1 - z g(x,y)
    \end{pmatrix} = (\bs \sigma_Z ,  \bs \sigma_X).
\end{equation}
In the following section, we will see that this class of models are particularly amenable to termination on the (001)-boundary. 
We note that Haah's code [Eq.~(\ref{eq:Haahscode})] can be re-cast in this form, with $h(x,y) = 1 + x + y + xy + x^2 +y^2$ and $g(x,y) = (1 + x + y)$~\cite{Yoshida2013}.

The bulk conservation laws of these models are not as simply expressed as in the toric code or X-Cube model.
For instance, one might naively try to write the conservation law,
\begin{equation} \label{type II infinite sum}
\Xi(x,y,z) \bs \sigma_X  = 0, \\
\end{equation} 
where we define the pattern $\Xi$ via an infinite sum,
\begin{equation}
\Xi(x,y,z) = \sum_{i,j = -\infty}^{\infty} \bigg( z g(x,y) \bigg)^i \bigg( 1 - h(x,y) \bigg)^j. \\
\end{equation}
This is in analogy to the conservation laws of 2D stabilizer models, replacing $x \rightarrow z g(x,y)$ and $y \rightarrow 1 - h(x,y)$.
Crucially, the pattern $\Xi$ obeys the conditions,
\begin{equation} \label{eq: Xi g h}
\begin{split}
    \Xi(x,y,z) & \cdot \left( 1 - z g(x,y) \right)  = 0, \\
    \Xi(x,y,z) & \cdot h(x,y)  = 0,
\end{split}
\end{equation}
which can be verified formally by re-indexing the summation as in Eq.~(\ref{infinite sum evaluate}). 
These conditions guarantee that Eq.~(\ref{type II infinite sum}) holds, since $\bs \sigma_X$ contains components equal to $h$ and $1-zg$.
Nonetheless, our expression for $\Xi$ is unwieldy since it contains negative powers of the polynomials $g(x,y)$ and $1-h(x,y)$, which are not easily defined\footnote{One can obtain a precise definition of $\Xi$ in the specific case of periodic boundary conditions with $L = n^\ell$ unit cells in each direction, where $n$ is the qudit dimension and $\ell$ is an integer. To do so, one exploits the fact that $(a+b)^n = a^n + b^n$ over $\mathbbm{Z}_n$ and defines $\Xi = (1-z g(x,y))^{n^\ell-1} h(x,y)^{n^\ell-1}$.
The conditions Eq.~(\ref{eq: Xi g h}) then follow because the multiples contain a factor of $(1-z g(x,y))^{n^\ell}$ or $h(x,y)^{n^\ell}$, each of which is zero with the given boundary conditions.}.

We can express the conservation laws more elegantly using the formalism introduced in the previous subsection.
We have two conservation laws,
\begin{equation}
    \bigg( \mathfrak{c} \,,\, \bs \sigma_X \bigg), \,\,\,\,\,\, \bigg( \bar{\mathfrak{c}}  \,,\,  \bs \sigma_Z \bigg)
\end{equation}
over the ideal $\mathfrak{c} = (1-zg(x,y),h(x,y))$ and the spatially-inverted ideal $\bar{\mathfrak{c}} = (1-\bar z g(\bar x,\bar y),h(\bar x,\bar y))$, respectively.
In the above, to be precise we should replace $\bs \sigma_X$ and $\bs \sigma_Z$ with the conservation law modules $\mathcal{C}_{\mathfrak{c}} = \text{span}(\bs \sigma_X) / \mathfrak{c} \, \text{span}(\bs \sigma_X)$ and $\mathcal{C}_{\bar{\mathfrak{c}}} = \text{span}(\bs \sigma_Z) / \bar{\mathfrak{c}} \, \text{span}(\bs \sigma_Z)$.
Note that the first conservation law corresponds to the infinite pattern $\Xi(x,y,z)$ described the previous paragraph, while the second conservation law corresponds to its spatial inversion, $\Xi(\bar x,\bar y,\bar z)$.
We see that with this notation, the conservation laws of fractal stabilizer models can be written just as simply as those of the toric code.

\subsection{Boundary operator algebra and fractal constraints}

We now  construct the operator algebra and constraints of the boundary Hilbert space.
We restrict to the (001)-boundary, which allows a particularly simple truncation for the models we consider.
We find truncated boundary operators,
\begin{equation}
\tbs \Sigma=\begin{pmatrix}
    1  &0\\
    h(\bar x,\bar y) &0\\
    \hline
    0& 0 \\
   0& g(x,y)
    \end{pmatrix},
\end{equation}
which feature an adjacency matrix,
\begin{equation} \label{type II AC}
\bs A =\begin{pmatrix}
    0& h(x,y) g(x,y) \\
    h(\bar x,\bar y) g(\bar x,\bar y)  &0\\
    \end{pmatrix}.
\end{equation}
We note that this is the same adjacency matrix obtained from the two-dimensional model,
\begin{equation}
\bs \Sigma_{\text{2D}} =\begin{pmatrix}
    h(\bar x,\bar y) &0\\
    \hline
   0& g(x,y)
    \end{pmatrix},
\end{equation}
which obeys a fractal subsystem symmetry~\cite{Devakul2019}.

To obtain the constraints on the boundary operator algebra, we must terminate the bulk conservation laws on the boundary.
Consider the expression for the conservation law as an infinite sum in Eq.~(\ref{type II infinite sum}). 
Boundary termination consists of eliminating all terms with negative powers of $z$.
This is again particularly simple for the (001)-boundary in the models we consider.
We obtain:
\begin{equation} \label{type II Z sum}
\begin{split}
\sum_{j = -\infty}^{\infty} & \bigg( 1 - h(x,y) \bigg)^j \tilde{\bs \sigma}_Z \\
& = \left[ \bar z \sum_{i=1}^{\infty} \sum_{j=-\infty}^{\infty} \bigg( z g(x,y) \bigg)^i \bigg( 1 - h(x,y) \bigg)^j \right] \bs \sigma_Z \\
& \eqgs 0. \\
\end{split}
\end{equation}
The LHS is equal to an infinite product of truncated boundary operators, and the RHS of bulk stabilizers.
Similarly, the $\bs \sigma_X$ constraint is given by
\begin{equation} \label{type II X sum}
\begin{split}
\sum_{j = -\infty}^{\infty} & \bigg( 1 - h(\bar x,\bar y) \bigg)^j \tilde{\bs \sigma}_X \\ 
& = \left[ \sum_{i=-\infty}^{1} \sum_{j=-\infty}^{\infty} \bigg( \bar z g(\bar x,\bar y) \bigg)^i \bigg( 1 - h(\bar x,\bar y) \bigg)^j \right] \bs \sigma_X \\
& \eqgs 0. \\
\end{split}
\end{equation}
As we did for the bulk conservation laws, we can express the boundary constraints more elegantly using polynomial ideals.
We have two boundary constraints:
\begin{equation}
    \bigg( \mathfrak{b} \,,\, \tbs \sigma_X \bigg), \,\,\,\,\,\, \bigg( \bar{\mathfrak{b}}  \,,\,  \tbs \sigma_Z \bigg),
\end{equation}
over the ideal $\mathfrak{b} = (h(x,y))$ and its spatial-inversion $\bar{\mathfrak{b}} = (h(\bar x,\bar y))$, respectively.
The first constraint signifies that the boundary operator $\tbs \sigma_X$ obeys $\Xi_{\mathfrak{b}} \cdot \tbs \sigma_X \eqgs 0$ for any infinite pattern $\Xi_{\mathfrak{b}}$ than annihilates all elements in the ideal $\mathfrak{b}$ (and similar for the second contraint and $\tbs \sigma_Z$).

\subsection{Mutual-\ob}

We now introduce the \obs{} for fractal stabilizer models.
These will allow us to demonstrate that their boundary operator algebra cannot be realized in any LTPS.

To begin, recall that the boundary constraints necessarily commute with every  boundary operator. 
Similar to previous sections, this enforces:
\begin{equation} \label{type II AC constraints}
\begin{split}
\bs A \cdot \bs{b}_X  
 & \in \mathfrak{b} \mathbbm{Z}^{\tilde{T}}_n[x,y], \\ 
\bs A \cdot \bs{b}_Z  
& \in \bar{\mathfrak{b}} \mathbbm{Z}^{\tilde{T}}_n[x,y]
\end{split}
\end{equation}
where we define $\tilde{T} = (K-1)T$.
These are easily verified from Eq.~(\ref{type II AC}).
Note that we can invert the second constraint to obtain,
\begin{equation} \label{type II AC constraints 2}
\bs{b}_Z^\dagger  \cdot \bs A  \in \mathfrak{b} \mathbbm{Z}^{\tilde{T}}_n[x,y], \\ 
\end{equation}
so that it involves the same ideal as the first.
Any of the above equations imply
\begin{equation} 
\bs{b}_Z^\dagger  \cdot \bs A \cdot \bs{b}_X  \in \mathfrak{b} \mathbbm{Z}_n[x,y]. \\ 
\end{equation}

To motivate the mutual-\ob, we first observe that the adjacency matrix obeys stronger requirements in a 2D LTPS.
Consider a LTPS with operators $\bs{\Sigma}_{\text{TP}}$.
If these operators have a conservation law for the ideal $\mathfrak{b}$, we have
\begin{equation}
\begin{split}
\bs \Sigma_{\text{TP}} \cdot \bs{b}_X  & \in \mathfrak{b} \mathbbm{Z}^{\tilde{T}}_n[x,y], \\
\bs{b}_Z^\dagger \cdot \bs \Sigma_{\text{TP}}^\dagger & \in \mathfrak{b} \mathbbm{Z}^{\tilde{T}}_n[x,y],
\end{split}
\end{equation}
where we have again inverted the second equation.
These imply that we can write:
\begin{equation}
    \bs{b}_Z^\dagger\bs{A}_{\text{TP}} \bs{b}_X = \bs{b}_Z^\dagger \bs \Sigma_{\text{TP}}^\dagger \bs{\lambda} \bs \Sigma_{\text{TP}} \bs{b}_X = b_1 b_2 \bs{p}_2^\dagger \bs{\lambda} \bs{p}_1.
\end{equation}
where $b_1, b_2 \in \mathfrak{b}$ and $\bs{p}_1,\bs{p}_2 \in \mathbbm{Z}_n[x,y]$.
This implies that $\bs{b}_Z^\dagger\bs{A} \bs{b}_X$ is contained in the \emph{square} of the original ideal,
\begin{equation} \label{A b2}
    \bs{b}_Z^\dagger\bs{A}_{\text{TP}} \bs{b}_X \in \mathfrak{b}^2 \mathbbm{Z}^{\tilde{T}}_n[x,y],
\end{equation}
where $\mathfrak{b}^2 = (\{ b_i b_j | b_i, b_j \in \mathfrak{b} \})$.
In the present case, we have $\mathfrak{b} = ( h(x,y) )$ and thus $\mathfrak{b}^2 = ( h(x,y)^2 )$, so Eq.~(\ref{A b2}) states that the adjacency matrix element has a zero of multiplicity two at $h(x,y) = 0$.

This suggests that we define the mutual-\ob{} by taking the quotient over the squared ideal.
Namely, we define
\begin{equation}
    \tilde{b}(Z,X) = \pi_{\mathfrak{b}^2}(\bs{b}_Z^\dagger\bs{A} \bs{b}_X),
\end{equation}
which lives in the quotient ring
\begin{equation}
\tilde{b}(Z,X) \in \mathfrak{b} \mathbbm{Z}_n[x,y] / \mathfrak{b}^2 \mathbbm{Z}_n[x,y] \cong \mathbbm{Z}_n[x,y] / \mathfrak{b} \mathbbm{Z}_n[x,y].
\end{equation}
The latter isomorphism holds for ideals, $\mathfrak{b}$, that are generated by algebraically independent polynomials\footnote{To show this, suppose $\mathfrak{b} = (b_1,\ldots,b_m)$ with $\{b_i\}$ algebraically independent. We define an isomorphism $\phi$ between the two modules as $\phi(p) = b_1 p  b_i b_j = b_1 p $. This is well-defined, since $\phi(p+b) = b_1 p + b_1 b \sim_{\mathfrak{b}^2} b_1 p = \phi(p)$. It is also clearly surjective. To verify that is also injective, suppose $\phi(p_1) \sim_{\mathfrak{b}^2} \phi(p_2)$. Then $b_1 p_1 = b_1 p_2 + \sum_{i,j} b_i b_j r_{ij}$. Since $b_1$ must divide the RHS and $\{b_i\}$ are algebraically independent, we can set $r_{ij} = 0$ unless $i = 1$, giving $b_1 p_1 = b_1 p_2 + \sum_{j} b_1 b_j r_{1j}$. This implies $p_1 = p_2 + \sum_{j} b_j r_{1j}$, and therefore $p_1 \sim_{\mathfrak{b}} p_2$.}.
In the fractal spin liquid models of Eq.~(\ref{LCA 2}) with $\mathfrak{b} = ( h(x,y) )$, we have
\begin{equation}
    \bs{b}_Z^\dagger\bs{A} \bs{b}_X = h(x,y) g(x,y),
\end{equation}
which gives,
\begin{equation}
    \tilde{b}(Z,X)  = \left[ g(x,y) \right] \in  \mathbbm{Z}_n[x,y] / \mathfrak{b} \mathbbm{Z}_n[x,y],
\end{equation}
where $\left[ g(x,y) \right]$ denotes the equivalence class containing $g(x,y)$.
The mutual-\ob{} is non-zero as long as $h(x,y)$ does not divide $g(x,y)$\footnote{If $h(x,y)$ does divide $g(x,y)$, one can easily show that the original model, Eq.~(\ref{LCA 2}), is equivalent to a stack of 2D Hamiltonians in the $xy$-plane, i.e. it has trivial topological order in the $z$-direction}.
%
%%%
It is illuminating to see how the \obs{} of the toric code and X-Cube model appear in this formalism.
We begin with the toric code.
The boundary constraints are
\begin{equation} \label{type II constraint}
    \bigg\{ \bs{b}_X = \begin{pmatrix} 0 \\ 1 \end{pmatrix} , \mathfrak{b} \bigg\},
    \bigg\{ \bs{b}_Z =\begin{pmatrix} 1 \\ 0 \end{pmatrix} , \mathfrak{b} \bigg\},
\end{equation}
with $\mathfrak{b} = \bar{\mathfrak{b}}$ since $(1+x)$ and $(1+\bar x)$ are related by multiplication by the monomial $x$.
The commutator between the two constraints is,
\begin{equation}
    \bs{b}_Z^\dagger\bs{A} \bs{b}_X = (1+x),
\end{equation}
which, when quotientied over $\mathfrak{b}^2 = ( (1+x)^2 )$, gives,
\begin{equation}
\begin{split}
    \tilde{b}(Z,X) & = [ D_x[\bs{b}_Z^\dagger\bs{A} \bs{b}_X](1)  ] \\
    & = [1] \in \mathbbm{Z}_n[x] / \mathfrak{b} \mathbbm{Z}_n[x].
\end{split}
\end{equation}
In the first expression we use the fact that $[ D_x[ (1+x) f(x) ] ] = [ f(x) ]$ to write the equivalence class of $\bs{b}_Z^\dagger\bs{A} \bs{b}_X$ by $(1+x)$ divided by $(1+x)$ in terms of the Hasse derivative.
Note that the quotient ring obeys $\mathbbm{Z}_n[x] / \mathfrak{b} \mathbbm{Z}_n[x] \cong \mathbbm{Z}_n$, in which case we can write $\tilde{b}(Z,X) = D_x[\bs{b}_Z^\dagger\bs{A} \bs{b}_X](1) \in \mathbbm{Z}_n$.

%%%
% X-Cube ideals
%%%
We now turn to Type-I fracton models, focusing on the (001)-boundary of the X-Cube model for concreteness.
The boundary constraints are:
\begin{equation} \label{type I constraint}
\begin{split}
    %\bigg\{ 
    \bigg\{ \bs{b}_X = \begin{pmatrix} 0 \\ 1 \end{pmatrix} &, \mathfrak{b}_x \bigg\}, 
    \bigg\{ \bs{b}_X = \begin{pmatrix} 0 \\ 1 \end{pmatrix} , \mathfrak{b}_y \bigg\}, \\
    \bigg\{ \bs{b}_Z =\begin{pmatrix} 1 \\ 0 \end{pmatrix} &, \mathfrak{b}_x \bigg\},
    \bigg\{ \bs{b}_Z =\begin{pmatrix} 1 \\ 0 \end{pmatrix} , \mathfrak{b}_y \bigg\} 
    %\bigg\}
\end{split}
\end{equation}
with inversion symmetric ideals, $\mathfrak{b}_x = \bar{\mathfrak{b}}_x = (1+x), \mathfrak{b}_y = \bar{\mathfrak{b}_y} = (1+y)$.
These differ from the constraints present in the toric code, as well as the fractal models we consider, since each interaction term, $\tilde{\bs{\sigma}}_{Z}, \tilde{\bs{\sigma}}_{X}$, participates in \emph{two} boundary constraints instead of one.
Comparing to the the fractal models, this reflects two different mechanisms of achieving fractonic excitations: in Type-I theories, bulk fractons cannot move due to their simultaneous participation in multiple independent planar conservation laws, while in fractal models, bulk fractons cannot move due to the fractal patterning of a single conservation law.

The rectangular mutual-\ob{} for the X-Cube boundary can be re-cast in our current framework by considering the product of the two ideals, $\mathfrak{b} = \mathfrak{b}_x \mathfrak{b}_y$.
To see this, suppose that in a general theory a boundary operator $\bs{b}_\alpha$ participates in two constraint patterns, $\mathfrak{b}_1$ and $\mathfrak{b}_2$, and that another boundary operator $\bs{b}_\beta$ participates in the inverse patterns, $\overline{\mathfrak{b}}_1$ and $\overline{\mathfrak{b}}_2$. We assume that the constraints are independent, so that $\mathfrak{b}_1 \cap \mathfrak{b}_2 = \mathfrak{b}_1 \mathfrak{b}_2$.
The constraints imply that the commutator contains at least one factor each from $\mathfrak{b}_1$ and $\mathfrak{b}_2$, i.e.~$\bs{b}_\beta^\dagger \bs{A}_{\text{TP}} \bs{b}_\alpha \in ( \mathfrak{b}_1 \mathfrak{b}_2 )$.
In a LTPS, we would have further that the commutator contains two factors each from $\mathfrak{b}_1$ and $\mathfrak{b}_2$, i.e.~$\bs{b}_\beta^\dagger \bs{A}_{\text{TP}} \bs{b}_\alpha \in ( \mathfrak{b}_1^2 \mathfrak{b}_2^2 )$.
This exactly resembles the scenarios in the previous paragraphs, with $\mathfrak{b} = \mathfrak{b}_1 \mathfrak{b}_2$.

The commutator on the X-Cube boundary is
\begin{equation}
    \bs{b}_Z^\dagger \bs{A} \bs{b}_X = (1+x)(1+y).
\end{equation}
Following our standard procedure, this gives the mutual-\ob{},
\begin{equation}
\begin{split}
    \tilde{b}(Z,X) & = \left[ D_x[D_y[\bs{b}_Z^\dagger \bs{A} \bs{b}_X]] \right] \\
    & = [1] \in \mathbbm{Z}_n[x,y] / \mathfrak{b} \mathbbm{Z}_n[x,y].
\end{split}
\end{equation}
Note that the mutual-\ob{} as defined above lives in $\mathbbm{Z}_n[x,y] / \mathfrak{b} \mathbbm{Z}_n[x,y]$.
This is different than the $\mathbbm{Z}_n$ classification we previously found.
This difference arises because here we only considered quotienting by local tensor product spaces, whereas previously we also quotiented over stacks of boundaries of two-dimensional theories.

\subsection{Self-\ob}

To address the self-\ob{} of models with fractal conservation laws, we consider models beyond CSS codes.
Specifically, we consider a twisted version of the fractal spin liquid models~\cite{Tantivasadakarn20}, which hosts emergent fermionic quasiparticles of the form
\begin{equation} 
\bs \sigma=\begin{pmatrix}
    1+ \bar z  g(\bar x,\bar y)  & (t_g+zg)h\\
    h(\bar x,\bar y) &\bar t_h (1+zg)\\
    \hline
    0& h(x,y) \\
   0& 1+  z g(x,y)
    \end{pmatrix}.
\end{equation}
Here, we work over $\mathbbm{Z}_2$ and assume that $h(x,y)$ and $g(x,y)$ consist of even and odd number of terms, respectively.
With these assumptions, we then define the polynomials $t_h(x,y)$ and $t_g(x,y)$ such that $t_h +\bar t_h = h \bar h$ and $t_g + \bar t_g = 1+g\bar g$\footnote{For Haah's code, one choice is $t_h= x(1+x+y+\bar y^2)+y(1+y+\bar x+\bar x^2) +x^2 \bar y^2 $ and $t_g= x+y+ x\bar y$.}. 
The bulk conservation laws of the model are identical to those in our previous CSS model, 
\begin{equation}
    \bigg( \mathfrak{c} \,,\, \bs \sigma_X \bigg), \,\,\,\,\,\, \bigg( \bar{\mathfrak{c}}  \,,\,  \bs \sigma_Z \bigg)
\end{equation}
where $\mathfrak c = (1+zg(x,y),h(x,y))$.

On the (001)-boundary, we obtain the truncated boundary operators,
\begin{equation} 
\tbs \Sigma=\begin{pmatrix}
    1  & gh\\
   \bar h &\bar t_h g\\
    \hline
    0& 0 \\
   0&  g
    \end{pmatrix},
\end{equation}
which satisfy identical constraints as in the previous model, again with $\mathfrak b = (h(x,y))$.
However, the commutation of the boundary operators now becomes
\begin{align}
    \bs A &= \begin{pmatrix}
   0 & hg\\
   \bar h\bar g & h\bar h g \bar g
    \end{pmatrix}.
\end{align}
Note that the second column and first row of $\bs{A}$ contain factors of $h$, and the first column and second row contain factors of $\overline{h}$, as required by the boundary constraints.

To analyze the self-\ob{}, we focus on the inner product $\bs{b}_X^\dagger \bs{A} \bs{b}_X = h \overline{h} g \overline{g}$.
Note that this is required to contain a factor of $h \overline{h}$ due to the constrains on $\bs{b}_X$ and $\bs{b}_X^\dagger$.
Utilizing the definition $t_h + \overline{t}_h = h \overline{h}$, we can decompose the inner product as
\begin{align}
    \bs{b}_X^\dagger \bs{A} \bs{b}_X = a^+(x,y) - a^+(\overline{x},\overline{y}),
\end{align}
with $a^+(x,y) = t_h g \overline{g}$.
Note that $a^+(x,y)$ is only well-defined modulo addition by a symmetric polynomial, $s(x,y) = s(\bar x, \bar y)$.

Following our usual procedure, we now consider how the same commutator would behave if the constraints were realized naturally in a LTPS.
We will find that in this case, the positive part of the commutator must contain a factor of $h \overline{h}$.
To see this, we first decompose $\bs{\Sigma} \bs{b}_X = h \bs{p}$ and $\bs{b}_X^\dagger  \bs{\Sigma}^\dagger = \overline{h} \bs{p}^\dagger$ for some vector $\bs{p}$.
This implies that the commutator  can be written as $h \overline{h} \bs{p}^\dagger \bs{A} \bs{p}$.
Decomposing $\bs{p}^\dagger \bs{A} \bs{p} = p^+(x,y) - p^+(\overline{x},\overline{y})$, we have that $a^+ = h \overline{h} p^+$.
This motivates us to define the self-\ob{} via
\begin{equation}
\begin{split}
    \tilde{q}(X) & = \pi_{\mathfrak{b} \overline{\mathfrak{b}},S} ( a^+(x,y) ) \\
    & \in \mathbbm{Z}_2[x,y]/ \left( \mathfrak{b} \overline{\mathfrak{b}} \mathbbm{Z}_2[x,y] + S_{\mathbbm{Z}_2}[x,y] \right)
\end{split}
\end{equation}
where the quotient $\pi_{\mathfrak{b} \overline{\mathfrak{b}},S}$ is performed over the addition of the submodule $\mathfrak{b} \overline{\mathfrak{b}} \mathbbm{Z}_2[x,y]$ and the submodule of symmetric polynomials, which we denote $S_{\mathbbm{Z}_2}[x,y]$.
The quotient over the first submodule implies that $\tilde{q}$ is trivial in any LTPS, while the quotient over symmetric polynomials is necessary for $\tilde{q}$ to be independent of our choice of $a^+(x,y)$.

\begin{figure*}
\centering
\includegraphics[width=\textwidth]{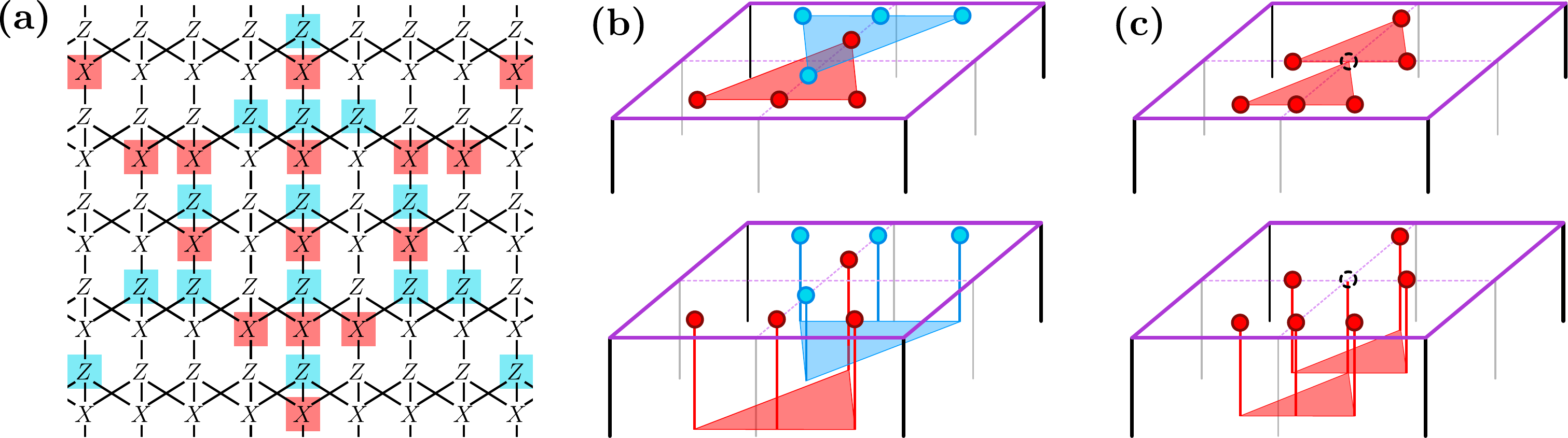}
\caption{
The (001)-boundary of the Fibonacci prism model.
\textbf{(a)} Frustration graph and constraints of local boundary operators.
\textbf{(b)} Patch operators formed from the $Z$ (blue) and $X$ (red) boundary operators. The mutual-\ob{} is equal to the commutator of the patch operators (top), which is in turn related to the mutual statistics of bulk quasiparticles (bottom).
\textbf{(c)} The self-\ob{} is equal to the commutator of two patch operators (top), and is similarly related to the self statistics of bulk quasiparticles (bottom).
} 
\label{fig: fibonacci}
\end{figure*}

Although it is not obvious from the quotient space above, we can show that $\tilde{q}$ in fact takes values in $\mathbbm{Z}_2$.
Note that the commutator $\bs{b}_X^\dagger \bs{A} \bs{b}_X$ can be written as $f h \bar h$ for some polynomial $f(x,y)$, owing to the boundary constraint.
In the above model, we have $f(x,y) = g(x,y) g(\bar x, \bar y)$.
Since the commutator is anti-symmetric, the polynomial $f$ is as well, $f(\bar x, \bar y) = - f(x,y)$. (To avoid confusion, we note that an anti-symmetric polynomial is symmetric, and vice versa, over $\mathbbm{Z}_2$.)
Now, recall that by taking tensor products with a LTPS we can modify $f$ via $f \rightarrow f + p - \bar p$ for any polynomial $p(x,y)$.
We can choose $p$ such that it cancels every term in $f$ aside from the constant term (i.e.~the term proportional to $x^0 y^0$).
The invariant $\tilde{q}$ corresponds precisely to this term, and therefore lies in $\mathbbm{Z}_2$.
To see this, note that if the constant term in $f$ is zero, then the commutator is zero up to tensor products with a LTPS and hence $\tilde{q}$ is also zero.
On the other hand, if the constant term is one, then $\tilde{q} = [ t_h  ]$.
To show that this equivalence class is not the zero equivalence class, we need to show that $t_h$ cannot be written as the sum of a symmetric polynomial $s$ and a multiple $r$ of $h \bar h$.
If this were the case, we would have $t_h + \bar t_h = s + \bar s + (r + \bar r) h \bar h$. However, this contradicts the definition $t_h + \bar t_h = h \bar h$, since $s + \bar s$ is zero (because $s$ is symmetric) and the constant term in $r + \bar r$ is also zero.
Hence, a non-zero constant term in $f$ implies a non-zero $\tilde{q}$, and so $\tilde{q}$ is a $\mathbbm{Z}_2$ invariant.

\subsection{Example: Fibonacci prism model}

We now show how the obstructor invariants appear in a particularly simple fractal model: the Fibonnaci prism model.
The Fibonnaci prism model is defined by the bulk stabilizers,
\begin{equation}
\bs \sigma=\begin{pmatrix}
    1+ \bar z   & 0\\
    1+\bar y(1+ x+\bar x)&0\\
    \hline
    0&1+y (1+ x+\bar x) \\
   0& 1+ z
    \end{pmatrix},
\end{equation}
which correspond to the choices $g(x,y)=1$ and $h(x,y)=1+y(1+ x+\bar x)$.
We again work over $\mathbbm{Z}_2$ for simplicity.
The simple form of the stabilizers with respect to both the $y$ and $z$ coordinates will allow us to connect the mutual-\ob{} on the boundary to the commutation of patch operators, and, in turn, to the bulk statistics.
To address the self-\ob{}, we will also consider a `fermionic' generalization of the Fibonnaci prism model [see Eq.~(\ref{eq: fermion fibonnaci bulk stabilizers})].

The bulk conservation laws of the Fibbonaci prism model are
\begin{equation}
\begin{split}
    \sum_{i,j} (1+ x+\bar x)^i \, y^i \, z^j \bs \sigma_X & = 0, \\
    \sum_{i,j} (1+ x+\bar x)^i \, \overline{y}^{i} \, \overline{z}^{j} \bs \sigma_Z & =0,
\end{split}
\end{equation}
These correspond to the ideals $\mathfrak{c} = \left( 1 + y(1+x+\overline{x}), 1 + z\right)$ and $\overline{\mathfrak{c}}$, respectively.
The truncated operators on the (001)-boundary are
\begin{equation}
\tbs \sigma=\begin{pmatrix}
    1  & 0\\
    1+\bar y(1+ x+\bar x)&0\\
    \hline
    0&0 \\
   0& 1
    \end{pmatrix},
\end{equation}
with an adjacency matrix
\begin{equation}
\bs A = \begin{pmatrix}
   0&  1+y(1+ x+\bar x)\\
 1+\bar y(1+ x+\bar x) &0
    \end{pmatrix}.
\end{equation}
The constraints imposed on the boundary by the bulk conservation laws are
\begin{equation}
\begin{split}
    \sum_{i} (1+ x+\bar x)^i \, y^i \tbs \sigma_X &\eqgs 0, \\
   \sum_{i} (1+ x+\bar x)^i \, \overline{y}^{i} \tbs \sigma_Z  & \eqgs 0,
\end{split}
\end{equation}
which correspond to the ideals $\mathfrak{b} = \left( 1 + y(1+x+\overline{x})\right)$ and $\overline{\mathfrak{b}}$, respectively.
The frustration graph and constraints are depicted in Fig.~\ref{fig: fibonacci}(a).

From the results of the previous section, the mutual-\ob{} is given by $[1] \in \mathbbm{Z}_2/\mathfrak{b} \mathbbm{Z}_2$.
We can connect this invariant to the commutation of fractal patch operators on the boundary.
Consider the patch operators shown in Fig.~\ref{fig: fibonacci}(b), which correspond to an upward facing triangle for the $\tbs \sigma_Z$ operators and a downward facing triangle for $\tbs \sigma_X$.
We will consider the commutation of the two patch operators when the tip of each triangle lies within the bulk of the opposing triangle, as in Fig.~\ref{fig: fibonacci}(b).
A straightforward calculation gives,
\begin{equation}
\begin{split}
    \langle \,  \sum_{i=0}^{\infty} & (1+ x+\bar x)^i \, \overline{y}^{i} \tbs \sigma_Z  \, , \, \sum_{j=0}^{\infty} (1+ x+\bar x)^j \, y^j \tbs \sigma_X \, \rangle \\
    & = \sum_{i=0}^{\infty} \sum_{j=0}^{\infty} y^{i+j} (1+ x+\bar x)^{i+j} \bs{b}_Z^\dagger \bs{A} \bs{b}_X \\
    & = \sum_{k=0}^{\infty} k \, y^{k} (1+ x+\bar x)^{k} \bs{b}_Z^\dagger \bs{A} \bs{b}_X. \\
\end{split}
\end{equation}
Now, recall that $\bs{b}_Z^\dagger \bs{A} \bs{b}_X = 1 + y(1+x+\bar x)$.
(More generally, $\bs{b}_Z^\dagger \bs{A} \bs{b}_X$ must contain at least a factor of $1 + y(1+x+\bar x)$ in order to obey the boundary constraint.)
Hence, the commutator above simplifies to
\begin{equation}
\begin{split}
    \sum_{k=0}^{\infty} k \, y^{k+1} & (1+ x+\bar x)^{k+1} + \sum_{k=0}^{\infty} k \, y^{k} (1+ x+\bar x)^{k} \\
    & = \sum_{k=0}^{\infty} y^{k} (1+ x+\bar x)^{k}. \\
\end{split}
\end{equation}
This is an semi-infinite pattern that encodes which translations of the upward facing triangle anti-commute with the downward facing triangle.
This is analogous to the semi-infinite sum, $\sum_{i=0}^\infty x^i$, we previously encountered when looking at the commutation of two semi-infinite one-dimensional boundary strings.
As in the one-dimensional case, the prefactor, $1$, of the semi-infinite pattern is precisely the mutual-\ob{}.

The commutator of the patch operators, and hence the mutual-\ob{}, are easily related to the mutual statistics of the bulk quasiparticles of the Fibbonaci prism model.
We show this in Fig.~\ref{fig: fibonacci}(b).

To discuss the self-\ob{}, we consider a generalization of the Fibonnaci prism model that features emergent fermionic quasiparticles in the bulk~\cite{Tantivasadakarn20,Shirley2020}. The stabilizers of this model are
\begin{equation} \label{eq: fermion fibonnaci bulk stabilizers}
\bs \sigma=\begin{pmatrix}
    1+ \bar z   &z (1+y (1+ x+\bar x))\\
    1+\bar y(1+ x+\bar x)&( \bar x^2 + y(1+x+\bar x)) (1+z)\\
    \hline
    0&1+y (1+ x+\bar x) \\
   0& 1+ z
    \end{pmatrix},
\end{equation}
which satisfy the same conservation laws as in the previous model. The truncated operators on the (001)-boundary are
\begin{equation}
\tbs \sigma=\begin{pmatrix}
    1  &  1+y (1+ x+\bar x)\\
    1+\bar y(1+ x+\bar x)& \bar  x^2 + y(1+x+\bar x) \\
    \hline
    0&0 \\
   0& 1
    \end{pmatrix},
\end{equation}
and the adjacency matrix is
\begin{align}
\bs A &= \begin{pmatrix}
   0&  h\\
 \bar h & h \bar h \end{pmatrix},
%      \bs A_{ZX} &= \begin{pmatrix}
%   0&  1+y(1+ x+\bar x)\\
%  0 &x^2 +y(1+x+\bar x)
%     \end{pmatrix}
\end{align}
with $h(x,y) = 1 + y(1+x+\bar x)$.

This features a non-zero self-\ob{}, since $\bs{b}_X^\dagger \bs{A} \bs{b}_X = h \bar h$ (see the discussion in the final paragraph of the preceding section).
The non-zero self-\ob{}  leads to a non-zero commutation of the patch operators shown in Fig.~\ref{fig: fibonacci}(c).
The two patch operators correspond to the sums,
\begin{equation}
    \sum_{j=0}^{2^A-1} y^j (1+x+\bar x)^j \,\,\,\,\,\,\,\, , \,\,\,\,\,\,\,\, y^{2^A} \sum_{j=0}^{2^A-1} y^j (1+x+\bar x)^j,
\end{equation}
respectively, where $A$ is an integer that determines the length of the patch operator.
Their commutation corresponds to the constant $x^0 y^0$ term in the following:
\begin{equation}
\begin{split}
    \left\langle   y^{2^A}  \sum_{i=0}^{2^A-1} y^i (1+x+\bar x)^i \tbs \sigma_X  ,   \sum_{j=0}^{2^A-1} y^j (1+x+\bar x)^j \tbs \sigma_X  \right \rangle \\
      = y^{-2^A} \sum_{i=0}^{2^A-1}\sum_{j=0}^{2^A-1} y^{j-i} (1+ x+\bar x)^{j-i} \bs{b}_X^\dagger \bs{A} \bs{b}_X 
\end{split}
\end{equation}
The terms with $y^0$ arise from the $i = 0, j = 2^A-1$ term in the sum, multiplied by the $y^1$ term in $\bs{b}_X^\dagger \bs{A} \bs{b}_X$.
Isolating these terms gives,
\begin{equation}
    (1+x+\bar x)^{2^A} = 1 + x^{2^A} + \bar x^{2^A},
\end{equation}
which indeed contains a non-zero term for $x^0$.
Hence the two patch operators anti-commute.
Note that if the self-\ob{} were zero, then we could write $\bs{b}_X^\dagger \bs{A} \bs{b}_X = (p-\bar p) h \bar h$ for some polynomial $p(x)$.
Assuming that $p$ does not depend on $y$ for simplicity, the above expression would be modified to $(1 + x^{2^A} + \bar x^{2^A})(p(x) - p(\bar x))$ for some polynomial $p(x)$. 
This contains no constant term and hence would lead to the patch operators commuting.

Similar to the mutual-\ob{}, the commutator of the patch operators corresponding to the self-\ob{} can easily be seen to be equal to the self statistics of the fermionic Fibbonaci prism model's bulk quasiparticles.
This is depicted in Fig.~\ref{fig: fibonacci}(c).

\section{Outlook}\label{sec:outlook}

In this work, we have established a general bulk-boundary correspondence for topological stabilizer codes.
We show that the anyon data of the bulk topological order, in the form of bulk conservation laws, gives rise to constraints on the boundary operator algebra.
In many cases, these constraints take the form of emergent symmetries---and, for fracton orders, emergent \emph{subsystem} symmetries---on the boundary theory.
From these constraints, we define invariants that codify the obstructions to realizing the boundary theory via local degrees of freedom.
These obstructions are directly inherited from the non-trivial braiding statistics of the bulk topological order.
Leveraging the polynomial formalism, we apply our framework to three-dimensional fracton phases of matter, providing a cohesive understanding of the bulk-boundary correspondence for the most well-known fracton stabilizer codes.

We remark that there is a curious relation between the frustration graphs of the boundaries of CSS codes (in which the graph is bipartite) and cluster states defined on such graphs, which can be interpreted as an SPT phase. Indeed, a similar observation has been made by interpreting the entanglement negativity of topological orders as arising from the partition function of a cluster state SPT phase at the bipartition, which arises from the exact same frustration graph~\cite{LuVijay22}. Along these lines, on the boundaries of fracton stabilizer codes, we note that our mutual-obstructor invariant (which distinguishes e.g. the boundary of the X-Cube model from that of a stack of toric codes) is closely reminiscent of the invariant used to detect strong subsystem-symmetry-protected topological phases in two dimensions~\cite{DevakulWilliamsonYou2018}.

It would also be interesting to study stabilizer models that do not have bulk topological order, yet have anyonic excitations on their boundaries. For example, Pauli stabilizer models have been constructed for all Walker-Wang models associated to modular Abelian anyon theories~\cite{BCFV14,HaahFidkowskiHastings18,Haah21,Shirley22}.

In the context of fracton phases, our results show that a more general framework than in two-dimensional topological order is needed to encapsulate the bulk-boundary correspondence.
For instance, unlike conventional topological order, certain exchange statistics of the bulk topological order can only be detected along certain orientations of the boundary termination.
For example, the self statistics of bulk fractons in the X-Cube model only affect the ($IJK$)-boundary terminations, where all of $I$, $J$, and $K$ are non-zero. It would be interesting to determine what are the minimal boundary terminations necessary to reconstruct the bulk fracton order.

Finally, although we have focused our discussion on $\ZZ_2$ stabilizer codes for simplicity, our framework is easily adapted to qudit stabilizer codes as well as fermions~\cite{VijayHaahFu2015,Tantivasadakarn20,Ellison22}. We illustrate this in Appendix~\ref{app:Majorana}, where we apply our framework to the bulk-boundary correspondence in the Majorana color code, and Appendix~\ref{app:DS}, where we outline the bulk-boundary correspondence for the Pauli double-semion model.

Looking further, in higher dimensions, there exist an abudance of topological and fracton orders that can be realized by stabilizer codes. 
For instance, recent work has unveiled so-called hybrid fracton orders~\cite{BulmashBarkeshli2019,PremWilliamson2019,StephenGarre-RubioDuaWilliamson2020,AasenBulmashPremSlagleWilliamson20,TJV1,TJV2}, which exhibit a non-trivial mixture of mobile and immobile excitations. Extending our framework to explore the bulk-boundary correspondence of such phases of matter is an intriguing topic for future work.

\section*{Acknowledgments}
We are grateful to Wenjie Ji and Ari Turner for helpful discussions. 
This work was supported in part by multiple awards from the NSF including the QII-TAQS program, the STAQ II program, and the Center for Ultracold Atoms. 
T.S. and N.T. thank the Les Houches summer school on Dynamics and Disorder in Quantum Many Body Systems Far from Equilibrium, where this project was initiated.
T.S. and N.T. are supported by the Walter Burke Institute for Theoretical Physics at Caltech.
T.S. also acknowledges support via an NSF GRFP.
A.V. is supported by a Simons Investigator grant, NSF-DMR 2220703, and the Simons Collaboration on Ultra-Quantum Matter, which is a grant from the Simons Foundation (618615, A.V.)
N.Y.Y. acknowledges support from a Simons Investigator Award.

\bibliography{references.bib}

\newpage
\widetext
%%%
%%%
%%%
\appendix
%%%
%%%
%%%

\section{Mathematical properties of the \obs{}} \label{app:quadraticform}

In this Appendix, we prove that the self-\obs{} $\tilde{q}(\alpha)$ and mutual-\obs{} $\tilde{b}(\alpha,\beta)$ define a quadratic and associated bilinear form, by deriving properties 1, 2, 3 in Section~\ref{sec:obstructor} of the main text.

We begin with Property 1: $\tilde{q}(\alpha^n) = n^2 \tilde{q}(\alpha)$.
The patch operators of the conservation law $\alpha^n$ are equal to the $n^{\text{th}}$ power of the patch operators of the conservation law $\alpha$.
The patch operators of the latter obey $\hat{R}^\alpha_i \hat{L}^\alpha_i = e^{2\pi i \tilde{q}(\alpha)/d} \hat{L}^\alpha_i \hat{R}^\alpha_i$ by definition.
By successively applying this identity to the commutator of $(\hat{R}^\alpha_i)^n$ and $(\hat{L}^\alpha_i)^n$, we can derive Property 1:
\begin{equation}
    \exp \left( \frac{2\pi i}{d} \tilde{q}(\alpha^n) \right) = (\hat{R}^\alpha_i)^n (\hat{L}^\alpha_i)^n (\hat{R}^{\alpha,\dagger}_i)^n (\hat{L}^{\alpha,\dagger}_i)^n = \exp \left( \frac{2\pi  i}{d} n^2 \tilde{q}(\alpha) \right).
\end{equation}

We now turn to Property 2: $\tilde{b}(\alpha\gamma,\beta) = \tilde{b}(\alpha,\beta)+\tilde{b}(\gamma,\beta)$.
This follows directly from the fact that Pauli operators commute up to a phase:
\begin{equation}
    \hat{R}^\alpha_i \hat{R}^\gamma_i \hat{L}^\beta_j = 
    e^{\frac{2\pi i}{d} \tilde{b}(\gamma,\beta)}
    \hat{R}^\alpha_i \hat{L}^\beta_j \hat{R}^\gamma_i = 
    e^{\frac{2\pi i}{d} [ \tilde{b}(\alpha,\beta) + \tilde{b}(\gamma,\beta)]}
    \hat{L}^\beta_j \hat{R}^\alpha_i  \hat{R}^\gamma_i,
\end{equation}
from which Property 2 directly follows.

Finally, we turn to Property 3: $\tilde{b}(\alpha,\beta) = \tilde{q}(\alpha \beta)-\tilde{q}(\alpha)-\tilde{q}(\beta)$.
We will prove the re-arranged expression $\tilde{q}(\alpha \beta) = \tilde{q}(\alpha)+\tilde{q}(\beta) - \tilde{b}(\alpha,\beta)$.
The self-\ob{} $\tilde{q}(\alpha \beta)$ is equal to the commutator of the string $\hat{L}^\alpha_j \hat{L}^\beta_j$ with the string $\hat{R}^\alpha_j \hat{R}^\beta_j$.
This receives four contributions: First, from the commutator of $\hat{L}^\alpha_j$ with $\hat{R}^\alpha_j$, which equals the self-\ob{} $\tilde{q}(\alpha)$.
Second, from the commutator of $\hat{L}^\beta_j$ with $\hat{R}^\beta_j$, which equals $\tilde{q}(\beta)$.
Finally, we receive two `cross-terms' corresponding the commutator of $\hat{L}^\alpha_j$ with $\hat{R}^\beta_j$, and of $\hat{L}^\beta_j$ with $\hat{R}^\beta_j$.
For Property 3 to hold, we need to show that the final two contributions combine to equal negative the mutual-\ob{}, $-\tilde{b}(\alpha,\beta)$.

To do so, we first introduce a new notation: we denote a finite string from sites $i$ to $j$ corresponding to a constraint $\alpha$ as $\hat{S}^\alpha_{i,j}$.
The right and left strings above correspond to taking the second endpoint to infinity, or the first endpoint to negative infinity, respectively.
We also define the commutator $C(\alpha,i,j ; \beta,k,l)$ of two strings via
\begin{equation}
    \hat{S}^\alpha_{i,j} \hat{S}^\beta_{k,l} \equiv e^{\frac{2\pi i}{d} C(\alpha,i,j ; \beta,k,l) } \hat{S}^\beta_{k,l} \hat{S}^\alpha_{i,j}.
\end{equation}
In what follows, we will always assume that endpoints of all strings are either equal, or separated from one another by distances greater than $K$. In this case the commutators of strings are independent of the strings' precise endpoints.

In this new notation, the aforementioned two contributions of interest are equal to the sum of commutators $C(\alpha,i,j ; \beta,j,k) + C(\beta,i,j ; \alpha,j,l)$.
Here we define a new leftmost endpoint $i < j - K$ and two new right endpoints, $k > j + K$ and $l > k + K$.
This particular arrangement of endpoints is chosen to be convenient for what follows; the commutators  are independent of the endpoints as long as $i < j-K$ and $k,l > j+K$.
We now perform several re-arrangements that lead to our desired Property 3.
First, we apply the equality $C(\alpha,i,j ; \beta,j,k) = -C(\alpha,j,l ; \beta,j,k)$ to the first commutator.
This follows because the string $\hat{S}^\alpha_{i,l}$ commutes with the string $\hat{S}^\beta_{j,k}$, since the latter string is fully contained within the former string.
The former string is equal to the product of our original string from $i$ to $j$ with the new string from $j$ to $l$, $\hat{S}^\alpha_{i,l} = \hat{S}^\alpha_{i,j} \hat{S}^\alpha_{j,l}$.
The fact that the commutator of the product is zero allows us to exchange the commutator of $\hat{S}^\alpha_{i,j}$ with that of $\hat{S}^\alpha_{j,l}$ up to a minus sign.
Second, we eliminate this minus sign by switching the order of the arguments, $-C(\alpha,j,l ; \beta,j,k) = C( \beta,j,k ; \alpha,j,l)$.
Third, we note that, after these re-arrangements, the original sum of commutators has become $C(\beta,j,k ; \alpha,j,l) + C(\beta,i,j ; \alpha,j,l)$.
Since the two commutators share a second argument, we can product together their first arguments to form the string $\hat{S}^\beta{i,k} = \hat{S}^\beta{i,j} \hat{S}^\beta{j,k}$.
The sum of commutators is equal to the single commutator $C(\beta,i,k ; \alpha,j,l) = -C(\alpha,j,l ; \beta,i,k)$.
The latter is precisely the mutual-\ob{} $\tilde{b}(\alpha,\beta)$, which proves Property 3.

\section{Equality of boundary \obs{} and bulk statistics}\label{sec: bulk boundary computation}

In this Appendix, we explicitly derive the equality between the self- and mutual-\obs{} on the boundary and the self and mutual statistics of quasiparticles in the bulk, for translation invariant 2D stabilizer models.
This explicit computation makes precise the pictorial derivation in the main text.

We consider a translationally-invariant 2D stabilizer model described by the stabilizer matrix $\bs \sigma(x,y)$, which contains powers of $x$ and $y$ between $0$ and $K$.
The conservation laws of $\bs \sigma(x,y)$ correspond to vectors, $\bs c$, that obey
\begin{equation} \label{bulk claw}
\sum_{i = -\infty}^{\infty} \sum_{j = -\infty}^{\infty}  x^i y^j \bs \sigma(x,y) \bs c = 0, \text{          i.e. } \bs \sigma(1,1) \bs c = 0.
\end{equation}
Here, we have restricted to planar conservation laws, which is valid for 2D topological stabilizer codes.
We have also assumed the conservation law has period one in both directions, which can be ensured by taking an appropriately large unit cell.
Writing $\bs d(x,y) \equiv \bs \sigma(x,y) \bs c$, this condition allows one to decompose:
\begin{equation} \label{2D decomposition}
\bs d = (x-1) \bs d_x + (y-1) \bs d_y.
\end{equation}

The topological excitations of the model are given by string operators that anti-commute with stabilizers at either end of the string.
As is familiar from the toric code, these string operators can in fact be generated from the conservation laws of the model.
To see this, take a conservation law, $\bs c$, and consider the product of the $L_x \times L_y$ grid of stabilizers, $\bs d = \bs \Sigma \bs c$. In the polynomial notation, this product takes the form:
\begin{equation} \label{grid to lines}
\Sigma_x^{0,L_x} \Sigma_y^{0,L_y} \bs d = (x^{L_x} - 1) \Sigma_y^{0,L_y} \bs d_x + (y^{L_y} - 1) \Sigma_x^{0,L_x} \bs d_y
\end{equation}
where for convenience we define the polynomial, $\Sigma_x^{a,b} \equiv \sum_{i=a}^{b-1} x^i$, which obeys $(x-1) \Sigma_x^{a,b} = x^{b} - x^{a}$ (and similar for $\Sigma_y^{a,b}$).
On the right hand side, we have applied Eq.~(\ref{2D decomposition}) to reduce the two-dimensional product of stabilizers to a product of one-dimensional strings around the grid (the first term corresponds to the right and left edge of the grid, the second term to the top and bottom edge).
Now, consider isolating a single string, e.g. along bottom edge of the grid, $\bs S_x = - \Sigma_x^{0,L_x} \bs d_y$.
Since the grid was formed by a product of stabilizers, $\bs S_x$ commutes with all stabilizers except those within a distance $K$ of its ends, which lie at $(0,0)$ and $(L_x,0)$.
At these points, $\bs S_x$ (potentially) creates a topological excitation.
Similarly, the string $\bs S_y = - \Sigma_y^{0,L_y} \bs d_x$ creates the same type of excitation at its ends, $(0,0)$ and $(0,L_y)$.

\begin{figure}
\centering
\includegraphics[width=0.7\columnwidth]{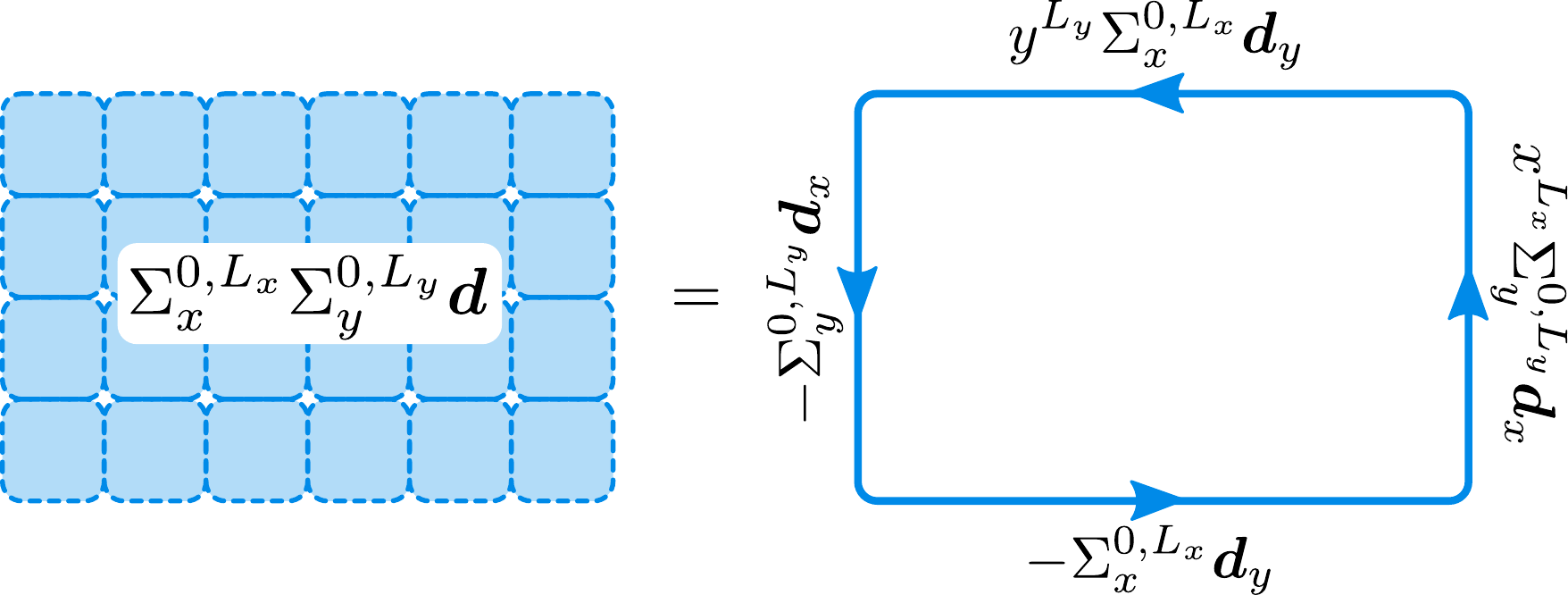}
\caption{
Depiction of Eq.~(\ref{2D decomposition}).
(Left) The product of a two-dimensional grid of bulk stabilizers, $\bs d = \bs \Sigma \bs c$, where $\bs c$ represents a bulk conservation law [Eq.~(\ref{bulk claw})], is equal to (Right) the product of one-dimensional string operators along the grid boundary.
}
\label{fig: conservation law to string}
\end{figure}

\emph{Mutual statistics in bulk.}---We now compute the  mutual statistics between topological excitations.
Consider the mutual statistics of two excitations, $\alpha$ and $\alpha'$, generated as above from conservation laws, $\bs d$, $\bs d'$.
The mutual statistics can be computed via the commutator of a horizontal string $H$ that creates $\alpha$ excitations with an intersecting vertical string $V$ that creates $\alpha'$ excitations (Fig.~\ref{fig: bulk equals boundary statistics}):
\begin{equation}
\exp \left(i \frac{2\pi}{d} b(\alpha,\alpha') \right) = H V H^\dagger V^\dagger.
\end{equation}
We take each string to be infinite to ensure they intersect, and abbreviate, $\Sigma_x \equiv \Sigma_x^{-\infty,\infty}$.
In polynomial notation, we have:
\begin{equation} \label{bulk K bulk d}
b(\alpha,\alpha') = [ \langle \Sigma_x \bs d_y, \Sigma_y \bs d_x' \rangle ]_{0,0} = \langle \bs d_y, \bs d_x' \rangle(1,1) = \sum_{i,j} \left[ \langle \bs d_y, \bs d_x' \rangle \right]_{i,j},
\end{equation}
where we use $\langle \Sigma_x \bs d_y, \Sigma_y \bs d_x' \rangle = \langle \bs d_y, \bs d_x' \rangle(x,y) \cdot \overline{\Sigma}_x \Sigma_y = \langle \bs d_y, \bs d_x' \rangle(1,1) \cdot \overline{\Sigma}_x \Sigma_y$, and $[\overline{\Sigma}_x \Sigma_y]_{0,0} = 1$.

\emph{Self statistics in bulk.}---We can similarly compute the self statistics.
The self statistics of an excitation $\alpha$ are determined by the outcome of the three-prong exchange process in Fig.~\ref{fig: bulk equals boundary statistics}:
\begin{equation}
\exp \left(i \frac{2\pi}{d} q(\alpha) \right) = A B C A^\dagger B^\dagger C^\dagger = (A B) (C B) (A B)^\dagger (C B)^\dagger.
\end{equation}
To compute this, we first define the ``left corner'' string, $\Sigma_x^{-\infty,0} \bs d_y + \Sigma_y^{-\infty,0} \bs d_x$, corresponding to $A B$, and the ``right corner'' string, $-\Sigma_x^{0,\infty} \bs d_y + \Sigma_y^{-\infty,0} \bs d_x$, corresponding to $CB$.
The self statistics are then determined by the commutator of the two:
\begin{equation}
\begin{split}
q(\alpha) & = [ \langle \Sigma_x^{-\infty,0} \bs d_y + \Sigma_y^{-\infty,0} \bs d_x, - \Sigma_x^{0,\infty}\bs d_y + \Sigma_y^{-\infty,0} \bs d_x \rangle]_{0,0} \\
& =  - [ \langle  \bs d_y, \bs d_y\rangle \cdot x \Sigma_x^{0,\infty} \Sigma_x^{0,\infty}]_{0,0} + [ \langle  \bs d_y,  \bs d_x \rangle \cdot x \Sigma_x^{0,\infty} \Sigma_y^{-\infty,0}]_{0,0} - [ \langle \bs d_x,  \bs d_y\rangle \cdot y \Sigma_y^{0,\infty} \Sigma_x^{0,\infty}]_{0,0} \\
\end{split}
\end{equation}
where the fourth cross term is the commutator of an operator with itself, and thus zero.
To simplify the first term in this expression, we note that $x \Sigma_x^{0,\infty} \Sigma_x^{0,\infty} = \sum_{k>0} k x^k$, which gives:
\begin{equation}
\begin{split}
- [ \langle  \bs d_y, \bs d_y\rangle \cdot x \Sigma_x^{0,\infty} \Sigma_x^{0,\infty}]_{0,0} & = - [ \langle  \bs d_y, \bs d_y\rangle \cdot \sum_{k>0} k x^k]_{0,0}\\
& = - \sum_{k>0} k \cdot [ \langle  \bs d_y, \bs d_y\rangle]_{-k,0}\\
& = \sum_{k>0} k \cdot [ \langle  \bs d_y, \bs d_y\rangle]_{k,0}\\
\end{split}
\end{equation}
For the second term, we have:
\begin{equation}
\begin{split}
[ \langle  \bs d_y,  \bs d_x \rangle \cdot x \Sigma_x^{0,\infty} \Sigma_y^{-\infty,0}]_{0,0} & =  \sum_{i < 0, \, j > 0} [ \langle  \bs d_y,  \bs d_x \rangle ]_{i,j}
\end{split}
\end{equation}
and for the third term:
\begin{equation}
\begin{split}
- [ \langle \bs d_x,  \bs d_y\rangle \cdot y \Sigma_y^{0,\infty} \Sigma_x^{0,\infty}]_{0,0} & =  - \sum_{i \leq 0, \, j < 0} [ \langle  \bs d_x,  \bs d_y \rangle ]_{i,j} 
=  \sum_{i \geq 0, \, j > 0} [ \langle  \bs d_y,  \bs d_x \rangle ]_{i,j} 
\end{split}
\end{equation}
Combining the three terms, we have:
\begin{equation} \label{bulk S bulk d}
\begin{split}
q(\alpha) & = \sum_{k>0} k \cdot [ \langle  \bs d_y, \bs d_y\rangle]_{k,0}
+ \left(  \sum_{i < 0, \, j > 0} +  \sum_{i \geq 0, \, j > 0} \right) [ \langle  \bs d_y,  \bs d_x \rangle ]_{i,j} \\
& = \sum_{k>0} k \cdot [ \langle  \bs d_y, \bs d_y\rangle]_{k,0}
+ \sum_{i, \, j > 0} [ \langle  \bs d_y,  \bs d_x \rangle ]_{i,j} \\
\end{split}
\end{equation}
We will now turn to the boundary commutator and show that they are equal to these expressions.

We begin with a brief note on notation.
In the main text and the above, we used subscripts to denote the coefficients of the term corresponding to a given power of $x$ (in 1D) or $x,y$ (in 2D).
Below, we will often want to isolate all the terms with a fixed power of $x$ or a fixed power of $y$, in 2D.
In what follows, to make it clear which variable we are isolating, we replace the subscript $k$ with a subscript $x^k$ or $y^k$.

We begin by writing down the operators of the boundary Hilbert space, using the truncation procedure introduced in the main text.
To do so, we first decompose the bulk stabilizers power-by-power in $y$, $\bs \sigma(x,y) = \sum_{k=0}^{K-1} y^k \cdot [\bs \sigma]_{y^k}(x)$.
We then obtain one set of boundary operators for each of $K$ translations of $\bs \sigma$,
\begin{equation}
\tbs \sigma_j(x) \equiv \text{trunc}\left( \bar y^j \bs \sigma \right) = \sum_{k=j}^{K-1} y^{k-j} \cdot [\bs \sigma]_{y^k}(x)
\end{equation}
with $j \in [1,\ldots,K-1]$.

In what follows, we will be focused on boundary operators that correspond to truncations of bulk conservation laws.
For a conservation law, $\bs d(x,y) = \bs \sigma(x,y) \bs c$, the associated boundary constraint involves a product over all terminations, $j$:
\begin{equation}
\begin{split}
\tbs d & \equiv \sum_{j=1}^{K-1} \tbs \sigma_j \bs c \\
& = \sum_{j=1}^{K-1} \sum_{k=j}^{K-1} y^{k-j} \cdot [\bs \sigma]_{y^k} \bs c \\
& = \sum_{l=0}^{K-2} y^l \cdot \left( \sum_{m=l+1}^{K-1} [\bs \sigma]_{y^m} \bs c \right), \\
\end{split}
\end{equation}
where in the final expression we have collected powers of $y$.
We can rewrite this using the operators $\bs d_x, \bs d_y$:
\begin{equation} \label{boundary d bulk d}
\begin{split}
\tbs d & = \sum_{l=0}^{K-2} y^l \cdot \left( \sum_{m=l+1}^{K-1} [(y-1) \bs d_y + (x-1) \bs d_x]_{y^m} \right) \\
& = \sum_{l=0}^{K-2} y^l \cdot \left( \sum_{m=l+1}^{K-1} [ \bs d_y ]_{y^{m-1}} - [ \bs d_y ]_{y^m}  + (y-1) [\bs d_x]_{y^m} \right) \\
& = \sum_{l=0}^{K-2} y^l \cdot \left(  [ \bs d_y ]_{y^l}  + (x-1) \sum_{m=l+1}^{K-1} [\bs d_x]_{y^m} \right) \\
\end{split}
\end{equation}
where we note that $[\bs d_y]_{y^{K-1}} = 0$ by assumption.
Finally, we note that translations under $y$ are not physical for boundary operators, so we implicitly always take the $y^0$-term of the boundary commutator.

\emph{Mutual statistics on boundary.}---We now compute the mutual statistics.
Consider two boundary strings, one, $\Sigma_x^{-\infty,A} \tbs d$, coming from the left and corresponding to a conservation law, $\bs c$, and the other, $ - \Sigma_x^{0,\infty} \tbs d'$, coming from the right and corresponding to a conservation law, $\bs c'$.
We assume the separation between the ends of the two strings is larger than the stabilizers themselves, $A > 2K$. 
As in the main text, the commutator of the two strings is related to the derivative of $\bs c \tbs A \bs c' = \langle \tbs d , \tbs d' \rangle$ (Fig.~\ref{fig: bulk equals boundary statistics}):
\begin{equation}
\begin{split}
\tilde{b}(\alpha,\alpha') & = [ \langle \Sigma_x^{-\infty,A} \tbs d, - \Sigma_x^{0,\infty} \tbs d' \rangle ]_{x^0} \\
& = - [ x^{-A+1} \Sigma_x^{0,\infty} \Sigma_x^{0,\infty} \langle \tbs d, \tbs d' \rangle ]_{x^0} \\
& = - [ x^{-A+1} \left( \sum_{k \geq 0} (k+1) x^{k} \right) \langle \tbs d, \tbs d' \rangle ]_{x^0} \\
& = -\sum_{l}(-l + A) \cdot [ \langle \tbs d, \tbs d' \rangle ]_{x^l} \\
& = \sum_{l} l \cdot [ \langle \tbs d, \tbs d' \rangle ]_{x^l} \\
& = D_x[\langle \tbs d , \tbs d' \rangle](1). \\
\end{split}
\end{equation}
where in the fourth line we use $A > 2K$ to extend the summation range to infinity, and in the fifth line we use $\sum_l  [ \langle \tbs d, \tbs d' \rangle ]_{x^l} = \langle \tbs d, \tbs d' \rangle(1) = 0$, since $\tbs d$ corresponds to a boundary constraint.

We can re-express the boundary commutator in terms of bulk operators using Eq.~(\ref{boundary d bulk d}).
We have:
\begin{equation}
\begin{split}
\langle \tbs d , \tbs d' \rangle & =  \sum_l  \left\langle  [ \bs d_y ]_{y^l}  + (x-1) \sum_{m=l+1}^K  [\bs d_x]_{y^m}, \,\, [ \bs d_y' ]_{y^l}  + (x-1) \sum_{m'=l+1}^K [\bs d_x']_{y^{m'}} \right\rangle \\
\end{split}
\end{equation}
and thus
\begin{equation} \label{boundary K bulk d}
\begin{split}
\tilde{b}(\alpha,\alpha') & =  \left[ \sum_{l} D_x \left[ \left\langle [ \bs d_y ]_{y^l}, [ \bs d_y' ]_{y^l} \right\rangle \right] 
+ \sum_{m > l} \left\langle [ \bs d_y ]_{y^l}, [\bs d_x']_{y^m} \right\rangle
- \sum_{m > l} \left\langle [\bs d_x]_{y^m}, [ \bs d_y' ]_{y^l} \right\rangle \right](1) \\
& =   \sum_{k} k \cdot \left[ \left\langle  \bs d_y, \bs d_y' \right\rangle \right]_{k,0} 
+ \sum_{k, \, l > 0}  [ \left\langle \bs d_y, \bs d_x' \right\rangle ]_{k,l}
- \sum_{k, \, l < 0} [ \left\langle \bs d_x, \bs d_y' \right\rangle ]_{k,l}.
\end{split}
\end{equation}

\begin{figure}
\centering
\includegraphics[width=0.45\columnwidth]{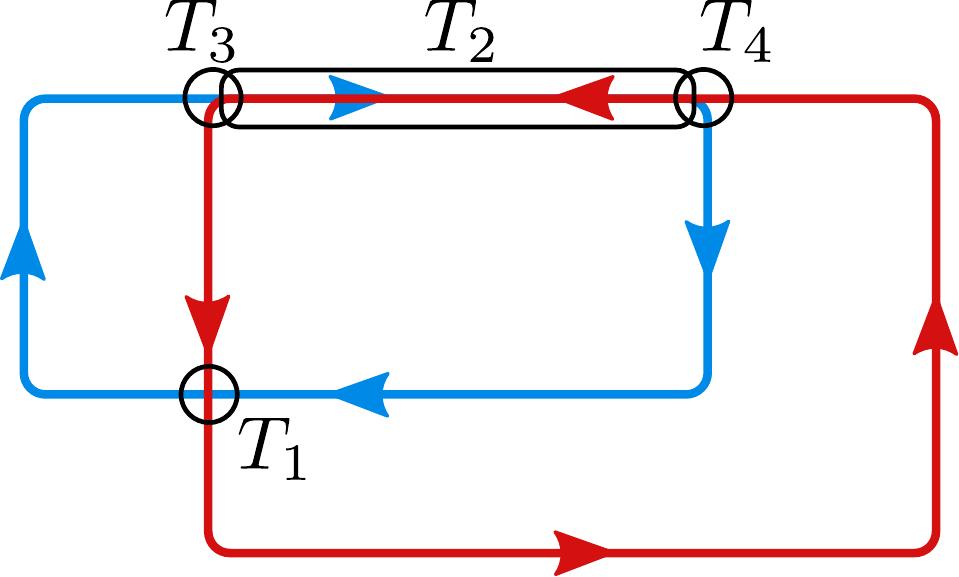}
\caption{
Depiction of the commutator in Eq.~(\ref{T14 grid comm}), with the components comprising each term, $T_1,T_2,T_3,T_4$, circled.
}
\label{fig: bulk comm for mutual}
\end{figure}

To demonstrate that Eq.~(\ref{boundary K bulk d}) is in fact equal to Eq.~(\ref{bulk K bulk d}), we must invoke the commutation of the bulk stabilizers $\bs d, \bs d'$.
Specifically, we consider the commutator depicted in Fig.~\ref{fig: bulk comm for mutual}, of a grid of bulk stabilizers, $\bs d$, with a grid of bulk stabilizers, $\bs d'$.
The commutator is zero since the bulk stabilizers mutually commute.
Using Eq.~(\ref{grid to lines}), we can decompose the commutator as a sum of four terms:
\begin{equation} \label{T14 grid comm}
\begin{split}
[\langle - \Sigma_x^{H_1,H_3} \Sigma_y^{-V_1,0} \bs d, \, \Sigma_x^{H_2,H_4} \Sigma_y^{-V_2,0} \bs d' \rangle]_{0,0} & = T_1 + T_2 + T_3 + T_4 = 0,  
\end{split}
\end{equation}
where the terms are defined as (see Fig.~\ref{fig: bulk comm for mutual}):
\begin{equation}
\begin{split}
T_1 & = [\langle y^{-V_1} \Sigma_x^{H_1,H_3} \bs d_y, \, - x^{H_2} \Sigma_y^{-V_2,0} \bs d'_x \rangle]_{0,0} \\
T_2 & = [\langle - \Sigma_x^{H_1,H_3} \bs d_y   , \, \Sigma_x^{H_2,H_4} \bs d'_y \rangle]_{0,0} \\
T_3 & = [\langle - \Sigma_x^{H_1,H_3} \bs d_y   , \, - x^{H_2} \Sigma_y^{-V_2,0} \bs d'_x \rangle]_{0,0} \\
T_4 & = [\langle - x^{H_3} \Sigma_y^{-V_1,0} \bs d_x   , \, \Sigma_x^{H_2,H_4} \bs d'_y \rangle]_{0,0}. \\
\end{split}
\end{equation}
Here, and throughout the following derivations, we take the grid corners to be separated by greater than twice the stabilizer support, i.e. we assume:
\begin{equation} \label{big grid assumption}
H_{\alpha+1}-H_\alpha > 2K, \,\,\,\,\,\, V_2-V_1, V_1 > 2K.
\end{equation}

It is now straightforward to show that the first term is equal to minus the bulk statistics, Eq.~(\ref{bulk K bulk d}), while the second through fourth terms are equal to the three terms of the boundary statistics, Eq.~(\ref{boundary K bulk d}).
Since the four terms sum to zero, this proves:
\begin{equation}
b(\alpha,\alpha') = \tilde{b}(\alpha,\alpha').
\end{equation}
In more detail, for the first term, we have:
\begin{equation}
\begin{split}
T_1 & = [\langle y^{-V_1} \Sigma_x^{H_1,H_3} \bs d_y, \, - x^{H_2} \Sigma_y^{-V_2,0} \bs d'_x \rangle]_{0,0} \\
& = -\left[ \sum_{i = -H_3+H_2+1}^{H_2-H_1} \sum_{j = -V_2+V_1}^{V_1-1} x^i y^j \cdot \langle \bs d_y, \,  \bs d'_x \rangle \right]_{0,0} \\
& = -\sum_{i,j}  \left[ \langle \bs d_y, \,  \bs d'_x \rangle \right]_{i,j} \\
& = -b(\alpha,\alpha'). \\
\end{split}
\end{equation}
where in the third line we use Eq.~(\ref{big grid assumption}) to extend each summation range to $(-\infty,\infty)$.
For the second term, we have:
\begin{equation}
\begin{split}
T_2 & = [\langle - \Sigma_x^{H_1,H_3} \bs d_y   , \, \Sigma_x^{H_2,H_4} \bs d'_y \rangle]_{0,0} \\
& = - \left[ \sum_{i=-H_3+1}^{-H_1} \sum_{j=H_2}^{H_4} x^{i+j} \langle \bs d_y, \, \bs d'_y \rangle \right]_{0,0} \\
& = - \left[ x^{-H_3+H_2} \sum_{i=1}^{H_3-H_1} \sum_{j=0}^{H_4-H_2} x^{i+j} \langle \bs d_y, \, \bs d'_y \rangle \right]_{0,0} \\
& = -\sum_{k} k \cdot \left[ \langle \bs d_y, \, \bs d'_y \rangle \right]_{0,-k} \\
& = \sum_{k} k \cdot \left[ \langle \bs d_y, \, \bs d'_y \rangle \right]_{0,k}
\end{split}
\end{equation}
For the third term:
\begin{equation}
\begin{split}
T_3 & = [\langle - \Sigma_x^{H_1,H_3} \bs d_y   , \, - x^{H_2} \Sigma_y^{-V_2,0} \bs d'_x \rangle]_{0,0}. \\
& = \left[ \sum_{i=-H_3+H_2+1}^{H_2-H_1} \sum_{j=-V_2}^{-1} x^i y^j \cdot \langle \bs d_y   , \,\bs d'_x \rangle \right]_{0,0} \\
& =  \sum_{i, \, j > 0} \left[ \langle \bs d_y   , \,\bs d'_x \rangle \right]_{i,j}. \\
\end{split}
\end{equation}
Finally, for the fourth term:
\begin{equation}
\begin{split}
T_4 & = [\langle -x^{H_3} \Sigma_y^{-V_1,0} \bs d_x   , \, \Sigma_x^{H_2,H_4} \bs d'_y \rangle]_{0,0} \\
& = -\left[  \sum_{i=-H_3+H_2}^{H_4-H_3} \sum_{j=1}^{V_1} x^i y^j \cdot \langle \bs d_x   , \,\bs d'_y \rangle \right]_{0,0} \\
& =  - \sum_{i, \, j<0} \left[ \langle \bs d_x   , \,\bs d'_y \rangle \right]_{i,j}. \\
\end{split} 
\end{equation}
Comparing to Eq.~(\ref{boundary K bulk d}), we see that indeed $\tilde{b}(\alpha,\alpha') = T_2 + T_3 + T_4$, and thus $b(\alpha,\alpha') = \tilde{b}(\alpha,\alpha')$.

\emph{Self statistics on boundary.}---We can also write down the analogous expression for the self statistics (Fig.~\ref{fig: bulk equals boundary statistics}):
\begin{equation}
\begin{split}
\tilde{q}(\alpha) & = [ \langle \Sigma_x^{-\infty,0} \tbs d, -\Sigma_x^{0,\infty} \tbs d \rangle ]_{x^0} \\
 & = - [\Sigma_x^{0,\infty} \Sigma_x^{0,\infty}  \cdot \langle \tbs d, \tbs d \rangle ]_{x^0} \\
 & = - [\sum_{k>0} (k+1) x^k  \cdot \langle \tbs d, \tbs d \rangle ]_{x^0} \\
 & = - \sum_{k>0} k \cdot [\langle \tbs d, \tbs d \rangle ]_{x^{-k}} \\
  & = \sum_{k>0} k \cdot [\langle \tbs d, \tbs d \rangle ]_{x^k}. \\
\end{split}
\end{equation}
Using Eq.~(\ref{boundary d bulk d}), we have:
\begin{equation} \label{boundary S bulk d}
\begin{split}
\tilde{q}(\alpha) & = \sum_{k>0} k \cdot \sum_l  \left[\left\langle [ \bs d_y ]_{y^l}  + (x-1) \sum_{m=l+1}^{K-1}  [\bs d_x]_{y^m}, \,\, [ \bs d_y ]_{y^l}  + (x-1) \sum_{m=l+1}^{K-1}  [\bs d_x]_{y^{m'}} \right\rangle \right]_{x^k}. \\
& = \sum_{k>0} k \cdot \left[ \sum_l \langle [ \bs d_y ]_{y^l}, [ \bs d_y ]_{y^l} \rangle \right]_{x^k} 
+ \sum_{k>0} k \cdot  \left[ (x-1) \sum_{m>l} \langle [\bs d_y]_{y^l}, [\bs d_x]_{y^m} \rangle  + h.c. \right]_{x^k} \\
& = \sum_{k>0} k \cdot \left[ \sum_l \langle [ \bs d_y ]_{y^l}, [ \bs d_y ]_{y^l} \rangle \right]_{x^k} 
+ \sum_{k>0} k \cdot  \left[ (x-1) \sum_{m>l} \langle [\bs d_y]_{y^l}, [\bs d_x]_{y^m} \rangle \right]_{x^k} \\
& = \sum_{k>0} k \cdot \left[ \sum_l \langle [ \bs d_y ]_{l}, [ \bs d_y ]_{y^l} \rangle \right]_{x^k} - \left[ \sum_{m>l} \langle [\bs d_y]_{y^l}, [\bs d_x]_{y^m} \rangle \right](1) \\
& = \sum_{k>0} k \cdot \left[ \langle \bs d_y, \bs d_y \rangle \right]_{k,0} + \sum_{k, \, j > 0} \left[ \langle \bs d_y, \bs d_x \rangle \right]_{k,j} \\
\end{split}
\end{equation}
In going from the first to second line, we have used that  $(x-1) (\bar x - 1) \langle \sum_m [\bs d_x]_{y^m}, \sum_{m'} [\bs d_x]_{y^{m'}} \rangle$  vanishes after summing over $k$. 
Comparing this expression to Eq.~(\ref{bulk S bulk d}), we have:
\begin{equation}
\tilde{q}(\alpha) = q(\alpha).
\end{equation}

\section{Classification of 1D constraint algebras via \obs{}}\label{app: 1D LTPS}

In the main text, we demonstrated that the \obs{} are unchanged if one considers  tensor products of the constraints with conservation laws arising in a 1D LTPS.
In this Appendix we go further, and show that the \obs{} in fact fully characterize the polynomials $a_{\alpha \beta}(x) = [\bs c^\dagger_\alpha \bs A \bs c_\beta](x)$ up to such tensor products.

We begin with the mutual-\ob{}, $b(\alpha,\beta)$.
First, recall that we can write $a_{\alpha \beta}(x) = (x-1) d_{\alpha\beta}(x)$ since $a_{\alpha \beta}(1)$ is equal to zero if $\alpha, \beta$ are boundary constraints.
Now, consider a different polynomial, $a'_{\alpha \beta}(x) = (x-1) d'_{\alpha\beta}(x)$ with the same \ob{} as $a$, i.e.~$D_x[a'](1) = D_x[a](1) = b(\alpha,\beta)$.
The difference, $a-a'$, has an \ob{} of zero, $0 = D_x[a-a'](1) = [d-d'](1)$.
This implies that we can write $a-a' = (x-1)(\bar x-1) f$ for some polynomial $f$.
Now, consider a LTPS with conservation laws $\bs c_{\text{TP},\alpha} = \left( (x-1), 0 \right)^T$ and $\bs c_{\text{TP},\beta} = \left( 0, (x-1)f \right)^T$.
We have $\bs c^\dagger_{\text{TP},\alpha} \bs A_{\text{TP}} \bs c_{\text{TP},\beta} = (x-1)(\bar x-1) f$ and thus $\bs c'^\dagger_\alpha \bs A \bs c'_{\beta} = a'_{\alpha\beta}(x)$ for $\bs c' = \bs c + \bs c_{\text{TP}}$.
Hence the commutation polynomial $a_{\alpha \beta}(x)$ can be converted into any $a'_{\alpha\beta}(x)$ with the same \ob{} via tensor products with a 1D LTPS conservation law.

We now turn to the self-\ob{}, which involves the polynomial, $a_\alpha(x) = [\bs c^\dagger_\alpha \bs A \bs c_\alpha](x)$.
Recall that we can always write $a_\alpha(x) = a^+_\alpha(x) - a^+_\alpha(\bar x)$, where $a^+_\alpha(x)$ is defined up to addition by an arbitrary symmetric polynomial, $s(x) = s(\bar x)$.
Now consider another polynomial, $a'_\alpha(x)$ of the same form with the same self-\ob, $D_x[a'^+_\alpha](1) = D_x[a^+_\alpha](1) = q(\alpha)$.
This implies that the difference has derivative zero, $D_x[a^+-a'^+](1) = 0$.
Now, consider a LTPS with conservation law $\bs c_{\text{TP},\alpha} = \left( (x-1), (x-1) t(x) \right)^T$ for arbitrary $t(x)$.
We have $\bs c^\dagger_{\text{TP},\alpha} \bs A_{\text{TP}} \bs c_{\text{TP},\alpha} = (x-1)(\bar x-1) (t(x) - t(\bar x))$.
Note that the prefactor obeys $(x-1)(\bar x-1) = x - 2 + \bar x$. Allowing tensor products with an arbitrary LTPS conservation law thus effectively sets $x - 2 + \bar x \rightarrow 0$, or, equivalently, $\bar x \rightarrow 2 - x$ and $x^2 \rightarrow 2x-1$.
Sequential applications of the tensor product can thus reduce an arbitrary polynomial $a^+-a'^+$ to a polynomial of the form $d_0+ d_1 x$.
The constant $d_0$ can be eliminated via addition with the symmetric polynomial $s(x) = d_0$.
Meanwhile, $d_1$ must be zero because $D_x[a^+-a'^+](1) = 0$.
Hence we can construct a 1D LTPS conservation law obeying $\bs c^\dagger_{\text{TP},\alpha} \bs A_{\text{TP}} \bs c_{\text{TP},\alpha} = a(x) - a'(x)$.
The tensor product of this conservation law with $\bs c_\alpha$ thus provides a constraint with commutation polynomial $a'_{\alpha}(x)$.

\section{$(I1)$-boundaries of the Toric Code}\label{app:TCI1boundary}

In the main text, we explicitly derived the operator algebra and constraints of the (01)-boundary of the toric code.
As we outline in the main text, our framework easily extends to arbitrary boundaries, and the self- and mutual-\obs{} are independent of the choice of boundary termination.
In this Appendix, we demonstrate this explicitly by providing a short explicit analysis of the ($I$1)-boundaries of the toric code model, for integer $I \geq 1$.
We begin the simplest case of the (11)-boundary, and then proceed to more general ($I$1)-boundaries with $I \geq 2$.

To truncate the stabilizers along the (11)-boundary, we note that the (11)-boundary can be realized as the line $y^0$ after the transformation
\begin{align}
 x &\rightarrow w \equiv x y & y \rightarrow y.
\end{align}
The new coordinate $w$ runs parallel to the (11)-boundary.
The stabilizers in the new coordinates are
\begin{align}
    \bs \Sigma  =
 \begin{pmatrix}
     1+ \bar w \bar y&0 \\
    1+\bar y&0\\
     0 &  1+y \\
   0 &1+w y
   \end{pmatrix}.
\end{align}
The truncated boundary operators can be obtained by shifting each stabilizer so that the largest degree in $y$ is $y^0$ and setting $\bar y$ to zero. This gives boundary operators
\begin{align}
    \tbs \Sigma = \begin{pmatrix}
    1 & 0\\
    1 & 0\\
    0& 1 \\
    0& w
    \end{pmatrix}.
\end{align}
The adjacency matrix of the boundary operators is
\begin{align}
    \bs A =\begin{pmatrix}
    0 & 1+w\\
    1+w& 0
    \end{pmatrix},
\end{align}
just as in the (01)-boundary.

The bulk conservation laws induce the following two constraints on the boundary operator algebra:
\begin{align}
  \sum_i \sum_{j=1}^\infty w^iy^{j} \bs \sigma_Z(w,y) &= \sum_i \sum_{j=0}^\infty w^i y^j \begin{pmatrix}
     y +\bar w \\
   1+y\\
     0 \\
   0 
   \end{pmatrix} = \sum_i w^i  \begin{pmatrix}
     1 \\
    1\\
     0 \\
   0 
   \end{pmatrix} = \sum_i w^i \tbs \sigma_Z \eqgs 0, \\
     \sum_i \sum_{j=0}^\infty w^iy^{j} \bs \sigma_X(w,y) &= \sum_i \sum_{j=0}^\infty w^i y^j \begin{pmatrix}
    0 \\
  0\\
     1+y \\
   1+wy 
   \end{pmatrix} = \sum_i w^i  \begin{pmatrix}
     0 \\
    0\\
     1 \\
   1 
   \end{pmatrix} = \sum_i w^i \tbs \sigma_X \eqgs 0,
 \end{align}
again, just as in the (01)-boundary.
Since the boundary operator algebra of the (11)-boundary is precisely equal to that of the (01)-boundary, the self- and mutual-\obs{} are equal as well.

% I1 boundaries
We now turn to the more general case $I \geq 2$.
We define new coordinates
\begin{align}
 x &\rightarrow w = x y^I & y \rightarrow y,
\end{align}
where $w$ runs parallel to the ($I$1)-boundary.
The bulk stabilizers in the new basis are
\begin{align}
    \bs \Sigma  =
 \begin{pmatrix}
     1+ \bar w \bar y^I&0 \\
    1+\bar y&0\\
     0 &  1+y \\
   0 &1+w y^I
   \end{pmatrix}.
\end{align}
As before, the ($I$1)-boundary now consists of truncating all operators with powers of $y$ less than zero. 
For $I \ge 2$, we obtain $I$ independent $Z$-type truncated boundary operators and $I$ independent $X$-type operators:
\begin{align}
    \tbs \Sigma = \left ( \begin{array}{c|cccc|cccc|c}
    1 & y & y^2 & \cdots & y^{I-1} &0&0& \cdots &0 & 0\\
    1 & 1+y & y+ y^2& \cdots & y^{I-2} +y^{I-1} & 0&0& \cdots &0 & 0\\
    0&0 &0& \cdots &0 & 0 & 0 & \cdots &0 & 1 \\
    0&0 &0& \cdots &0 & w & wy & \cdots &wy^{I-2} & wy^{I-1}
    \end{array} \right).
\end{align}
The commutation of the truncated boundary operators is  described by an adjacency matrix
\begin{align}
    \bs A=\begin{pmatrix}
    \bs 0 & \bs A_{ZX}\\
    -\bs A_{ZX}^\dagger & \bs 0
    \end{pmatrix}
\end{align}
where $\bs A_{ZX}$ encodes the commutation of the $Z$-type operators with the $X$-type operators. % (recall that for $I=0$ and $I=1$ in the main text, we have $\bs A_{ZX} = (1+x)$). 
For $I \ge 2$, we have
\begin{align}
\bs A_{ZX} =\begin{pmatrix}
 x & 0 & 0 & 0 & 1 \\
 x & x & 0 & 0 & 0 \\
 0 & \ddots & \ddots & 0 & 0  \\
 0 & 0 & x& x& 0 \\ 
  0 & 0  & 0 & x & x
\end{pmatrix}_{I \times I}.
\end{align}

The bulk conservation laws, which involve products over all translations of bulk stabiliers, lead to boundary constraints that involve products over all $I$ truncated boundary operators (of each type $X$ and $Z$).
Namely, we find 
\begin{align}
    \sum_i w^i \tbs \Sigma \bs c_Z & \eqgs 0, \,\,\,\,\,\,\,\,\,\,\,\,\,\,\, \bs c_Z = (\underbrace{1,1,\ldots,1}_{I},\underbrace{0,0,\ldots,0}_{I})^T\\
    \sum_i w^i \tbs \Sigma \bs c_X & \eqgs 0, \,\,\,\,\,\,\,\,\,\,\,\,\,\,\, \bs c_X = (\underbrace{0,0,\ldots,0}_{I},\underbrace{1,1,\ldots,1}_{I})^T.
\end{align}
The self- and mutual-\obs{} are defined by inner products of the adjacency matrix with the constraint vectors $\bs c_Z, \bs c_X$.
We have:
\begin{align}
    \bs c_Z^T \bs A \bs c_Z & = \bs c_X^T \bs A \bs c_X = 0 \\
    \bs c_X^T \bs A \bs c_Z & = \bs c_Z^T \bs A \bs c_X = 1+x.
\end{align}
We see that the inner products with the constraint vectors are exactly equal to the those of the (01)-boundary, despite the fact that the ($I$1)-boundary operator algebra is quite different.
The equality of the self- and mutual-\obs{} directly follows.
We note that, in principle, the inner products above could have differed between boundaries as long as the \obs{} remained the same.

\section{Boundaries of higher-dimensional toric codes}\label{sec:higherdimtoriccode}

In this Appendix, we study the boundaries of 3D and 4D toric codes. We show that the obstruction to a LTPS is characterized by first derivatives of the boundary adjacency matrix, in contrast to the higher derivatives found for Type-I fracton models.
\subsection{3D Toric Code}
The stabilizers of the 3D toric code can be written as
\begin{align}
 \bs \Sigma=  \begin{pmatrix}
1+\bar x&0&0&0\\
1+\bar y&0&0&0\\
1+\bar z&0&0&0\\
0&0 &1+z & 1+y \\
0&1+z& 0 & 1+x\\
0&1+y& 1+x &0 
   \end{pmatrix}.
\end{align}
The excitations of the 3D toric code consist of point charges $e$ and flux loops $m$. The charges obey a global ($0$-form) conservation law that constrains the total number of charges to be even, while the fluxes obey a $1$-form conservation law that constrains them to form closed loops. Algebraically, these bulk conservation laws can be represented as elements of the kernel of $\bs \Sigma$ (that is, vectors that are zero after multiplication by $\bs \Sigma$).
We have
\begin{align}
 \bs \Sigma \cdot
\begin{pmatrix}
\sum_{ijk}x^iy^jz^k&0&0&0&0\\
0&\sum_{jk} y^jz^k&0&0&1+x\\
0&0&\sum_{ik}x^iz^k&0&1+y\\
0&0 &0 &\sum_{ij} x^iy^j&1+z
   \end{pmatrix} = 0.
\end{align}
The kernel corresponds to the span of the above columns. The final column is a local conservation law enforcing that the product of vertex operators is the identity. The truncated operators on the (001)-boundary are
\begin{align}
 \tbs \Sigma=  \begin{pmatrix}
1+\bar x&0&0\\
1+\bar y&0&0\\
1&0&0\\
0&0 &1 \\
0&1& 0 \\
0&0& 0 
   \end{pmatrix},
\end{align}
where the final stabilizer does not contribute a truncated boundary operator because it lies either entirely within or  out of the bulk. The adjacency matrix is
\begin{align}
     \bs A &=\begin{pmatrix}
0&1+y&1+x\\
1+\bar y&0&0\\
1+\bar x&0&0
   \end{pmatrix}.
\end{align}

The bulk conservation laws impose the following constraints on the boundary
\begin{align}
\tbs \Sigma \cdot
 \begin{pmatrix}
\sum_{ij} x^iy^j&0&0&0\\
0&\sum_{j} y^j&0&1+x\\
0&0&\sum_{i} x^i&1+y
   \end{pmatrix} \eqgs 0 
   \end{align}
 The last constraint is local since the two truncated boundary vertex operators product to a bulk vertex stabilizer. (We can view this as a gauge constraint, i.e. the subspace where the last constraint holds is the gauge-invariant subspace). 
 The first constraint is global. 
 The second and third constraints are one-dimensional constraints around cycles of the torus.
 However, they are not subsystem symmetry constraints, as they are deformable by applying the local gauge constraint. Therefore, these are 1-form constraints. The algebra of these operators can thus be realized in the 1-form symmetric sector of a $\ZZ_2$ gauge theory where the first constraint is automatically satisfied. Alternatively, via the Kramers-Wannier duality, it can also be realized via the $\ZZ_2$ symmetric sector of the 2D transverse-field Ising model, where the 1-form and local constraints are automatically satisfied.

The mutual-\ob{} on the boundary corresponds to the commutation of 2D patch operator $\hat{P}_m$ for the first boundary constraint (which can be viewed as creating an $m$ flux loop at its boundary) and a 1D patch operator $\hat{P}_e$ for the second or third boundary constraint (which can be viewed as creating  $e$ excitations at its ends).
The two patch operators anti-commute whenever one endpoint of the 1D patch operator lies within the 2D patch.
Let us take
\begin{equation}
\begin{split}
    \bs P_m  & = \sum_{i=-\infty}^{0} \sum_{j=-\infty}^{\infty} x^i y^j \tbs \sigma_1,  \\
    \bs P_e  & = \sum_{i=-A}^{\infty} x^i y^0 \tbs \sigma_3,
\end{split}
\end{equation}
for concreteness, where $A > K$.
Their commutator is
\begin{equation}
\begin{split}
    \langle \bs P_m, \bs P_e \rangle_{0,0} & = \left[ \sum_{i'=-A}^{\infty} \sum_{i=0}^{\infty} \sum_{j=-\infty}^{\infty} x^{i+i'} y^j \, \bs c_1^\dagger \bs A \bs c_3 \right]_{0,0} \\
    & = \left[ x^{-A} \sum_{i=0}^{\infty} \sum_{j=-\infty}^{\infty} (i)  x^{i} y^j \, \bs c_1^\dagger \bs A \bs c_3 \right]_{0,0} \\
    & = D_x[\bs c_1^\dagger \bs A \bs c_3](1,1) \\
    & = 1,
\end{split}
\end{equation}
where throughout we take the $x^0 y^0$ component (i.e.~no relative translation between the patch operators as defined). As anticipated, the patch operators anti-commute.
As in the examples in the main text, this anti-commutation can be related to the braiding of a point charge $e$ with a flux loop $m$ in the bulk.

\subsection{Fermionic 3D toric code}

We can perform a similar analysis for the fermionic 3D toric code, where the $e$-quasiparticle is an emergent fermion\cite{LevinWen2003,ChenKapustin2019,WalkerWang2012}. We analyze the Walker-Wang model,
\begin{align}
 \bs \Sigma=  \begin{pmatrix}
1+\bar x&\bar x + yz&0&0\\
1+\bar y&0&\bar y + xz&0\\
1+\bar z&0&0&\bar z + xy\\
0&0 &1+z & 1+y \\
0&1+z& 0 & 1+x\\
0&1+y& 1+x &0 
   \end{pmatrix}.
\end{align}
The conservation laws are
\begin{align}
\sum_{ijk} \bs \sigma_1x^iy^jz^k=\sum_{jk} (\bs \sigma_1+\bs \sigma_2)y^jz^k=\sum_{ik}  (\bs \sigma_1+\bs \sigma_3)x^iz^k =\sum_{ij}  (\bs \sigma_1+\bs \sigma_4)x^iy^j =0,
\end{align}
along with the local equalities,
\begin{equation}
    \bs \sigma_1 (1+xyz) +\bs \sigma_2(1+x) +\bs \sigma_3 (1+y)+\bs \sigma_4(1+z)=0.
\end{equation}

The truncated operators on the (001)-boundary are
\begin{align}
 \tbs \Sigma=  \begin{pmatrix}
1+\bar x&y&0&0\\
1+\bar y&0&x&0\\
1&0&0&xy\\
0&0 &1 &1+y \\
0&1& 0 &1+x\\
0&0& 0 &0 
   \end{pmatrix}.
\end{align}
The fourth column can be dropped, since
$\tbs \sigma_4 = xy \tbs \sigma_1  + (1+x) \tbs \sigma_2 + (1+y) \tbs \sigma_3$. To simplify upcoming computations, we perform a column operation to define a new set of generators for the boundary operators, via $\tbs \sigma_2 \rightarrow \tbs \sigma_2 + \tbs \sigma_1 $ and  $\tbs \sigma_3 \rightarrow \tbs \sigma_3 + \tbs \sigma_1$. This leads to the boundary operators
\begin{align}
 \tbs \Sigma=  \begin{pmatrix}
1+\bar x&1+\bar x+y&1+\bar x\\
1+\bar y&1+\bar y&1+\bar y+x\\
1&1&1\\
0&0 &1 \\
0&1& 0 \\
0&0& 0 
   \end{pmatrix}.
\end{align}

The conservation laws impose the same constraints on the boundary as in the 3D toric code:
\begin{align}
\tbs \Sigma \cdot
 \begin{pmatrix}
\sum_{ij} x^iy^j&0&0&0\\
0&\sum_{j} y^j&0&1+x\\
0&0&\sum_{i} x^i&1+y
   \end{pmatrix} \eqgs 0
\end{align}
However, the algebra of the operators are different. 
They have an adjacency matrix
\begin{align}
   \bs A = \begin{pmatrix}
0 & 1+y & 1+x \\
1 + \bar y & y + \bar y & 0 \\
1 + \bar x& 0 &x+ \bar x
   \end{pmatrix}.
\end{align}
This corresponds to the algebra of fermion bilinears on a square lattice, or equivalently, a flux-attached gauge theory with an anomalous $\ZZ_2$ 1-form symmetry constraint \cite{ChenKapustinRadicevic2018,Tantivasadakarn20}.  Indeed, this is natural from the viewpoint that they correspond to different input categories for the Walker-Wang model.

Let us demonstrate that the boundary of the fermionic 3D toric code is distinct from the ordinary 3D toric code. In the bulk, the two models differ in the self statistics of the $e$ particle. 
On the boundary, we thus turn to the self-\ob{} of the 1D constraints (which can be viewed as creating $e$ excitations at their ends).
We define the 1D patch operators
\begin{equation}
    \bs L_e = \sum_{i=-\infty}^0 x^i \tbs{\sigma}_3, \,\,\,\,\,\,\,\,\,\,
    \bs R_e = \sum_{i=1}^\infty x^i \tbs{\sigma}_3.
\end{equation}
Their commutation is
\begin{equation}
\begin{split}
    \Inner{\bs L_e, \bs R_e} & = \sum_{i=0}^\infty \sum_{i'=1}^\infty x^{i+i'} \Inner{\tbs{\sigma}_3,\tbs{\sigma}_3} \\
    & = \sum_{k=0}^\infty k \cdot x^k  \bs c_3^\dagger \bs{A} \bs c_3. \\
\end{split}
\end{equation}
Decomposing $\bs c_3^\dagger \bs{A} \bs c_3 = a^+(x,y) - a^+(\bar x, \bar y)$ with $a^+(x,y) = 1+x$, we have
\begin{equation}
\begin{split}
    \Inner{\bs L_e, \bs R_e}_{0,0} & =  D_x[a^+(x,y)](1,1) = 1. \\
\end{split}
\end{equation}
The two patch operators anti-commute, as expected since the bulk $e$ excitation is now a fermion.
The analogous patch operators clearly commute in the 3D toric code since the diagonal elements of the adjacency matrix are zero.

\subsection{(1,3) 4D toric code}
The $(1,3)$ toric code has stabilizers
\begin{align}
 \bs \Sigma=  \begin{pmatrix}
1+\bar x&0&0&0&0&0&0\\
1+\bar y&0&0&0&0&0&0\\
1+\bar z&0&0&0&0&0&0\\
1+\bar w&0&0&0&0&0&0\\
0&0 &1+z & 1+y&1+w&0&0 \\
0&1+z& 0 & 1+x&0&1+w&0\\
0&1+y& 1+x &0 &0&0&1+w\\
0&0&0&0&1+x&1+y&1+z
   \end{pmatrix}.
\end{align}
Here, the $e$-quasiparticle is a point charge with 0-form conservation law, and $m$ is a membrane with a $2$-form conservation law. Algebraically, these are
\begin{align}
   \bs \Sigma \cdot \begin{pmatrix}
\sum_{ijkl} x^iy^jz^kw^l&0&0&0&0&0&0\\
0&\sum_{jk}y^jz^k&0&0&0&0&0\\
0&0&\sum_{ik}x^iz^k&0&0&0&0\\
0&0&0&\sum_{ij}x^iy^l&0&0&0\\
0&0&0&0&\sum_{il}x^iw^l&0&0&\\
0&0&0&0&0&\sum_{jl}y^jw^l&0\\
0&0&0&0&0&0&\sum_{kl}z^kw^l
   \end{pmatrix} = 0.
\end{align}
We also have the local conservation laws,
\begin{align}
   \bs \Sigma \cdot \begin{pmatrix}
0&0&0&0\\
1+x&0&0&1+w\\
1+y&0&1+w&0\\
1+z&1+w&0&0\\
0&1+y&1+z&0\\
0&1+x&0&1+z\\
0&0&1+x&1+y
   \end{pmatrix} = 0.
   \end{align}
Taking the $(0001)$-boundary, the truncated operators are
 \begin{align}
 \tbs \Sigma=  \begin{pmatrix}
1+\bar x&0&0&0\\
1+\bar y&0&0&0\\
1+\bar z&0&0&0\\
1&0&0&0\\
0&1&0&0 \\
0&0&1&0\\
0&0&0&1\\
0&0&0&0
   \end{pmatrix}.
\end{align}
where we have removed columns 2,3, and 4 since they live entirely in the bulk.

The constraints on the boundary induced from the bulk conservation laws are
\begin{align}
\tbs \Sigma \cdot
 \begin{pmatrix}
\sum_{ijk} x^iy^jz^k&0&0&0&0&0&0\\
0&\sum_{i} x^i&0&0&1+y & 1+z &0\\
0&0&\sum_{j} y^j&0&1+x& 0 & 1+z\\
0&0&0&\sum_{k} z^k&0&1+x&1+y\\
   \end{pmatrix} \eqgs 0
\end{align}
The adjacency matrix is
 \begin{align}
 \bs A =  \begin{pmatrix}
0&1+x&1+y&1+z\\
1+\bar x&0&0&0\\
1+\bar y&0&0&0\\
1+\bar z&0&0&0
   \end{pmatrix}.
\end{align}
The above operator algebra is equivalent to a 2-form $\ZZ_2$ gauge theory with 2-form symmetry constraints, or equivalently the symmetric sector of the 3D transverse-field Ising model. 
The mutual-\obs{} corresponds to the commutator of a 3D patch operator for the first constraint with the 1D patch operators for the second through fourth constraints.
A straightforward calculation shows that these are equal to the derivatives
\begin{align}
  D_x[\bs{c}_1^\dagger \bs{A} \bs{c}_2](1,1,1)&=1, & D_y[\bs{c}_2^\dagger \bs{A} \bs{c}_3](1,1,1)&=1, &D_z[\bs{c}_1^\dagger \bs{A} \bs{c}_4](1,1,1)&=1.
\end{align}
Similar to before, the mutual-\obs{} can be related to braiding bulk $e$ particle with a bulk $m$ membrane.

\subsection{(2,2) 4D Toric Code}
The $(2,2)$ toric code has stabilizers
\begin{align}
\bs \Sigma =
\begin{pmatrix}
1+\bar y & 1+\bar x & 0 & 0 & 0 & 0 & 0 & 0 \\
1+\bar w & 0 & 0 & 1+\bar x & 0 & 0 & 0 & 0 \\
1+\bar z & 0 & 1+\bar x & 0 & 0 & 0 & 0 & 0 \\
 0 & 1+ \bar z & 1+\bar y & 0 & 0 & 0 & 0 & 0 \\
 0 & 1+\bar w& 0 &1+ \bar y& 0 & 0 & 0 & 0 \\
 0 & 0 & 1+\bar w & 1+ \bar z & 0 & 0 & 0 & 0 \\
 0 & 0 & 0 & 0 & 0 & 0 & 1+w & 1+z \\
 0 & 0 & 0 & 0 & 0 &1+z& 1+y & 0 \\
 0 & 0 & 0 & 0 & 0 & 1+w & 0 &1+y \\
 0 & 0 & 0 & 0 & 1+w & 0 & 0 &1+x \\
 0 & 0 & 0 & 0 & 1+z & 0 & 1+x& 0 \\
 0 & 0 & 0 & 0 & 1+y & 1+ x & 0 & 0 \\
\end{pmatrix}.
\end{align}
Both charge and flux excitations are loop operators and obey the following 1-form conservation laws,
\begin{align}
   \bs \Sigma \cdot \begin{pmatrix}
\sum_{jkl} y^jz^kw^l&0&0&0&1+\bar x\\
0&\sum_{ikl} x^iz^kw^l&0&0&1+\bar y\\
0&0&\sum_{ijl} x^iy^jw^l&0&1+\bar z\\
0&0&0&\sum_{ijk} x^iy^jz^k&1+\bar w\\
0&0&0&0&0\\
0&0&0&0&0\\
0&0&0&0&0\\
0&0&0&0&0
   \end{pmatrix}  = 0,\\
     \bs \Sigma \cdot  \begin{pmatrix}
   0&0&0&0&0\\
0&0&0&0&0\\
0&0&0&0&0\\
0&0&0&0&0\\
\sum_{jkl} y^jz^kw^l&0&0&0&1+x\\
0&\sum_{ikl} x^iz^kw^l&0&0&1+y\\
0&0&\sum_{ijl} x^iy^jw^l&0&1+z\\
0&0&0&\sum_{ijk} x^iy^jz^k&1+w
 \end{pmatrix} = 0.
\end{align}
 
The truncated operators on the $(0001)$-boundary are
 \begin{align}
\tbs \Sigma =
\begin{pmatrix}
1+\bar y & 1+\bar x & 0 & 0  & 0   & 0\\
1 & 0 & 0  & 0 & 0  & 0 \\
1+\bar z & 0 & 1+\bar x& 0  & 0 & 0 \\
 0 & 1+ \bar z & 1+\bar y  & 0 & 0  & 0 \\
 0 & 1& 0& 0 & 0  & 0 \\
 0 & 0 & 1   & 0 & 0  & 0 \\
 0 & 0 & 0 &  0 & 0 & 1&  \\
 0 & 0 & 0 &  0 &0& 0  \\
 0 & 0 & 0 &  0 & 1 & 0  \\
 0 & 0 & 0 &  1 & 0 & 0  \\
 0 & 0 & 0 &  0 & 0 & 0 \\
 0 & 0 & 0 &  0 & 0 & 0 \\
\end{pmatrix}
\end{align}
where we have thrown away the $4^\text{th}$ and $8^\text{th}$ columns.

The constraints induced on the boundary are
\begin{align}
\tbs \Sigma \cdot
 \begin{pmatrix}
\sum_{jk} y^jz^k&0&0&0&0&0&1+\bar x&0\\
0&\sum_{ik} x^iz^k&0&0&0&0&1+\bar y&0\\
0&0&\sum_{ij} x^iy^j&0&0&0&1+\bar z&0\\
0&0&0&\sum_{jk} y^jz^k&0&0&0&1+x\\
0&0&0&0&\sum_{ik} x^iz^k&0&0&1+y\\
0&0&0&0&0&\sum_{ij} x^iy^j&0&1+z\\
   \end{pmatrix} \eqgs 0.
   \label{eq:4DTCconstraints}
\end{align}
 The adjacency matrix is
\begin{align}
    \bs A = \begin{pmatrix}
    \bs 0 & \bs A_{ZX}\\
    -\bs A_{ZX}^\dagger & \bs 0
    \end{pmatrix}, \,\,\,\,\,\,\,\,\,\,\,\,\,\,\,\, &
    \bs A_{ZX} = \begin{pmatrix}
 0 & 1+z & 1+y \\
 1+z & 0 & 1+x\\
1+y& 1+x & 0 
    \end{pmatrix}.
\end{align}
The adjacency matrix gives rise to the frustration graph of the RBH lattice\cite{RaussendorfBravyiHarrington05}.
The operator algebra is equivalent to a $\ZZ_2$ gauge theory with 1-form symmetry constraints in 3D.

The mutual-\obs{} correspond to the commutation of 2D patch operators for the first through third constraints with 2D patch operators for the fourth through sixth constraints.
The patch operators anti-commute whenever their boundaries are linked.
To see this explicitly, consider two patch operators $\hat{P}_m$ in the $xz$ plane and $\hat{P}_e$ in the $xy-$ plane, constructed from the second and sixth constraints in Eq.~\ref{eq:4DTCconstraints} respectively,
\begin{equation}
\begin{split}
    \bs P_m & = \sum_{i=-\infty}^{0} \sum_{k=-\infty}^{\infty} x^i z^k \tbs \sigma_2, \\
    \bs P_e & = \sum_{i=-A}^{\infty} \sum_{j=-\infty}^{\infty} x^i y^j \tbs \sigma_6, \\
\end{split}
\end{equation}
where $A > K$ (here $K=2$).
The two patch operators are linked because boundary of the first patch operator pierces that of the second patch operator at a single point.
A straightforward evaluation of their commutator gives
\begin{equation}
    \langle \bs P_m , \bs P_e \rangle_{0,0} = D_x [ \bs c_2^\dagger \bs A \bs c_6 ](1,1,1) = 1.
\end{equation}
This corresponds to braiding a bulk $e$ membrane with a bulk $m$ membrane.

Let us also compute the self statistics of the dyonic excitation $em$ We take the patch operator
\begin{align}
    \bs L_{em} &= \sum_{i=-\infty}^0\sum_{j=-\infty}^\infty x^iy^j (\tbs \sigma_3 +\tbs \sigma_6),  & \bs R_{em} &= \sum_{i=0}^\infty \sum_{j=-\infty}^\infty x^iy^j (\tbs \sigma_3 +\tbs \sigma_6)
\end{align}
The commutator if these two operator evaluates to
\begin{align}
    \langle \bs L_{em} , \bs R_{em} \rangle_{0,0} = D_x [ \bs c_{36}^\dagger \bs A^{+} \bs c_{36} ](1,1,1) = 0.
\end{align}
where $\bs A^+ =   \begin{pmatrix}
    \bs 0 & \bs A_{ZX}\\
    \bs 0 & \bs 0
    \end{pmatrix}$ and $\bs c_{36}^\dagger = (0,0,1,0,0,1)$. Indeed, it has been observed that the dyonic loop excitation has trivial self statistics~\cite{Chen21}.

 \section{Boundaries of Majorana fermion codes}\label{app:Majorana}
 Stabilizer codes consisting of Majorana operators can also exhibit topological order\cite{BravyiTerhalLeemhuis2010,VijayHsiehFu2015,VijayHaahFu2015}, and similar anomalies can be calculated for the boundary of such systems. The geometric proof follows identically for the algebra of Majorana operators, and similar algebraic techniques can be applied when translation invariance is assumed\cite{VijayHaahFu2015} (also see \cite{Tantivasadakarn20} for a review). The position of the Majorana operators can similarly be expressed using polynomials in $\bs \sigma$ and commutation relations can be computed using the Euclidean inner product
 \begin{align}
     \inner{\bs \sigma_1, \bs \sigma_2}_F = \bs \sigma_1^\dagger \bs \sigma_2.
 \end{align}
In this appendix, we give an example calculation for the Majorana color code\cite{VijayHsiehFu2015}, which exhibits a $\ZZ_2$ topological order. 

 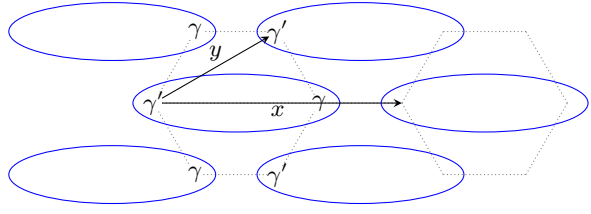
\begin{figure}
     \centering
     \begin{tikzpicture}[scale=0.55,>=stealth]
 \coordinate (0) at (0,0) {};
 \node[label=center:$\gamma$] (1) at (2,0) {};
 \node[label=center:$\gamma'$] (2) at (1,1.73) {};
 \node[label=center:$\gamma$] (3) at (-1,1.73) {};
 \node[label=center:$\gamma'$] (4) at (-2,0) {};
 \node[label=center:$\gamma$] (5) at (-1,-1.73) {};
 \node[label=center:$\gamma'$] (6) at (1,-1.73) {};
 \coordinate (7) at (4,0) {};
 \draw[-,densely dotted,color=gray] (1) -- (2) {};
 \draw[-,densely dotted,color=gray] (2) -- (3) {};
 \draw[-,densely dotted,color=gray] (3) -- (4) {};
 \draw[-,densely dotted,color=gray] (4) -- (5) {};
 \draw[-,densely dotted,color=gray] (5) -- (6) {};
 \draw[-,densely dotted,color=gray] (6) -- (1) {};
 \draw[-,densely dotted,color=gray] (4) -- (7) {};
 \draw[color=blue] (0)         ellipse (2.5 and 0.7);
 \draw[color=blue] (3,1.73)     ellipse (2.5 and 0.7);
 \draw[color=blue] (-3,1.73)     ellipse (2.5 and 0.7);
 \draw[color=blue] (-3,-1.73)     ellipse (2.5 and 0.7);
 \draw[color=blue] (3,-1.73)     ellipse (2.5 and 0.7);
 \draw[->] (4) -- (2);
 \draw[->] (4) -- (7);
 \node[label=center:$x$]  at (1,-0.2) {};
 \node[label=center:$y$]  at (-0.5,1.2) {};
 \coordinate (11) at (2+6,0) {};
 \coordinate (12) at (1+6,1.73) {};
 \coordinate (13) at (-1+6,1.73) {};
 \coordinate (14) at (-2+6,0) {};
 \coordinate (15) at (-1+6,-1.73) {};
 \coordinate (16) at (1+6,-1.73) {};
 \coordinate (17) at (4+6,0) {};
 \draw[-,densely dotted,color=gray] (11) -- (12) {};
 \draw[-,densely dotted,color=gray] (12) -- (13) {};
 \draw[-,densely dotted,color=gray] (13) -- (14) {};
 \draw[-,densely dotted,color=gray] (14) -- (15) {};
 \draw[-,densely dotted,color=gray] (15) -- (16) {};
 \draw[-,densely dotted,color=gray] (16) -- (11) {};
 \draw[color=blue] (6,0)     ellipse (2.5 and 0.7);
 \end{tikzpicture}
     \caption{Stabilizer of the Majorana color code, with the unit cell and choice of coordinates indicated.}
     \label{fig:MCC}
 \end{figure}

 A Majorana is placed on each vertex of the honeycomb lattice. The stabilizers consist of a product of six Majoranas around each hexagon. To represent this algebraically, we choose a unit cell consisting of two Majoranas denoted $\gamma$ and $\gamma'$, and choose the translation vectors as shown in Fig. \ref{fig:MCC}. The stabilizer is given by
 \begin{align}
     \bs \Sigma = \begin{pmatrix}
 1+\bar x y + \bar y  \\
 1+x \bar y +  y
 \end{pmatrix}
 \end{align}
 where the first and second row denotes the position of $\gamma$ and $\gamma'$ respectively. 

 The bulk has two global conservation laws given by
 \begin{align}
 \sum_{ij} \bs \Sigma (1+x\bar y) x^iy^{3j} =\sum_{ij} \bs \Sigma (1+y) x^iy^{3j}=0
 \end{align}
 which can be given the following geometric interpretation. The plaquettes of the honeycomb lattice can be three-colored so that adjacent plaquettes have different colors. Doing so, we notice that the product of all the stabilizers on a given color are all equal. Therefore, the conservation law is given by performing a product of stabilizers over two of the three colors.

 Now, let us truncate the operators along the (01)-boundary. Translating all terms so that the largest degree of $y$ is $y^0$, we have
 \begin{align}
     \bs \Sigma = \begin{pmatrix}
 \bar y+\bar x  + \bar y^2  \\
 \bar y+x \bar y^2 +  1
 \end{pmatrix}
 \end{align}
 Therefore, there are two truncated terms given by
 \begin{align}
     \tbs \Sigma = \begin{pmatrix}
 \bar x & 1+\bar x y\\
 1&1 + y
 \end{pmatrix}
 \end{align}
 Note that the remaining $y$ terms no longer correspond to translations. we find
 \begin{align}
   \inner{ \tbs \sigma,\tbs \sigma}_F = \begin{pmatrix}
 0 &1+\bar x\\
 1+x &0
 \end{pmatrix}
 \end{align}
 The bulk conservation laws induce the following constraints
 \begin{align}
     \sum_i x^i \tbs \sigma_1 = \sum_i x^i \tbs \sigma_2=0
 \end{align}
 which is analogous toric code constraints. It therefore follows that we obtain the same self and mutual-obstructor invariants.
 
 We note that the algebra of the boundary operators can be obtained by restricting to the fermion parity even subspace. Thus, we expect this subspace to have the same constraints as the boundary of the toric code since the parity even fermionic operators can be related to $\ZZ_2$ symmetric Pauli operators via the Jordan-Wigner transformation. This also agrees with the fact that the ground state of this model is actually equivalent to that of the $\ZZ_2$ (bosonic) toric code tensored with trivial fermionic degrees of freedom \cite{WangShirleyChen2019}.

 \section{Boundary of the double-semion model}\label{app:DS}

   \begin{figure}
     \centering
     \includegraphics[width=0.8\columnwidth]{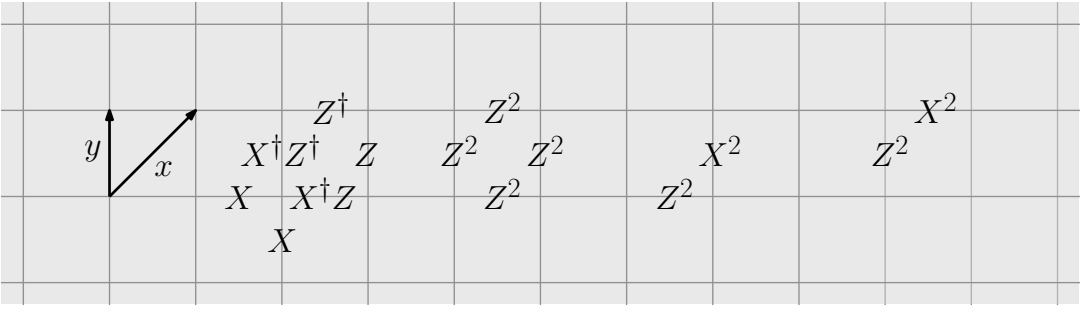}
     \caption{Choice of coordinates and stabilizers for the double-semion model.}
     \label{fig:DS}
 \end{figure}

 As a final interesting example, we apply our framework to the boundary of the $\mathbb Z_4$ double-semion model introduced in Ref.~\cite{Ellison22}. For convenience, we have chosen $x$ and $y$ to correspond to the vectors $(1,1)$ and $(0,1)$ respectively.  The stabilizers are shown in Fig.~\ref{fig:DS}. Algebraically, they can be written as
 \begin{align}
     \bs \Sigma = \begin{pmatrix}
1-y & 2(1-y) & 2 & 0\\
1-\bar x y & 2(1-\bar x y) & 0 & 2\bar x\\
-1+\bar x y &0&0&2\\
\bar x(1-y)& 0 & 2 &0
 \end{pmatrix}.
 \end{align}
 The bulk conservation laws are given by
 \begin{align}
     \bs \Sigma \cdot \begin{pmatrix}
\sum_{ij}x^iy^j&0&2\\
0&\sum_{ij}x^iy^j&1+\bar x\\
0&0& \bar x(1+y)\\
0&0&1+\bar x y
 \end{pmatrix} = 0.
 \end{align}
 The third column is a \emph{local} conservation law that encodes the fact that, when squared, the first stabilizer can be written as a product of the other three stabilizers.

We now turn the boundary operator algebra, focusing on the $(1,-1)$-boundary for convenience.
To truncate on this boundary, we consider translations of the bulk stabilizers along the $y$ coordinate and set $\bar y=0$. This gives boundary operators
 \begin{align}
     \tbs \Sigma = \begin{pmatrix}
-1 & 2  \\
-\bar x & 2\bar x\\
\bar x &0\\
-\bar x& 0 
 \end{pmatrix}.
 \end{align}
 The bulk conservation laws lead to the boundary constraints,
  \begin{align}
     \tbs \Sigma \cdot \begin{pmatrix}
\sum_{i}x^i&0\\
0&\sum_{i}x^i\\
0&0\\
0&0
 \end{pmatrix}=0.
 \end{align}

 The adjacency matrix is given by
 \begin{align}
\bs A = \begin{pmatrix}
x-\bar x & 2(1+x)\\
2(1-\bar x) & 0
 \end{pmatrix}
 \end{align}
 From this, we find the mutual-\obs{},
 \begin{align}
     \tbs b = D_x \bs A (1) = \begin{pmatrix}
2 & 2\\
2 & 0
 \end{pmatrix},
 \end{align}
and the self-\obs{},
\begin{align}
     \tilde q((1,0)) &=1,&\tilde q((0,1)) &=0, &\tilde q((1,1)) &=-1,
\end{align}
which correspond to the semion, boson, and anti-semion, respectively.

%\nolinenumbers

\end{document}